\documentclass[journal]{vgtc}        

\usepackage{mathptmx}
\usepackage{graphicx}
\usepackage{times}
\usepackage{algorithm}
\usepackage{algorithmic}
\usepackage{amsfonts}
\usepackage{booktabs}         
\usepackage{gensymb}
\usepackage{amsmath}
\usepackage{comment}
\usepackage{times} 
\usepackage{caption}
\usepackage{bm}
\usepackage{multirow}

\usepackage{color} 
\usepackage{anyfontsize}
\usepackage{soul}

\DeclareMathOperator*{\smooth}{smooth}
\DeclareMathOperator*{\gt}{GT}

\DeclareMathAlphabet{\altmathcal}{OMS}{cmsy}{m}{n}
\newenvironment{myitemize}{
\begin{itemize}
 \setlength{\itemsep}{1pt}
 \setlength{\parskip}{0pt}
 \setlength{\parsep}{0pt}}{\end{itemize}
}

\onlineid{9586}

\vgtccategory{IEEE PacificVis Journal Track}

\title{iVR-GS: Inverse Volume Rendering for Explorable Visualization via Editable 3D Gaussian Splatting}

\author{%
\authororcid{Kaiyuan Tang}{0009-0001-3512-0112},
\authororcid{Siyuan Yao}{0000-0002-4093-193X}, and
  \authororcid{Chaoli Wang}{0000-0002-0859-3619}
}
\authorfooter{
\item 
K.\ Tang, S.\ Yao, and C.\ Wang are with 
the Department of Computer Science and Engineering, University of Notre Dame, Notre Dame, IN, USA.
E-mail: \{ktang2, syao2, chaoli.wang\}@nd.edu.
}

\abstract{
In volume visualization, users can interactively explore the three-dimensional data by specifying color and opacity mappings in the transfer function (TF) or adjusting lighting parameters, facilitating meaningful interpretation of the underlying structure. However, rendering large-scale volumes demands powerful GPUs and high-speed memory access for real-time performance. While existing novel view synthesis (NVS) methods offer faster rendering speeds with lower hardware requirements, the visible parts of a reconstructed scene are fixed and constrained by preset TF settings, significantly limiting user exploration. This paper introduces inverse volume rendering via Gaussian splatting (iVR-GS), an innovative NVS method that reduces the rendering cost while enabling scene editing for interactive volume exploration. Specifically, we compose multiple iVR-GS models associated with basic TFs covering disjoint visible parts to make the entire volumetric scene visible. Each basic model contains a collection of 3D editable Gaussians, where each Gaussian is a 3D spatial point that supports real-time scene rendering and editing. We demonstrate the superior reconstruction quality and composability of iVR-GS against other NVS solutions (Plenoxels, CCNeRF, and base 3DGS) on various volume datasets. The code is available at \url{https://github.com/TouKaienn/iVR-GS}.
} 


\CCScatlist{ 
 \CCScat{K.6.1}{Management of Computing and Information Systems}%
{Project and People Management}{Life Cycle};
 \CCScat{K.7.m}{The Computing Profession}{Miscellaneous}{Ethics}
}

\begin{document}
\firstsection{Introduction}
\maketitle
Direct volume rendering (DVR) is essential for volume visualization (VolVis). 
By assigning colors and opacities to different voxel values via a transfer function (TF), DVR can generate high-quality visualization images that help users explore the underlying data, identify unique features, and discover new patterns.
Although effective, rendering large volumes requires advanced GPU and fast memory access for real-time performance.
Due to the stringent hardware requirements, large volume data often need to be rendered on high-end remote workstations or clusters, limiting user exploration on low-end local machines.

The recent advance of {\em novel view synthesis} (NVS) techniques can be promising for overcoming the resource limitation of DVR on large volume visualization.
In visualization, NVS falls into the category of generative AI for visualization, which includes {\em data generation} and {\em visualization generation}~\cite{Wang-TVCG23}. 
Data generation has been well investigated for scalar fields~\cite{Han-VIS20, Yao-CG23, Han-VI22, Han-TVCG23, Tang-CG24, Tang-PVIS24} and vector fields~\cite{Gu-PVIS22, Han-CGA19, Gu-CGA21, Guo-PVIS20, Han-CG22} across various tasks such as super-resolution generation, data translation, reconstruction, and completion. 
For visualization generation, researchers have designed various neural networks for synthesizing novel visualizations that bypass the conventional rendering pipeline~\cite{Berger-TVCG19, Hong-DNN-VolVis, He-InsituNet, Han-TVCG23}. 
Similarly, an NVS model can output images from unseen viewpoints after optimization by taking sparsely sampled multi-view images as input.
Furthermore, the low training cost and efficient inference of current NVS models make them a promising choice for reducing the cost of DVR when rendering large-scale volumes on low-end devices. 
For instance, Niedermayr et al.\ \cite{Niedermayr-arxiv24} designed a compact format of {\em 3D Gaussian splatting} (3DGS)~\cite{Kerbl-TOG23} that achieves real-time rendering of 3D anatomy structure on mobile devices through NVS.

Despite their effectiveness in reducing rendering costs, current NVS methods have several limitations that restrict their application to the VolVis scene. 
First, these methods are constrained by the visible parts of a preset TF used to generate multi-view training images.
After training, NVS models cannot represent the value ranges with zero opacity in the preset TF.
Meanwhile, forcing all value ranges of the preset TF visible can cause severe occlusion, resulting in poor visualization outcomes.
Second, the light (i.e., angle and magnitude) and TF (i.e., color and opacity) settings are baked into the model representations and can not be changed during inference, which limits interactive volume exploration.
Third, existing NVS models either offer fast rendering speeds but incur large model sizes or have compact models that suffer from slow rendering speeds.
An ideal NVS model should render in real-time but remain relatively lightweight to achieve optimal performance.

To address these limitations, we present iVR-GS, \underline{\bf i}nverse \underline{\bf v}olume \underline{\bf r}endering via editable 3D \underline{\bf G}aussian \underline{\bf s}platting.
We design a solution to overcome the restriction to the visible parts of a single TF by leveraging multiple basic NVS models, each optimized on a basic scene under a disjoint basic TF. 
Such models exhibit {\em composability}—the ability to accurately compose multiple basic scenes by composing the parameters of basic models.
By composing basic models into a composed model, we make all basic scenes visible in one model without extra optimization.
We build upon the {\em point-based} 3DGS framework and leverage Gaussian primitives to represent scenes, achieving superior composability compared to other {\em neural radiance field} (NeRF) methods and enabling real-time rendering.
Like other NVS methods, the original Gaussian primitive will bake the light and TF settings into its representation.
To enable scene editing, we extend the existing Gaussian structure into the {\em editable Gaussian} primitive by introducing additional attributes with extra objective functions and incorporating the Blinn-Phong reflection model to decompose the lighting of a basic scene.
Using numerous editable Gaussian primitives to reconstruct a scene, iVR-GS supports {\em interactive volume exploration} of the VolVis scene by adjusting the color, opacity, and lighting parameters. 
Moreover, users can also perform {\em inverse volume exploration}: given a reference image indicating the desired effect, the composed editable Gaussian model can automatically adjust its parameters to approximate the reference image and mimic its light and TF settings.
Due to the explicit nature of Gaussian representation, the model size can become substantial after composing.
To mitigate the storage cost, we employed {\em vector quantization} (VQ) to extract a more compact model format with no significant increase in computation time or loss of accuracy.

The contributions of iVR-GS can be summarized as follows.
First, our work overcomes the existing limitations of NVS methods on exploring VolVis scenes by utilizing composable and editable Gaussian primitives to represent scenes.
Second, we demonstrate the superior reconstruction quality and composability of iVR-GS compared with state-of-the-art NeRF-based composable NVS methods (Plenoxels and CCNeRF) on various volume datasets.
Third, we showcase scene relighting, editing, and inverse volume exploration results to indicate the flexibility of iVR-GS for VolVis scene exploration.

\vspace{-0.05in}
\section{Related Work}
\vspace{-0.025in}

{\bf Image-based methods for VolVis.}
Image-based rendering and modeling methods~\cite{IBM-survey} synthesize new images based on existing rendering results.
{\em Image-based rendering} (IBR) outputs novel images without involving a geometric 3D modeling process.
In the VolVis context, IBR is essential for exploring volume datasets that are too large to store or when the original volume data is inaccessible.
For example,
Tikhonova et al.\ \cite{Tikhonova-PVIS10} proposed a novel type of explorable image for VolVis, which supports image-space user editing of color and opacity without accessing the original volume.
Later, they extended this explorable image idea to time-varying volume datasets for selecting appropriate global TFs~\cite{Tikhonova-CGF10}.
Ahrens et al.\ \cite{Ahrens-SC14} proposed an image-based approach that facilitates large-scale in situ visualization of simulation results.

Although the explorable image concept enables interactive VolVis, it is view-dependent, requiring volume data access when generating images under unseen viewpoints.
Researchers have attempted to optimize a deep network on a pre-rendered image dataset to achieve a more flexible IBR.
For instance,
surrogate models, including InsituNet~\cite{He-InsituNet}, VDLSurrogate~\cite{Shi-VDLSurrogate}, and ParamsDrag~\cite{Li-VIS24}, replace the simulation process via training their deep networks on large-scale image datasets rendered with different simulation parameters.
Unlike surrogate models, which primarily tackle simulation parameters, other works focus on visualization parameters such as viewpoint or TF.
Berger et al.\ \cite{Berger-TVCG19} proposed a GAN-based network to synthesize DVR images by optimizing a large collection of images under various TFs and viewing parameters.
Hong et al.\ \cite{Hong-DNN-VolVis} designed DNN-VolVis to generate desired DVR images using the volume data and a reference image as input.

While achieving state-of-the-art performance, these works require a long training time and expensive hardware resources due to the extensive parameter size.
In contrast, our iVR-GS gains valuable insights from Niedermayr et al.\ \cite{Niedermayr-arxiv24} to represent the less-explored image-based modeling (IBM) approach, which includes a geometry modeling process. 
Through modeling a scene's geometric information with uniquely designed Gaussian primitives, iVR-GS efficiently renders images in real-time with lower computational resources while allowing users to interactively adjust the scene's color, opacity, and lighting.

\begin{figure*}[!ht]
\centering
\includegraphics[width=\linewidth]{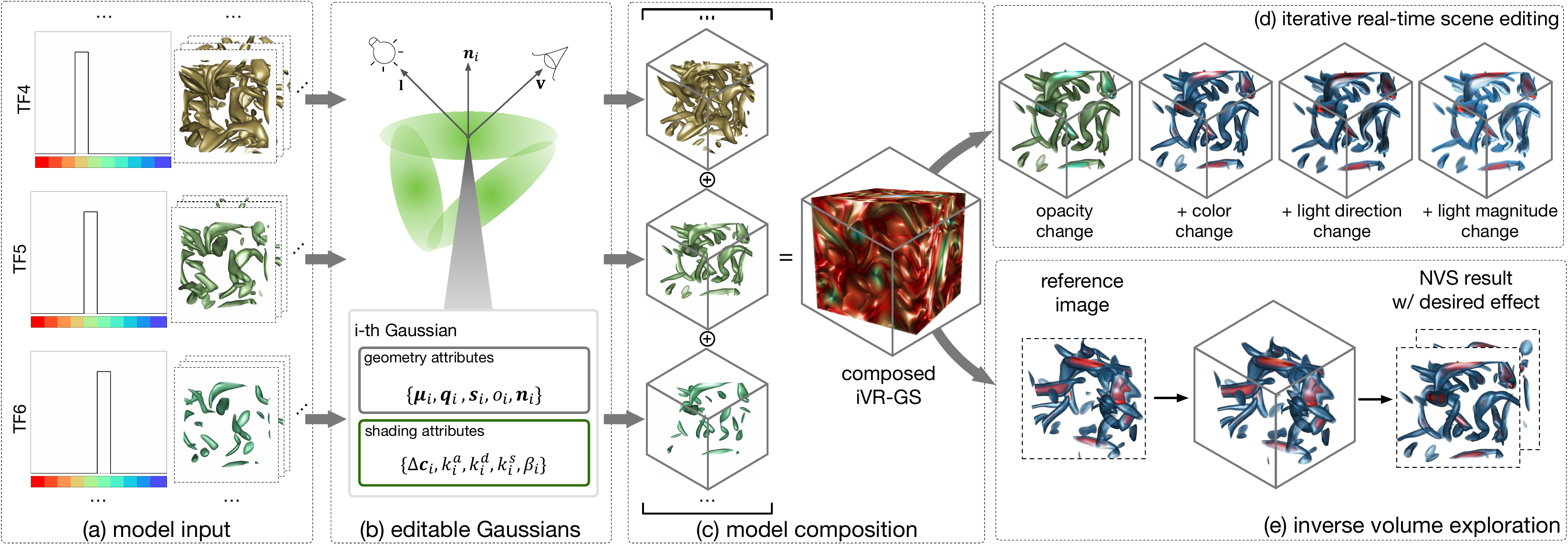}
\vspace{-.25in}
\caption{The workflow of iVR-GS. (a) For each basic scene corresponding to one basic TF with a unique visible part, we render a collection of multi-view images as the training dataset of this scene. (b) Then, we optimize a basic iVR-GS model associated with each basic scene with editable Gaussians to enable scene editing. (c) Next, we compose multiple basic iVR-GS models into the composed iVR-GS model to make the entire VolVis scene visible. The composed iVR-GS supports (d) iterative real-time scene editing and (e) inverse volume exploration.}
\label{fig:workflow}
\end{figure*}

{\bf Scene representation networks.}
When rendering a large volume, memory footprint can be a significant issue.
Recent advances in {\em scene representation networks} (SRNs) address this by optimizing small-footprint neural networks. 
SRNs provide a more compact representation of one volume, mapping input coordinates to corresponding voxel values and allowing efficient evaluation with random access.
For example, 
Weiss et al.\ \cite{Weiss-fVSRN} proposed fV-SRN, a dense-grid encoding method to support faster rendering speeds than a conventional fully connected network~\cite{Lu-CGF21}.
Wu et al.\ \cite{Wu-MHT} developed custom CUDA kernels to maximize the training speed of SRNs with multi-resolution hash encoding. 
Wurster et al.\ \cite{Wurster-APMGSRN} presented APMGSRN to fit and render large-scale data with multiple adaptive feature grids.
Although both iVR-GS and SRNs aim to reduce the hardware requirements for large-scale volume rendering, iVR-GS does not require access to the original volume data during training. Instead, it only needs access to VolVis images from sparsely sampled viewpoints, often rendered by high-performance workstations or clusters. 
Due to its low training hardware requirements, iVR-GS can be easily trained on most consumer-level GPU without depending on multi-GPU resources~\cite{Wurster-APMGSRN}, out-of-core strategies~\cite{Wu-MHT}, or access to the volume data. 

{\bf Novel view synthesis.}
Recently, NeRF~\cite{Mildenhall-NeRF} has drawn considerable attention for its effectiveness in 3D geometry modeling and high-fidelity NVS. 
The success of NeRF has resulted in numerous follow-up works~\cite{Thomas-InstantNGP,Barron-CVPR22, Barron-ICCV23,Hu-ICCV23,Liu-Siggraph24} that enhance reconstruction quality and speed. 
Some works enable the composition of multiple optimized NVS models.
For instance,
Liu et al.\ \cite{Liu-neurips20} proposed NSVF, which enables composing multiple optimized models via sharing the same {\em multilayer perceptron} (MLP).  
Fridovich-Keil et al.\ \cite{Fridovich-Keil-CVPR22} introduced Plenoxels; its explicit sparse voxel representation naturally supports the direct composition of several scenes.
Tang et al.\ \cite{Tang-neurips22} designed CCNeRF to achieve composition via concatenating model parameters. 
Lazova et al.\ \cite{Lazova-WACV23} presented Control-NeRF to support interpretable and controllable scene representation with a hybrid representation of feature volumes and a neural rendering network. 
Wang et al.\ \cite{Wang-ICMR23} developed RIP-NeRF to improve the rendering quality of fine-grained editing by leveraging a point-based radiance field representation.
Wang et al.\ \cite{Wang-arXiv24} proposed SCARF, a NeRF-based continual learning framework to learn multiple scenes incrementally.
Yuan and Zhao~\cite{Yuan-3DV24} introduced SlimmeRF, which supports instant size-accuracy trade-offs at test time. 
Despite significant progress, NeRF-based methods still suffer from low rendering speed or high training-time memory usage. 
To overcome these limitations, beyond NeRF, Kerbl et al.\ \cite{Kerbl-TOG23} presented point-based 3DGS that utilizes a set of Gaussian primitives to explicitly model the scene, enabling real-time rendering and efficient composition by composing the Gaussian primitives of different scenes. 
Our iVR-GS adopts such an architecture 
and augments the Gaussian primitives to enable editing of the scene's color, opacity, and light editing to support explorable visualization. 
For light editing, existing works~\cite{Yao-ECCV22,Zhang-ICCV23,Gao-arXiv23,Jin-CVPR23,Liang-CVPR24} primarily focus on the plausible relighting of real-world scenes based on the bidirectional reflectance distribution function (BRDF).
For example, Relightable 3DGS~\cite{Gao-arXiv23} augments 3D Gaussians with extra BRDF properties to produce plausible shadow effects in relighting. 
We also extend vanilla Gaussian attributes to support relighting. Unlike Relightable 3DGS,  we employ the Blinn-Phong reflection model to relight synthetic VolVis scenes accurately. 
One work close to our VolVis scene editing is StyleRF-VolVis~\cite{Tang-VIS24}. However, its light editing is limited to magnitude adjustment, and interactive rendering is supported with no capability for scene composition. 

{\bf Vector quantization.}
{\em Vector quantization} (VQ) is a data compression technique from signal processing that clusters large collections of high-dimensional data into a limited number of representations.
VQ has been applied in the context of deep learning for scientific visualization.
For example,
Lu et al.\ \cite{Lu-CGF21} leveraged k-means to extract a more compact representation of the MLP network. 
Tang and Wang~\cite{Tang-PVIS24} divided the volume datasets into multiple blocks, then clustered and assigned similar blocks for one local MLP to achieve efficient encoding.
Liu et al.\ \cite{Liu-JVIS24} employed k-means to reduce the parameter size of the encoding grid and MLP. 
VQ has also been utilized in NVS model compression~\cite{Li-CVPR23,Fan-arXiv23,Niedermayr-CVPR24,Papantonakis-CGIT24}.
It reduces the total file size by clustering network parameters of a NeRF model~\cite{Li-CVPR23} or attribute values of a 3DGS-based model~\cite{Fan-arXiv23,Niedermayr-CVPR24,Papantonakis-CGIT24} into a small code book.
This work uses a VQ-based compression method to ensure a compact model representation with negligible extra time cost.

\vspace{-0.05in}
\section{Preliminaries}

{\bf 3D Gaussian splatting.}
Unlike the NeRF architecture such as~\cite{Tang-VIS24}, our method builds upon 3DGS~\cite{Kerbl-TOG23}, which abandons computation-intensive network architectures and employs explicit 3D Gaussian primitives as rendering entities for faster rendering. Formally, it is defined as
\begin{equation}
	G(\mathbf{x}) = \exp\left(-\frac{1}{2}(\mathbf{x}-\bm{\mu})^\top\Sigma^{-1}(\mathbf{x}-\bm{\mu})\right),
\end{equation}
where $\mathbf{x}\in\mathbb{R}^{3\times1}$ denotes a spatial coordinate, $\bm{\mu}\in\mathbb{R}^{3\times1}$ and $\Sigma\in\mathbb{R}^{3\times3}$ represent, respectively, the spatial mean and covariance matrix. 
For the original 3DGS, each Gaussian is characterized by an opacity value $o\in[0, 1]$ and a view-dependent color $\mathbf{c}\in\mathbb{R}^{3\times1}$. To ensure meaningful interpretation during optimization, the covariance matrix $\Sigma$ is parameterized by a quaternion rotation vector $\mathbf{q}\in\mathbb{R}^{4\times1}$ and a scaling vector $\mathbf{s}\in\mathbb{R}^{3\times1}$. The view-dependent color $\mathbf{c}$ is further parameterized by {\em spherical harmonic} (SH) coefficients.

When rendering from a specific viewpoint, one can project 3D Gaussians onto the image plane as 2D Gaussians through rasterization. The mean of 2D Gaussians can be computed directly through the projection matrix. The projected 2D covariance matrices can be approximated as $\Sigma^{\prime}=\mathbf{J}\mathbf{W}\Sigma\mathbf{W}^\top\mathbf{J}^\top$, where $\mathbf{W}$ and $\mathbf{J}$ denote, respectively, the viewing transformation matrix and the Jacobian matrix of the projective transformation. 
After rasterization, the pixel color $\mathbf{C}$ is determined by summing the color and opacity of $N$ sequentially layered 2D Gaussians sorted by depth, i.e., 
\begin{equation}
	\mathbf{C}=\sum_{i\in N}T_i\alpha_i\mathbf{c}_i \quad \textrm{with} \quad T_i=\prod^{i-1}_{j=1}(1-\alpha_j),
\end{equation}
where the sampled $\alpha$ denotes the alpha value obtained by multiplying the Gaussian weight with the opacity $o$ learned per-primitive, and $T$ represents the accumulated transmittance. In summary, 3DGS utilizes a collection of 3D Gaussians to represent a scene, with the $i$-th Gaussian primitive $G_i$ parameterized with $\{\bm{\mu}_i, \mathbf{q}_i, \mathbf{s}_i, o_i, \mathbf{c}_i \}$.

{\bf Blinn-Phong reflection model.}
The Blinn-Phong reflection model is one of the most widely used lighting models for generating VolVis scenes.
In this paper, our VolVis multi-view image datasets are generated by ParaView using the NVIDIA IndeX plugin~\cite{IndeX},
which supports Blinn-Phong reflection.
As a widely used open-source renderer, NVIDIA IndeX supports one headlight (i.e., point light emitted from the camera) or orbital light (i.e., directional light specified by a user-defined angle) in the scene. 
Given a light direction $\mathbf{l}$, after Blinn-Phong shading, one sample voxel will emit color $\mathbf{c}^{\prime}$ to the camera, which can be expressed as the sum of the ambient ($\mathbf{c}_a$), diffuse ($\mathbf{c}_d$), and specular ($\mathbf{c}_s$) colors.
Specifically, after sampling voxel color $\mathbf{c}_v$ according to the voxel value and user-specified TF,
the ambient, diffuse, and specular terms are determined as
\begin{subequations}
\label{eqn:Blinn-Phong}
\begin{align}
	\mathbf{c}_a&=k_a\mathbf{I}_a, \\
	\mathbf{c}_d&=k_d\mathbf{I}_d |\mathbf{n} \cdot \mathbf{l}|,\\
	\mathbf{c}_s&=
	\begin{cases}
		k_s\mathbf{I}_s |\mathbf{n} \cdot \mathbf{h}|^{\beta},	& \text{if } |\mathbf{n} \cdot \mathbf{l}| > 0 \\
		0,		& \text{otherwise} 
	\end{cases}
\end{align}
\end{subequations}
where $(k_a, \mathbf{I}_a)$, $(k_d, \mathbf{I}_d)$, and $(k_s, \mathbf{I}_s)$ are, respectively, ambient, diffuse, and specular coefficients and light colors. 
The NVIDIA IndeX plugin assigns $\mathbf{c}_v$ to $\mathbf{I}_a$ and $\mathbf{I}_d$, and white color to $\mathbf{I}_s$. 
Additionally, $\beta$ is the shininess term. 
$\mathbf{n}$ is the normal vector determined by the gradient direction at the sample voxel position. $\mathbf{h}=\frac{\mathbf{v}+\mathbf{l}}{|\mathbf{v}+\mathbf{l}|}$ is the halfway vector between light direction $\mathbf{l}$ and viewing direction $\mathbf{v}$. 
During ParaView rendering, we can adjust the TF or the light angle to produce a collection of high-quality VolVis images representing the underlying volume data. 

\vspace{-0.05in}
\section{iVR-GS}
\label{sec:method}

Existing NVS methods~\cite{Tang-VIS24, Niedermayr-arxiv24} for a VolVis scene are constrained by the preset TF and can only reconstruct the visible parts of the scene (i.e., non-zero-opacity voxel value ranges in the TF), significantly hindering user exploration.
One feasible approach to overcome this is to compose multiple NVS models corresponding to basic TFs that span the entire value domain, each covering a disjoint visible part (e.g., an opacity bump in Figure~\ref{fig:workflow}(a)). 
Together, they make the entire VolVis scene visible. 
Some NVS methods~\cite{Fridovich-Keil-CVPR22, Tang-neurips22, Kerbl-TOG23} naturally support {\em composability}, i.e., composing multiple basic scenes via simply adding the parameters of corresponding basic models without requiring extra optimization.
Our iVR-GS adopts 3DGS~\cite{Kerbl-TOG23} to support such a capability via concatenating Gaussian parameters.

As illustrated in Figure~\ref{fig:workflow}, our framework takes multiple sets of multi-view images rendered with different basic TFs as input. 
Each set of multi-view images rendered with the same basic TF is utilized to optimize an iVR-GS for modeling the corresponding basic scene. 
In particular, each iVR-GS model leverages numerous editable Gaussian primitives, which we extend from the original 3DGS, to reconstruct the scene. 
Such editable Gaussians follow the Blinn-Phong shading equation, effectively decomposing the original color and lighting effects.
Once these basic iVR-GS models are optimized, we compose them to recover all visible parts of the VolVis scene. 
Using the editable Gaussian representation, we can leverage the composed iVR-GS model for iterative real-time scene editing of color, opacity, and lighting effects.
Furthermore, similar to DNN-VolVis~\cite{Hong-DNN-VolVis}, when provided with one rendered VolVis image as a reference, the composed iVR-GS can inversely reproduce the scene with the desired target effect.
We achieve inverse volume exploration by learning the color, opacity, and light transformation parameters for each basic iVR-GS model, supporting real-time 3D navigation of the VolVis scene.

The rest of Section~\ref{sec:method} is organized as follows. 
Section~\ref{subsec:ViewpointSelection} explains the viewpoint sampling strategy for generating the training dataset.
In Sections~\ref{subsec:base3DGS} and~\ref{subsec:editableGS}, we describe how to augment original 3DGS primitives and optimize them to formulate editable Gaussian primitives.
Section~\ref{subsec:VQ} discusses the application of VQ to compress editable Guassians to reduce storage cost, followed by the details about scene composition and editing as well as inverse volume exploration described in Sections~\ref{subsec:sceneEditing} and~\ref{subsec:inverseVolume}, respectively. 
Finally, we briefly introduce the visual interface developed for iVR-GS in Section~\ref{subsec:GUI}.

\begin{figure}[!ht]
 \begin{center}
 $\begin{array}{c@{\hspace{0.025in}}c@{\hspace{0.025in}}c}
 \includegraphics[width=0.315\linewidth]{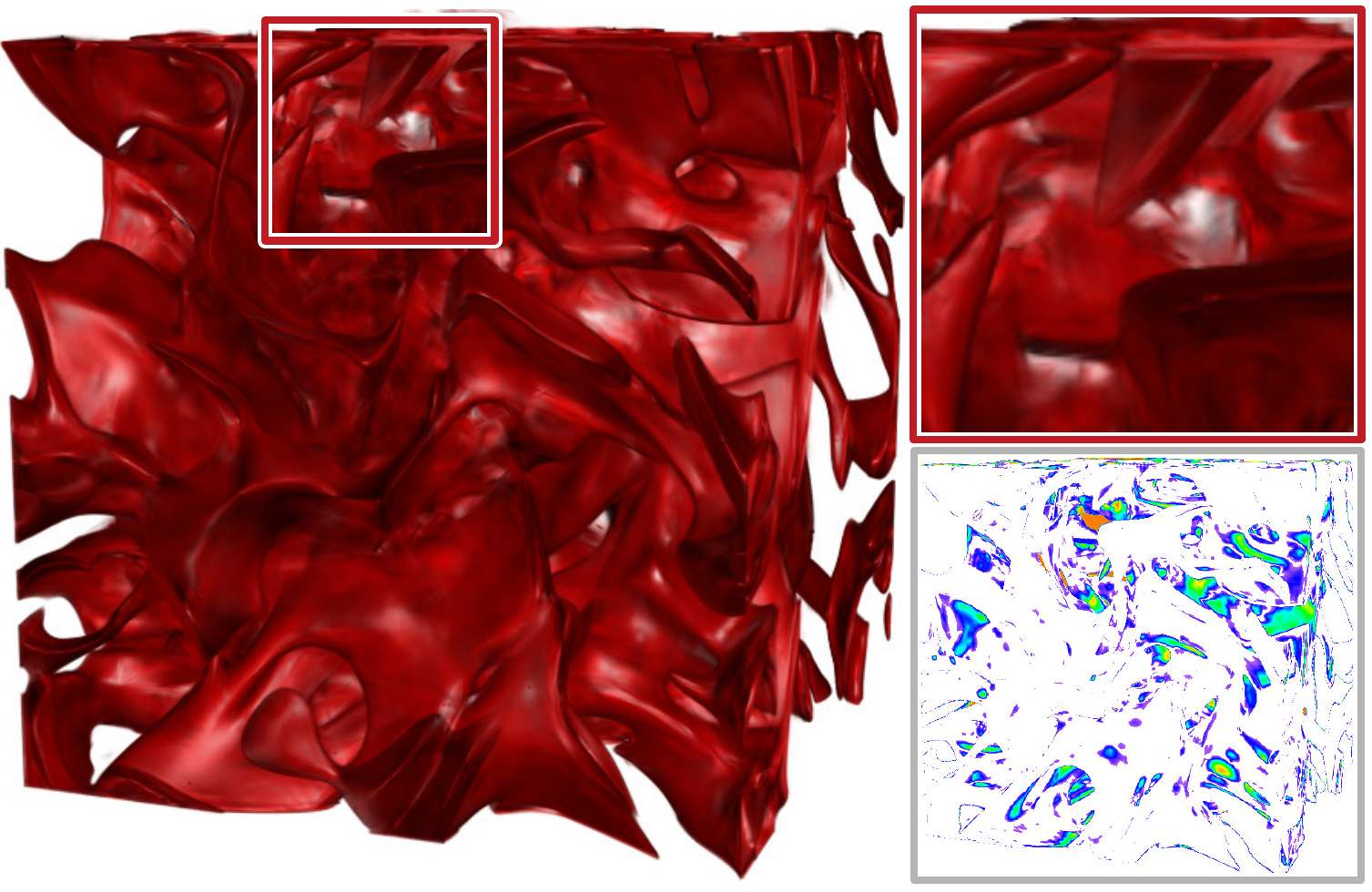}&
 \includegraphics[width=0.315\linewidth]{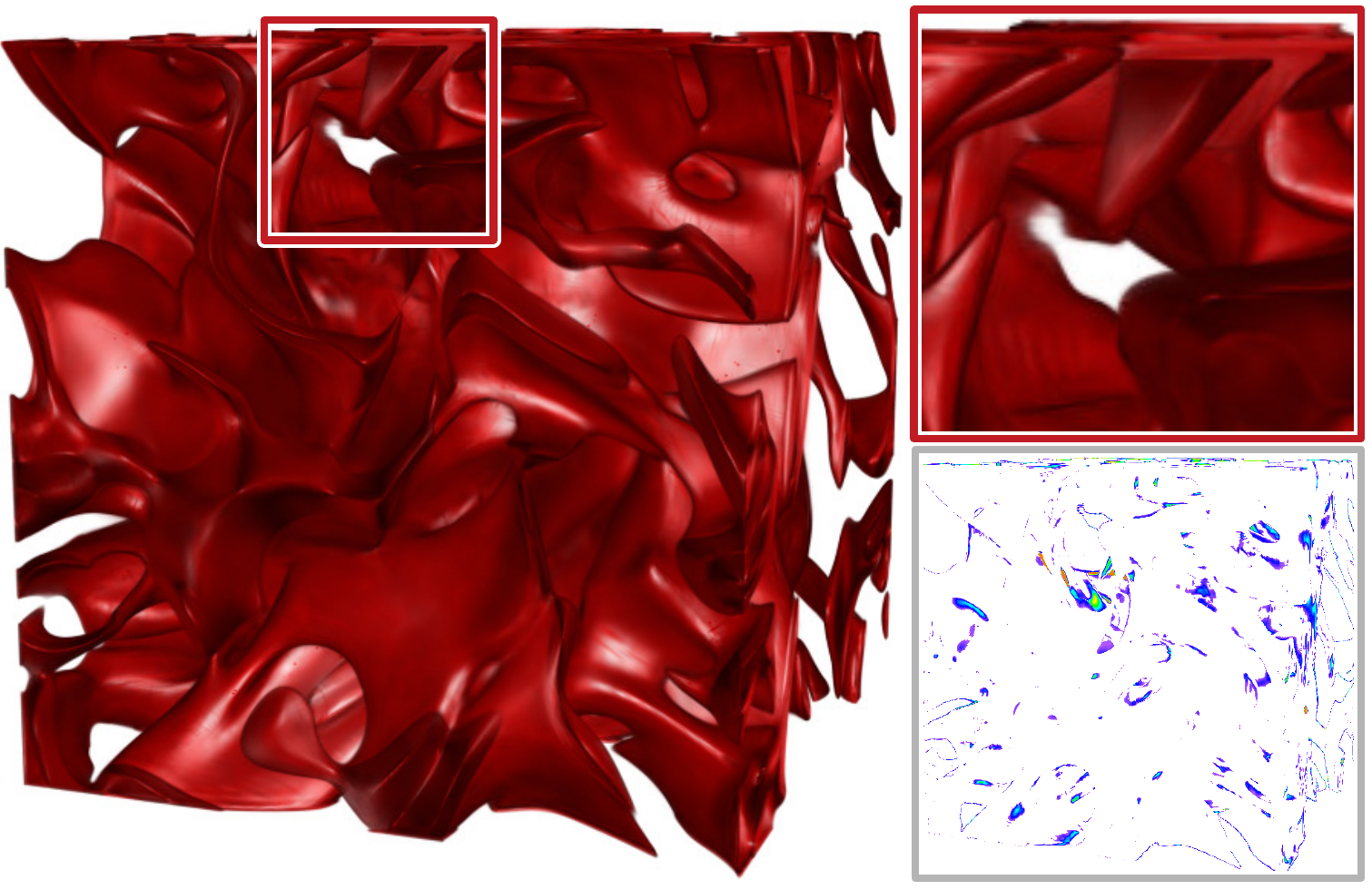}&
\includegraphics[width=0.315\linewidth]{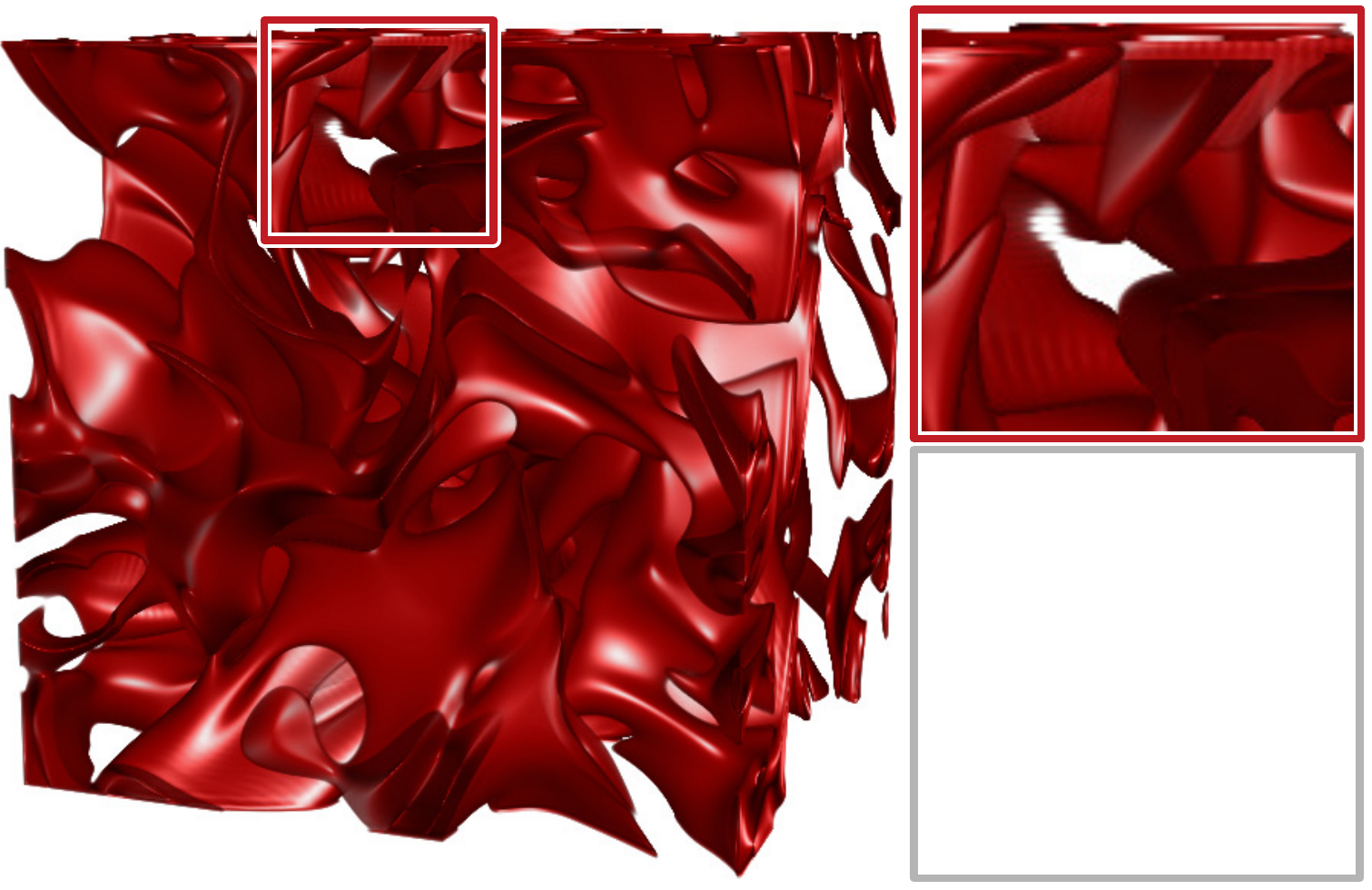}\\
\includegraphics[width=0.315\linewidth]{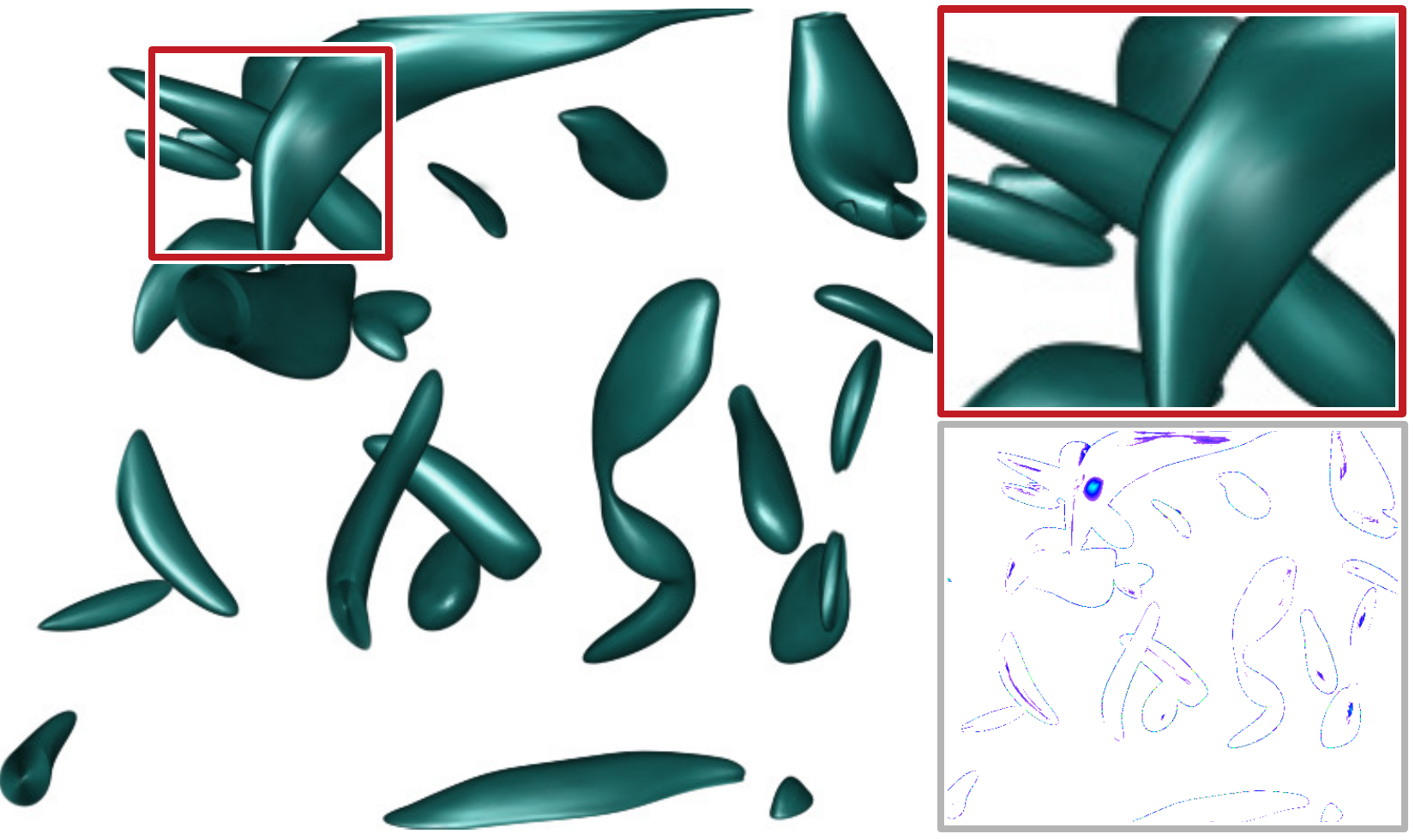}&
 \includegraphics[width=0.315\linewidth]{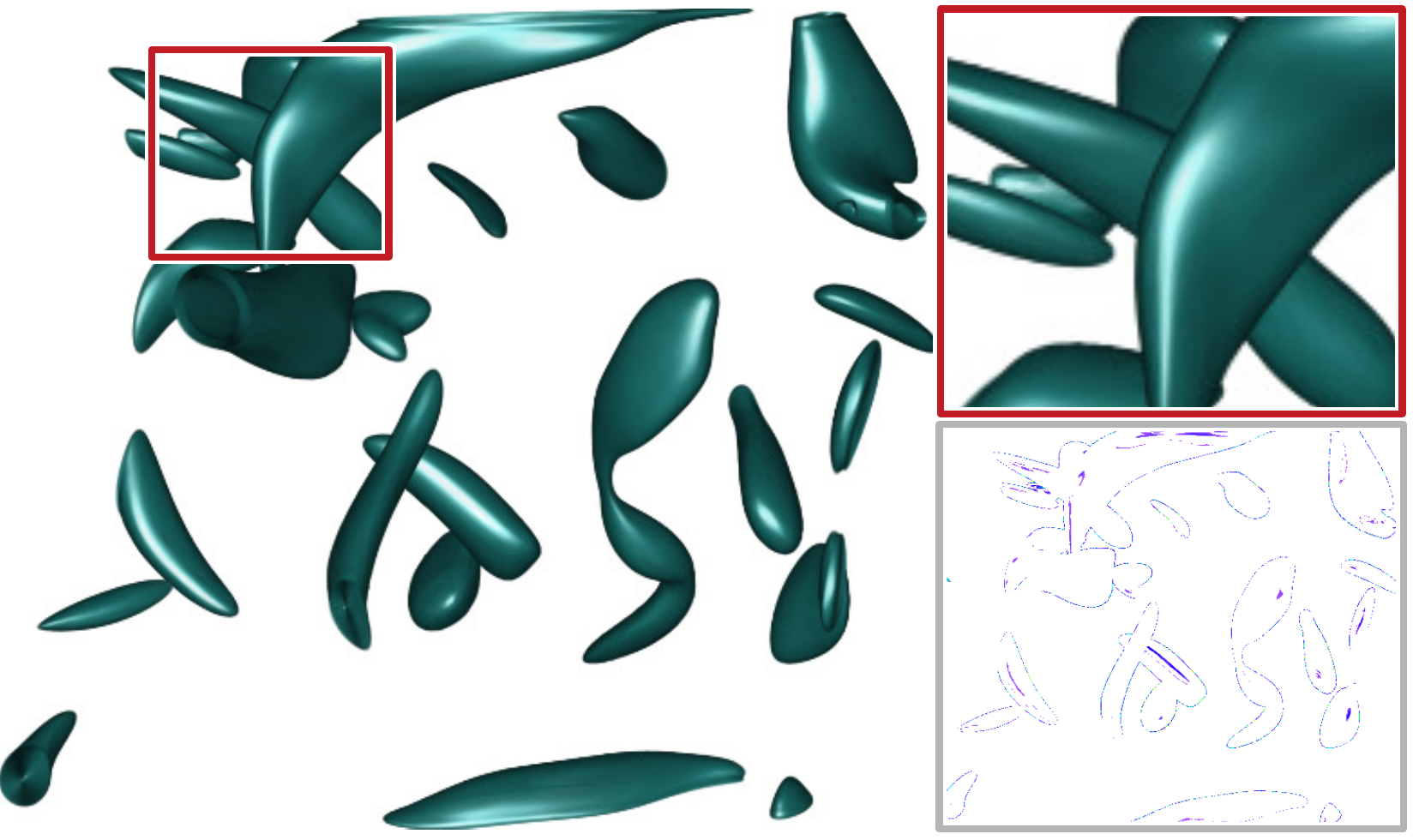}&
\includegraphics[width=0.315\linewidth]{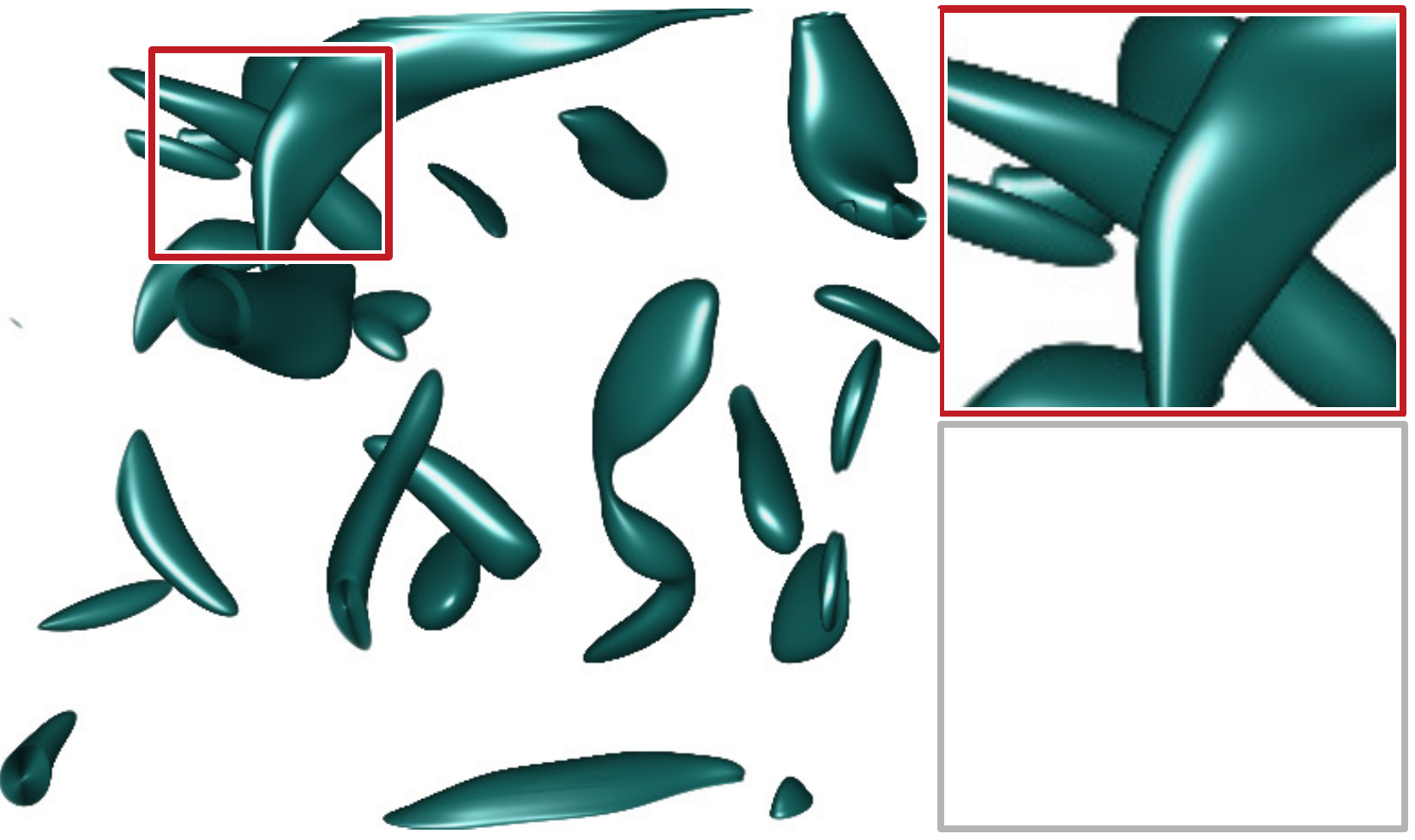}\\
\mbox{\footnotesize (a) 42 images} & \mbox{\footnotesize (b) 162 images} & \mbox{\footnotesize (c) GT}
\end{array}$
\end{center}
\vspace{-.25in} 
\caption{Comparing the reconstruction quality of iVR-GS trained with different numbers of multi-view images. The difference image shows the pixel-wise perceivable difference (purple/blue to cyan/green to yellow/red indicates low to medium to high) in the CIE LUV color space. Top and bottom: complex and simple scenes. 
} 
\label{fig:eval-viewpoints}
\end{figure}

\vspace{-0.05in}
\subsection{Viewpoint Sampling}
\label{subsec:ViewpointSelection}

Similar to~\cite{Niedermayr-arxiv24}, our scenario assumes VolVis images are rendered on high-end workstations or clusters, and users can train and infer the NVS models on their low-end local machines.
To generate multi-view images for iVR-GS training, designing a set of basic TFs to extract meaningful objects from the VolVis scene training is necessary.
While we use one-dimensional opacity TFs as basic TFs in this paper for simplicity, our framework is also compatible with high-dimensional TF specification techniques~\cite{Tzeng-TCVG04, WangYunhai-TVCG11, WangLei-TVCG11}, which can be leveraged to extract high-dimensional basic TFs automatically.

As illustrated in Figure~\ref{fig:eval-viewpoints}, the number of multi-view images required to achieve accurate reconstruction varies depending on the basic scene's complexity. 
Therefore, we evaluate the complexity of the basic scene and determine the number of viewpoints for training. 
We use icosphere sampling to choose 12 multi-view images for each basic TF as initial sample images. 
Inspired by~\cite{Ji-TVCG06}, we evaluate the complexity of each basic scene via a combination of color and opacity entropies calculated over initial sample images. The entropy score is defined as
\begin{equation}
	E = -\sum^{N_p}_{i=1} (p^{c}_i\log{p^c_i} + p^{\alpha}_i\log{p^{\alpha}_i}),
\end{equation}
where $N_p$ is the total number of pixels for all initial sample images, 
$p^{c}_i$ and $p^{\alpha}_i$ represent the $i$-th pixel's color and alpha value probabilities, respectively. 
We divide $E$ by the maximum entropy score among all basic scenes for normalization. 
After that, 
we uniformly sample varying numbers of additional views for each basic scene if necessary. 
Doing this reduces the total number of multi-view images required while incurring only negligible accuracy loss.

\vspace{-0.05in}
\subsection{Base 3DGS Optimization}
\label{subsec:base3DGS}

To correctly decompose the color and lighting from input multi-view images with editable Gaussians, each Gaussian must first accurately represent the scene's geometry.
Therefore, each basic iVR-GS will optimize a base 3DGS model that initializes Gaussian geometry attributes for optimizing subsequent shading attributes.
The base 3DGS is similar to the original 3DGS except we incorporate an extra normal attribute $\mathbf{n}$ for each Gaussian and optimize $\mathbf{n}$ with a normal consistency loss adopted from~\cite{Gao-arXiv23}.  
The normal consistency loss entails rendering depth and normal maps with base 3DGS for a specified viewpoint
\begin{equation}
\label{eqn:depth-normal-map}
	\{ \altmathcal{D}, \altmathcal{N}\} = \sum_{i\in N}T_i\alpha_i\{d_i, \mathbf{n}_i\},
\end{equation}
where $\altmathcal{D}$ and $\altmathcal{N}$ denote the rendered depth and normal maps of the base 3DGS represented scene, $d_i$ and $\mathbf{n}_i$ are the depth and normal of the $i$-th sampled Gaussian. 
After computing the depth and normal map, we calculate a pseudo normal $\altmathcal{\tilde{N}}$ from $\altmathcal{D}$ under the local planarity assumption. 
Finally, the normal consistency loss is determined as
\begin{equation}
\altmathcal{L}_n = || \altmathcal{N} - \altmathcal{\tilde{N}} ||_2.	
\end{equation}
It should be noted that although we optimize $\mathbf{n}$ during base 3DGS training, adjusting $\mathbf{n}$ does not affect the rendering results of base 3DGS.
Like the original 3DGS, the base 3DGS utilizes a collection of SH coefficients to restore the color information of each Gaussian.
The primary objective for optimizing $\mathbf{n}$ is to provide an initialization that can accelerate the convergence of editable Gaussian optimization.
Moreover, unlike IBR methods that only synthesize RGB colors, 3DGS reconstructs per-pixel alpha channel information.
As demonstrated in~\cite{Niedermayr-arxiv24}, optimizing the alpha channel can enhance the reconstruction quality.
Therefore, during training, we employ a combination of L1 and SSIM loss for base 3DGS output color and alpha channels along with $\altmathcal{L}_n$.
After the optimization of base 3DGS, the $i$-th Gaussian is parameterized with $\{\bm{\mu}_i, \mathbf{q}_i, \mathbf{s}_i, o_i, \mathbf{n}_i, \mathbf{c}_i \}$, where the view-dependent color $\mathbf{c}_i$ is represented by a set of SH coefficients.


\vspace{-0.05in}
\subsection{Editable Gaussian Optimization}
\label{subsec:editableGS}

For editable Gaussian optimization, we aim to decompose color and lighting from the scene and enable scene editing.
We adopt the Blinn-Phong reflection model described in Equation~\ref{eqn:Blinn-Phong} to compute the view-dependent color of each Gaussian instead of using SH coefficients like base 3DGS.
To do so, we assign additional shading attributes to each Gaussian primitive optimized in base 3DGS: an offset color $\Delta \mathbf{c} \in \mathbb{R}^3$, ambient, diffuse, and specular coefficients $\{k^a, k^d, k^s\}$, and a shininess term $\beta$.
In particular, the voxel color $\mathbf{c}_v$ will be expressed as $\mathbf{c}_p + \Delta \mathbf{c}$, where $\mathbf{c}_p \in \mathbb{R}^3$ is a palette color parameter that is shared by all Gaussians within a basic scene. $\mathbf{c}_p$ is fixed as the mean color value of input multi-view images during training and can be adjusted during inference to achieve scene recoloring. 
The offset color $\Delta \mathbf{c}$ is added to enhance the expressiveness of $\mathbf{c}_p$ for $\mathbf{c}_v$ representation.

In addition to applying the same loss as base 3DGS optimization during the training, we employ additional regularizations for optimizing the opacity term and shading attributes.
Like the original 3DGS, we remove low-opacity Gaussians periodically to reduce the model size.
Because the color represented by editable Gaussians depends on the normal attribute, the model tends to utilize more Gaussians to reconstruct compact surfaces than base 3DGS.
To neutralize this effect, we encourage the opacity of Gaussians to be as low as possible by adding an L1 regularization term. 
These low-opacity Gaussians can be more easily removed during training and thus reduce the final model size.
Additionally, 
for offset color $\Delta \mathbf{c}$, we add sparsity regularization to avoid $\Delta \mathbf{c}$ causing significant palette color shiftings. 
We define the sparsity regularization as the L1 loss of the rendered offset color map $\Delta \altmathcal{C} = \sum_{i \in N}T_i\alpha_i\Delta \mathbf{c}_i$.
For other shading attributes, we leverage bilateral smoothness~\cite{Yao-ECCV22, Zhang-ICCV23} to ensure these shading attributes will not change significantly in smooth-color regions.
Assuming $\mathbf{c}_{\gt}$ is the ground-truth (GT) color of multi-view images, and $\altmathcal{K}^a = \sum_{i \in N}T_i\alpha_ik^{a}_i$ is the rendered ambient coefficient map.
The bilateral smoothness regularization on the ambient coefficient is
\begin{equation}
	\altmathcal{L}_{\smooth}= |\bigtriangledown \altmathcal{K}^a|\exp{(-|\bigtriangledown\mathbf{c}_{\gt}|)}.
\end{equation}
Similarly, we employ bilateral smoothness regularization to other shading attributes, including $k^d$, $k^s$, and $\beta$.
When the training completes, we represent the scene with a collection of editable Gaussians, where the $i$-th editable Gaussian is parameterized with geometry attributes $\{\bm{\mu}_i, \mathbf{q}_i, \mathbf{s}_i, o_i, \mathbf{n}_i\}$ along with shading attributes $\{\Delta\mathbf{c}_i, k^{a}_i, k^{d}_i, k^{s}_i, \beta_i\}$.

\vspace{-0.05in}
\subsection{Compression via Vector Quantization}
\label{subsec:VQ}

To recover all visible parts of the VolVis scene, we compose multiple basic iVR-GS models into the composed iVR-GS model. 
The resulting composed model size is the summation of the sizes of all basic iVR-GS models.
The size of the composed iVR-GS model grows as the number of basic models increases.
Therefore, we apply VQ to extract a more compact form of editable Gaussians for each basic iVR-GS model before composition.

For one optimized iVR-GS model, similar to other VQ techniques for 3DGS compression~\cite{Fan-arXiv23,Niedermayr-CVPR24,Papantonakis-CGIT24}, we employ k-means clustering to generate a shared codebook for one attribute value across all Gaussian primitives.
Instead of storing the exact value of an attribute, each primitive then stores an index to the nearest value within the fixed-size codebook.
A shared codebook is used for vector attributes (e.g., scaling $\mathbf{s}$), with separate indices for each vector component.
We generate codebooks for all attributes of editable Gaussians except $\bm{\mu}$ and $\mathbf{n}$ as quantizing these two attributes can notably decrease the reconstruction quality.
The time cost of the VQ process is negligible, yet it provides roughly 4$\times$ compression without sacrificing rendering quality. 

\vspace{-0.05in}
\subsection{Scene Composition and Editing}
\label{subsec:sceneEditing}

After training and quantizing basic iVR-GS models, we save the exact values of attributes $\bm{\mu}$ and $\mathbf{n}$, and indices with shared codebooks of other attributes for each editable Gaussian. 
Additionally, palette color $\mathbf{c}_p$ is stored as a global parameter for each basic scene. 
Since each basic scene is represented by a set of explicit primitives, composability can be naturally accomplished by spatially composing all editable Gaussians.
In practice, this is implemented by decompressing each model from the quantized format and then appending attribute values of primitives from basic iVR-GS models into a single parameter list to form the composed iVR-GS model.
Leveraging editable Gaussians as representation primitives, the composed iVR-GS enables real-time scene editing by adjusting the editable Gaussian attributes of each basic iVR-GS model. 
Such scene editing can achieve plausible editing of color, opacity, light magnitude, and accurate editing of light direction. 
Specifically, we edit the color of any basic iVR-GS model by replacing its corresponding $\mathbf{c}_p$ with a new palette color $\mathbf{c}^{\prime}_p$.
The opacity is adjusted by directly scaling the opacity term.
For the lighting effect, we scale attributes $k^a$, $k^d$, $k^s$, and $\beta$ to adjust the magnitudes of these shading terms.
As each primitive includes a normal $\mathbf{n}$, a new VolVis scene with light direction change can be rendered following the Blinn-Phong equation shown in Equation~\ref{eqn:Blinn-Phong}.

\vspace{-0.05in}
\subsection{Inverse Volume Exploration}
\label{subsec:inverseVolume}

In some cases, rather than employing scene editing to the composed iVR-GS model from scratch to reach accurate and desired rendering results, users may wish the composed iVR-GS model could approximate the intended rendering setting automatically, given a reference image as the desired effect.
This capability saves users time by eliminating tedious and inefficient trial-and-error TF adjustments.
Given a reference image and its corresponding camera pose, we freeze the parameters of all editable Gaussian attributes and learn the transformation parameters of color, opacity, and lighting for each basic iVR-GS model.
After training, the transformed iVR-GS model can replicate the reference image and generate accurate NVS results with the desired effect. 

In particular, the color is optimized by making the global parameter $\mathbf{c}_p$ learnable for each basic iVR-GS model. 
For opacity, we multiply and optimize a scaling factor for each basic iVR-GS model to control their contribution to the VolVis scene.
Unlike color and opacity optimized for a basic iVR-GS, the lighting is modified globally for the composed iVR-GS.
We optimize the target light magnitude via basic iVR-GS attributes $\{k^a, k^d, k^s, \beta\}$ using two learnable parameter vectors ($\bm{\lambda}\in\mathbb{R}^{4\times1}$ and $\mathbf{b}\in\mathbb{R}^{4\times1}$), i.e., $\bm{\lambda}\{k^a,k^d,k^s,\beta\}+ \mathbf{b}$.
If the reference image employs orbital light, the composed iVR-GS model can also optimize the light direction by modeling it with two learnable parameters denoting polar and azimuthal light angles in the spherical coordinate system. 
Given a reference image, we freeze all primitive attributes and optimize only the transformation parameters using a combination of L1 and SSIM loss, allowing the transformed iVR-GS rendering to match the reference image, along with its corresponding color, opacity, and light settings.

\begin{figure}[!htb]
\centering
\includegraphics[width=\columnwidth]{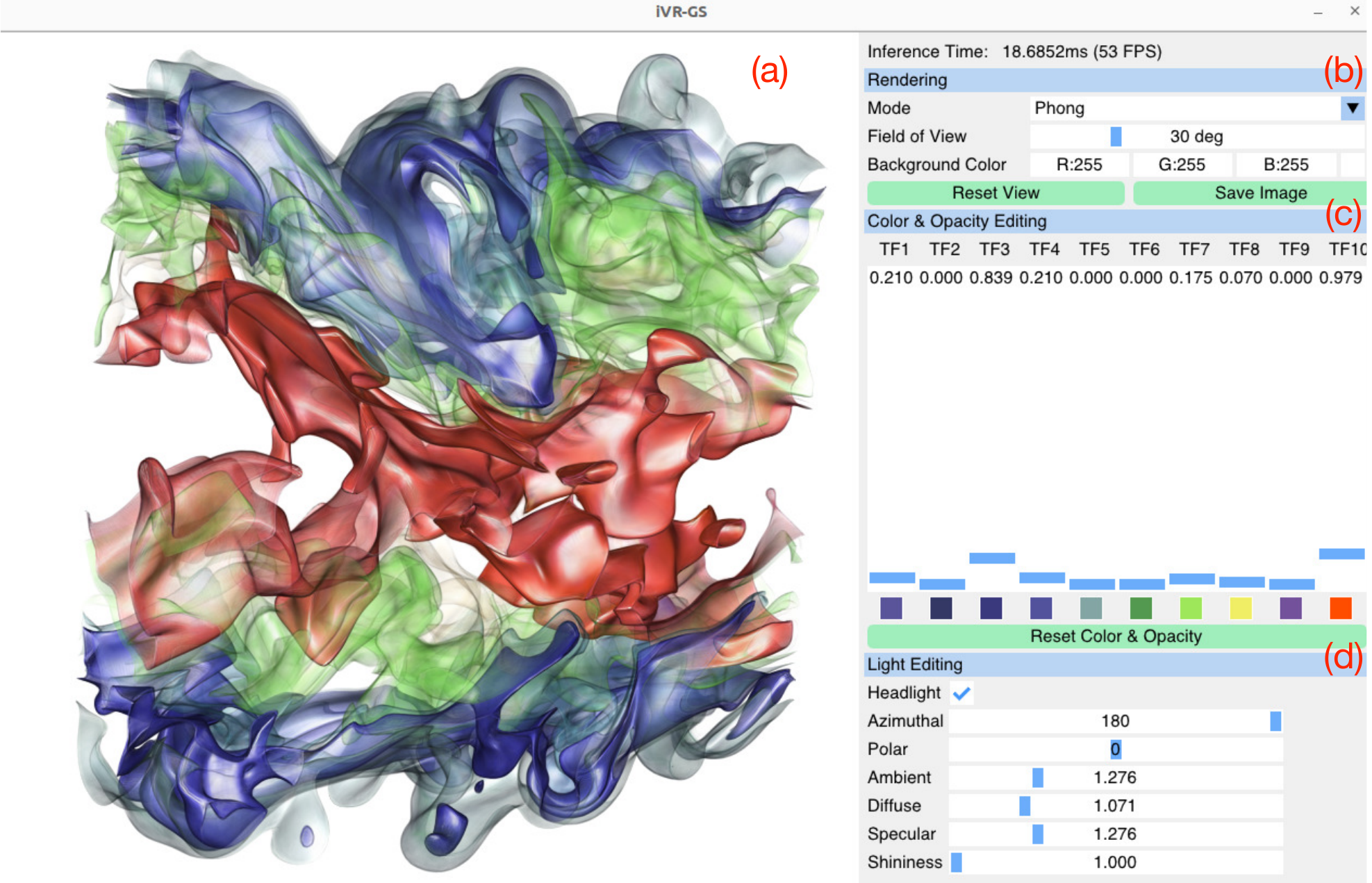}
\vspace{-.25in}
\caption{The screenshot of iVR-GS interface showing rendering results of the combustion dataset with ten basic TFs. (a) Rendering result of the iVR-GS. (b) Rendering options for selecting rendering modes, saving images, etc. (c) Color editors and sliders for color and opacity adjustment. (d) Light editing panel for lighting effect adjustment.}
\label{fig:GUI}
\end{figure}

\vspace{-0.1in}
\subsection{Interactive Interface}
\label{subsec:GUI}

Figure~\ref{fig:GUI} illustrates our interface for users to visualize the scene with interactive exploration. 
Our interface supports arbitrary scene rotating or editing in real-time (above 30 FPS).
In addition to the VolVis rendering, the interface supports rendering Gaussian attributes such as the normal map or other lighting terms (i.e., ambient, diffuse, and specular) by selecting different rendering modes.
We refer readers to the supplemental video for the recorded interaction with the interface.

\begin{figure*}[!ht]
 \begin{center}
 $\begin{array}{c@{\hspace{0.05in}}c@{\hspace{0.05in}}c@{\hspace{0.05in}}c@{\hspace{0.05in}}c}
 \includegraphics[width=0.185\linewidth]{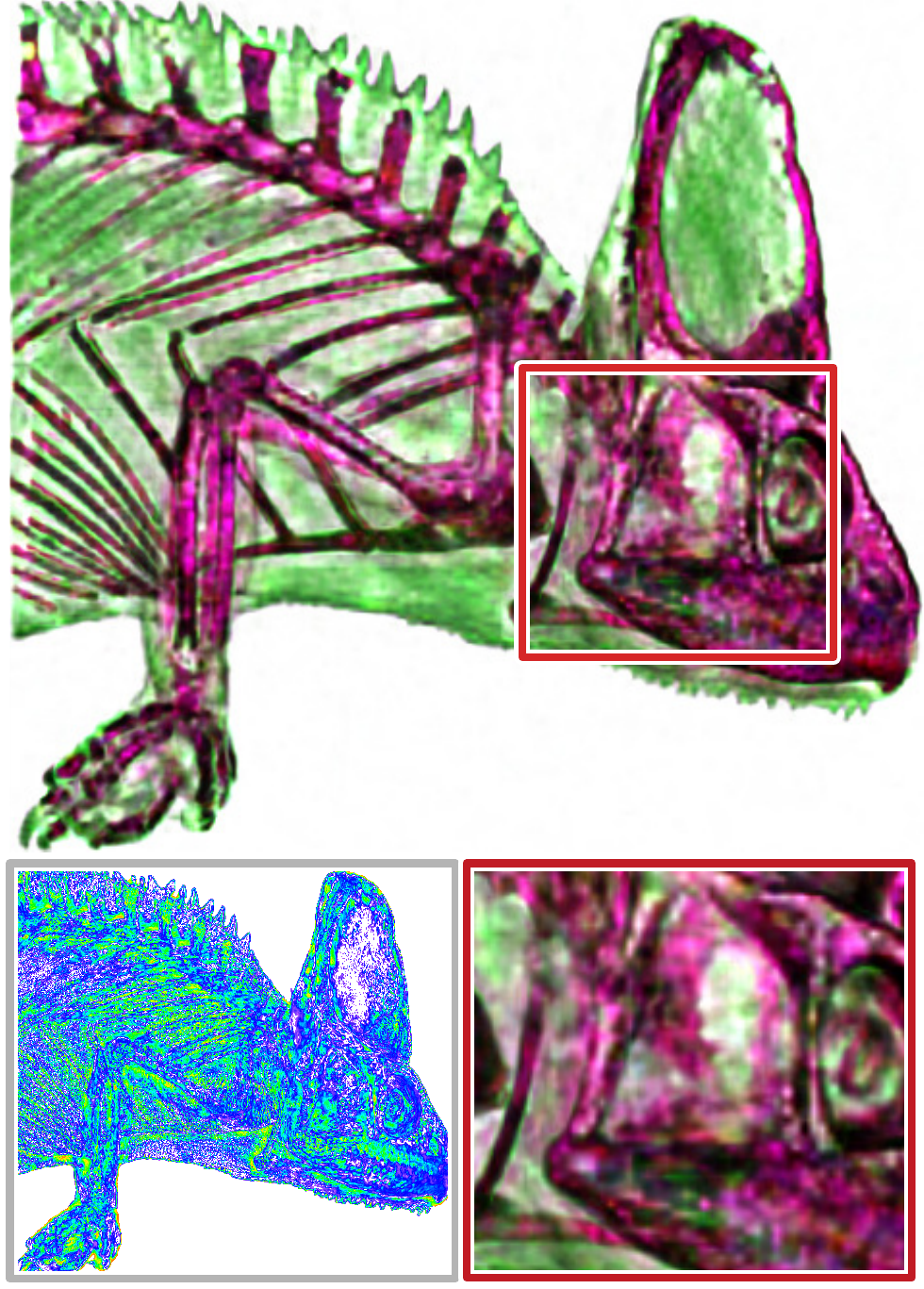}&
 \includegraphics[width=0.185\linewidth]{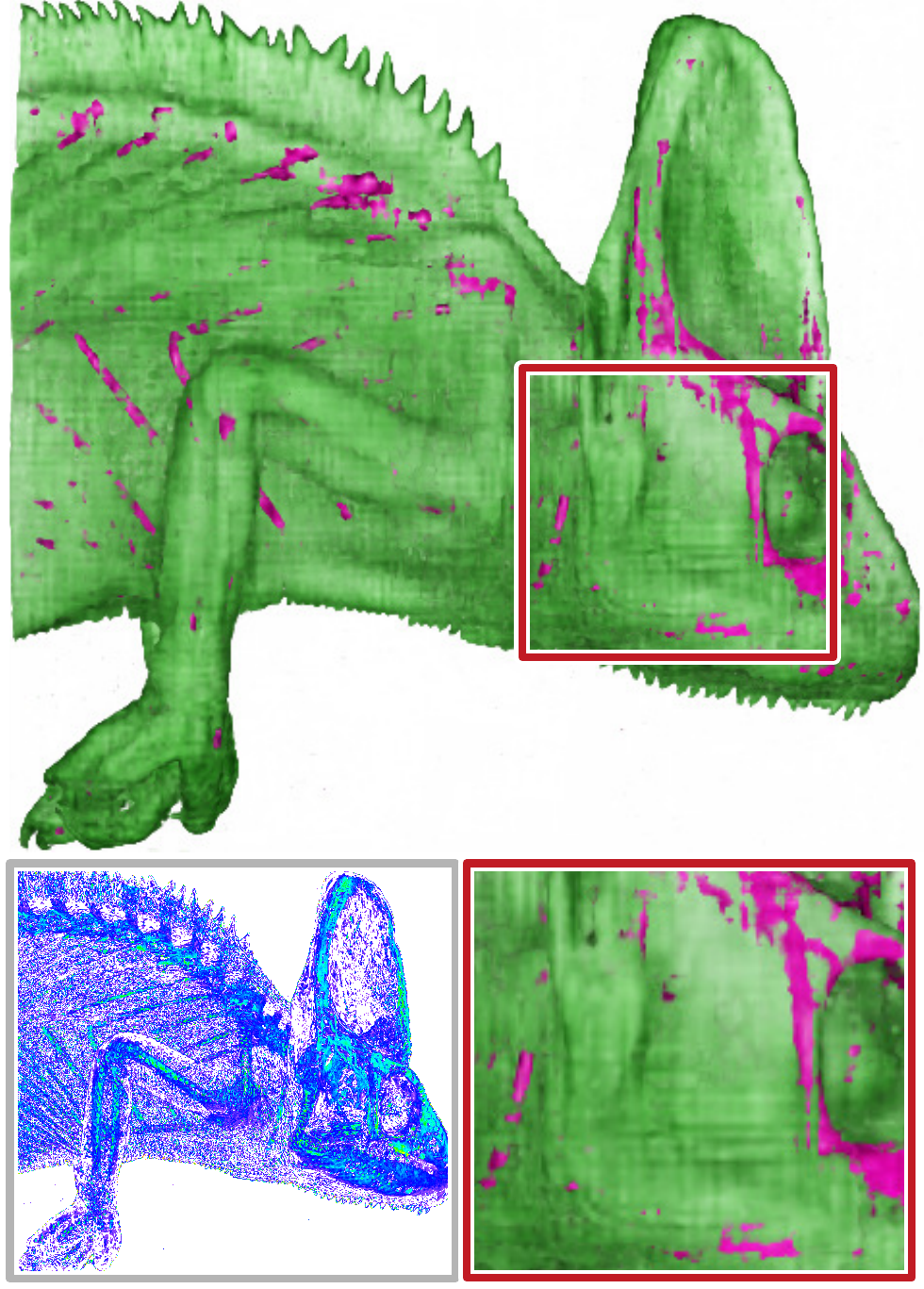}&
 \includegraphics[width=0.185\linewidth]{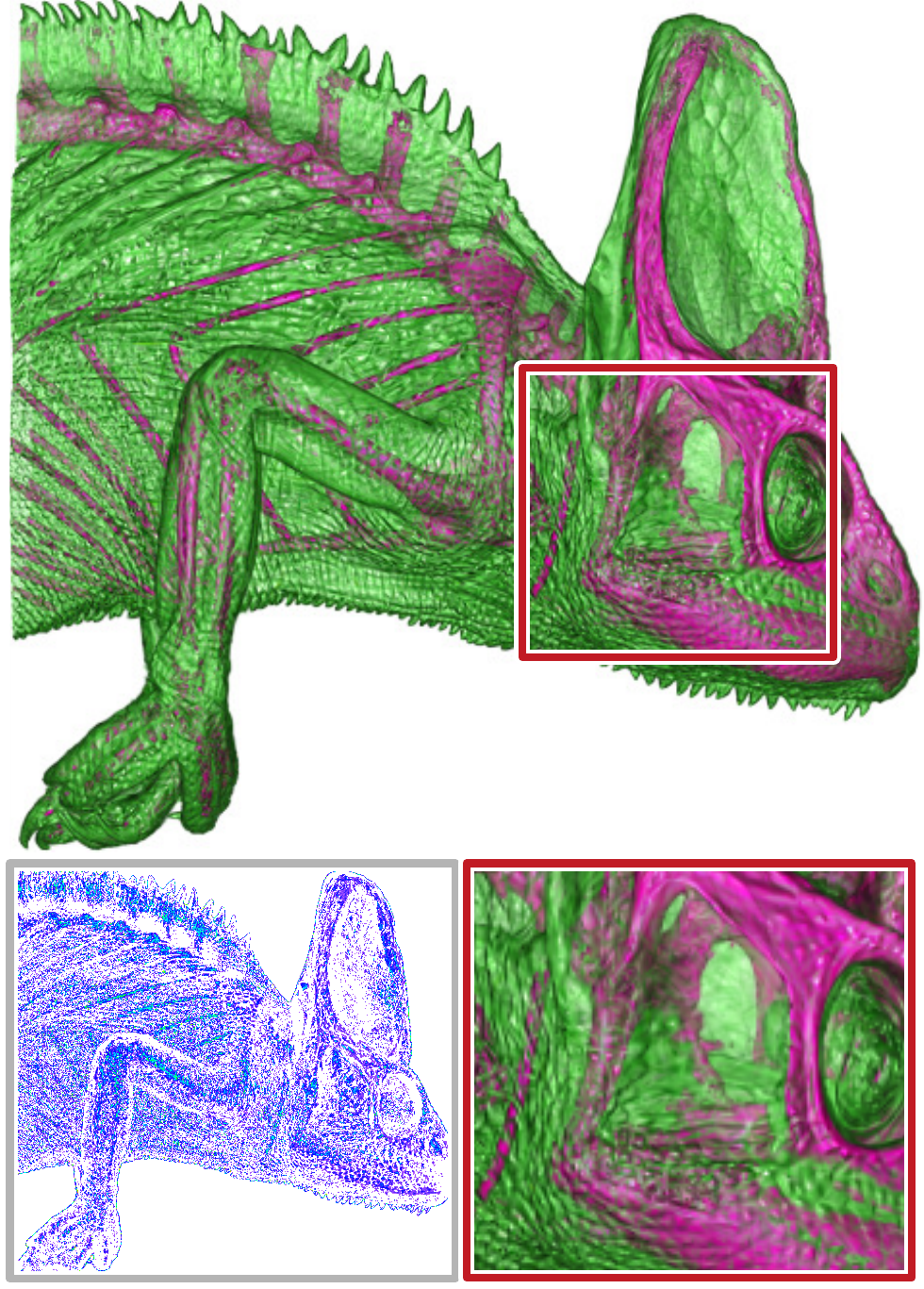}&
  \includegraphics[width=0.185\linewidth]{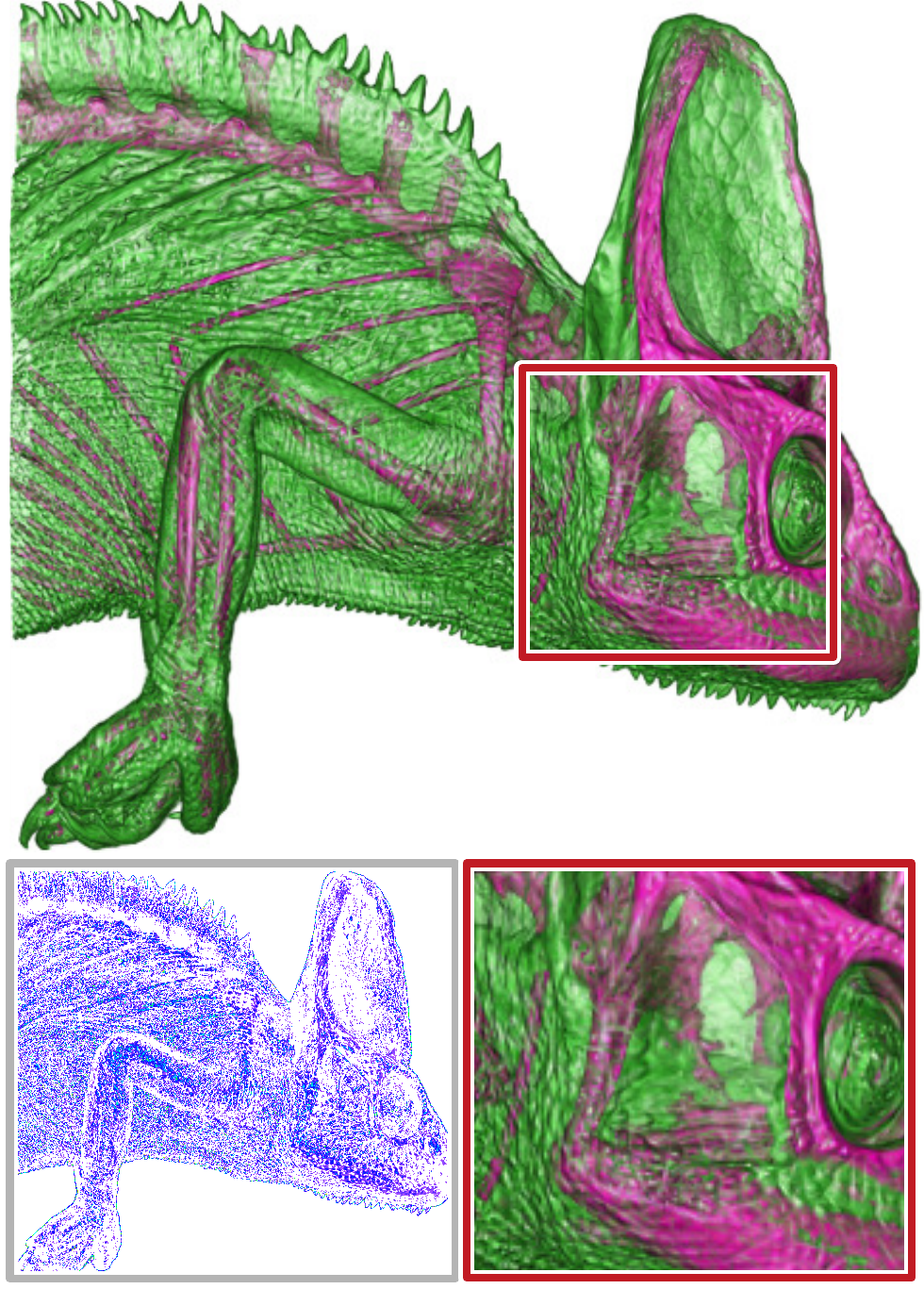}&
\includegraphics[width=0.185\linewidth]{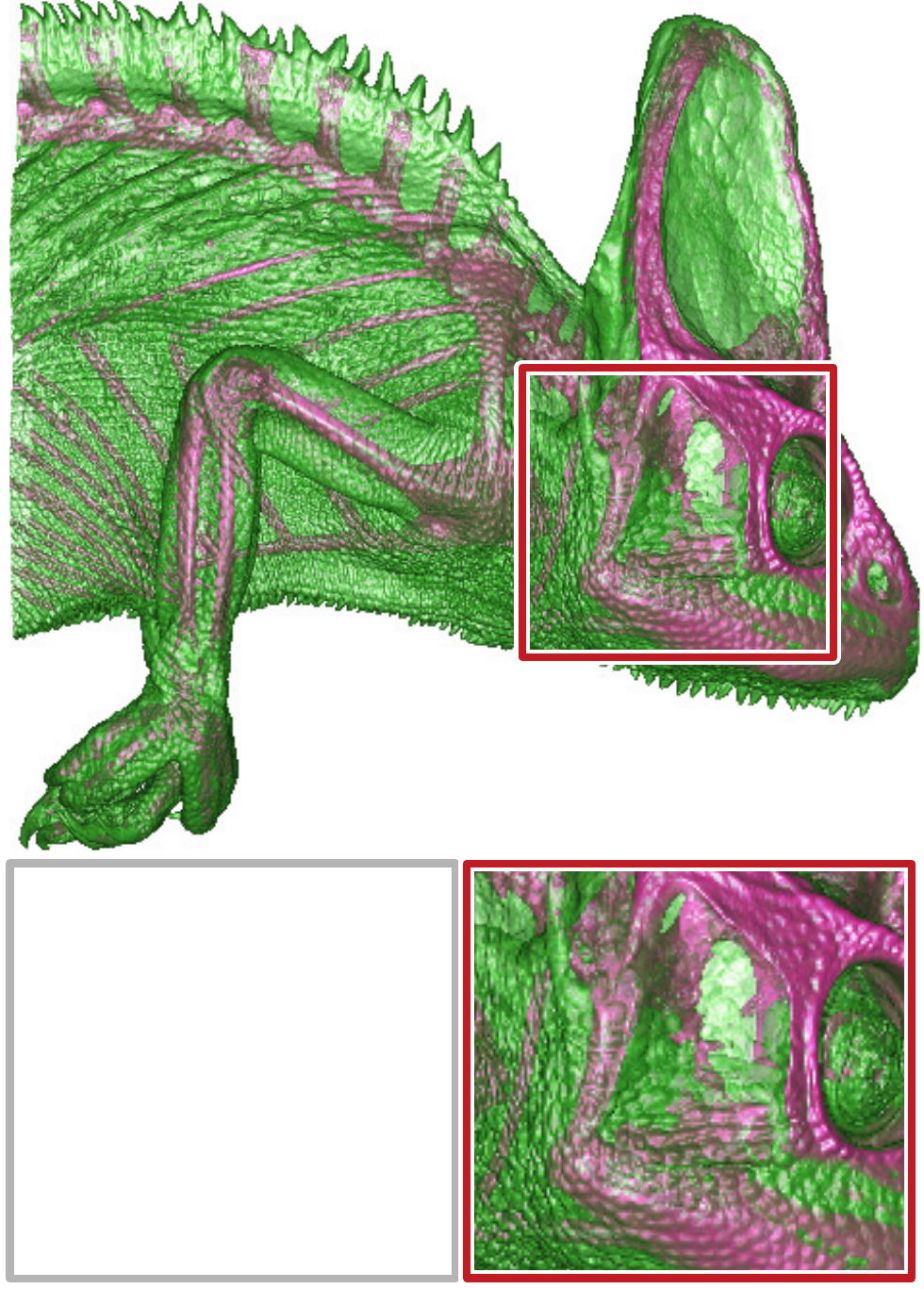}\\
 \includegraphics[width=0.185\linewidth]{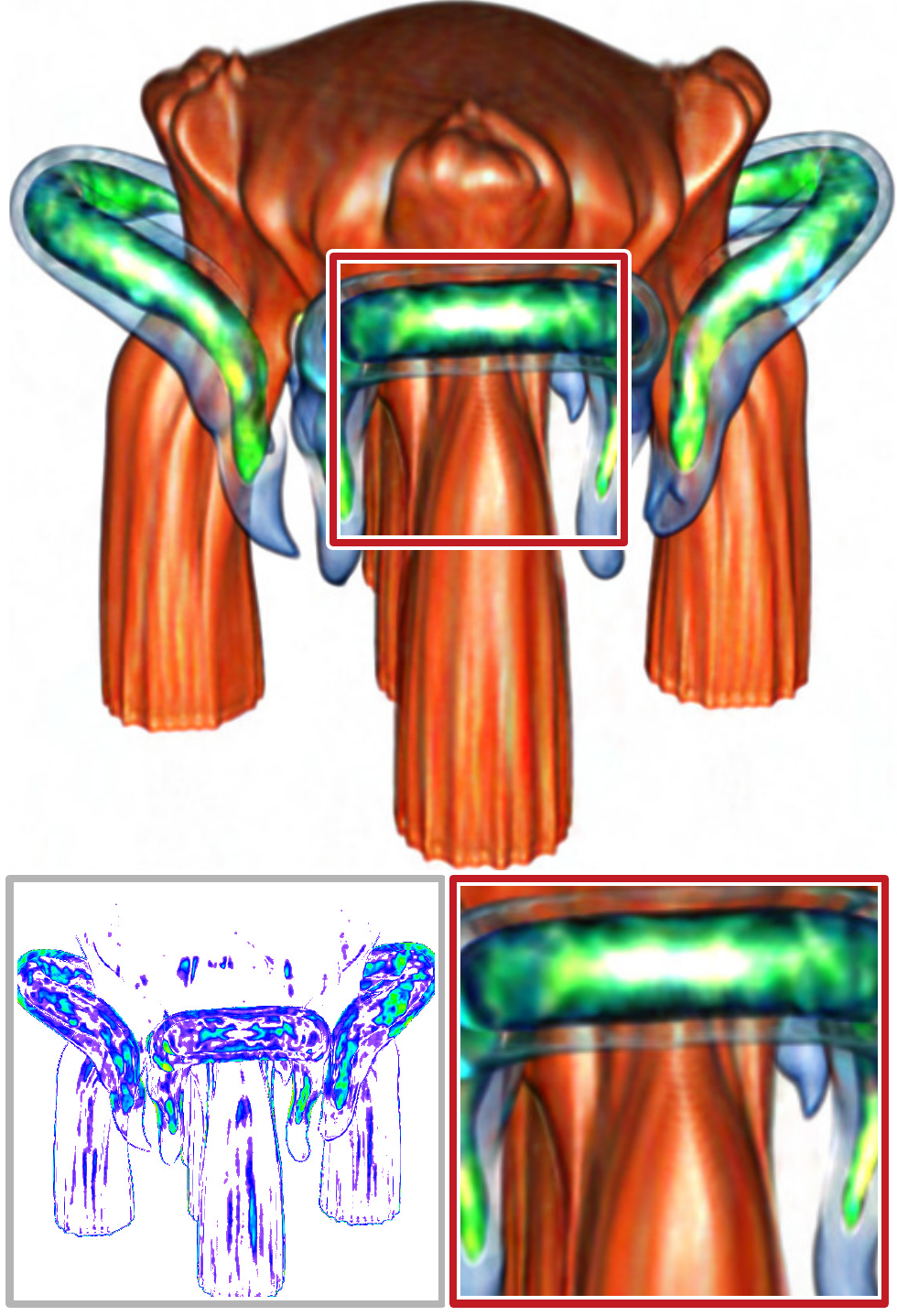}&
 \includegraphics[width=0.185\linewidth]{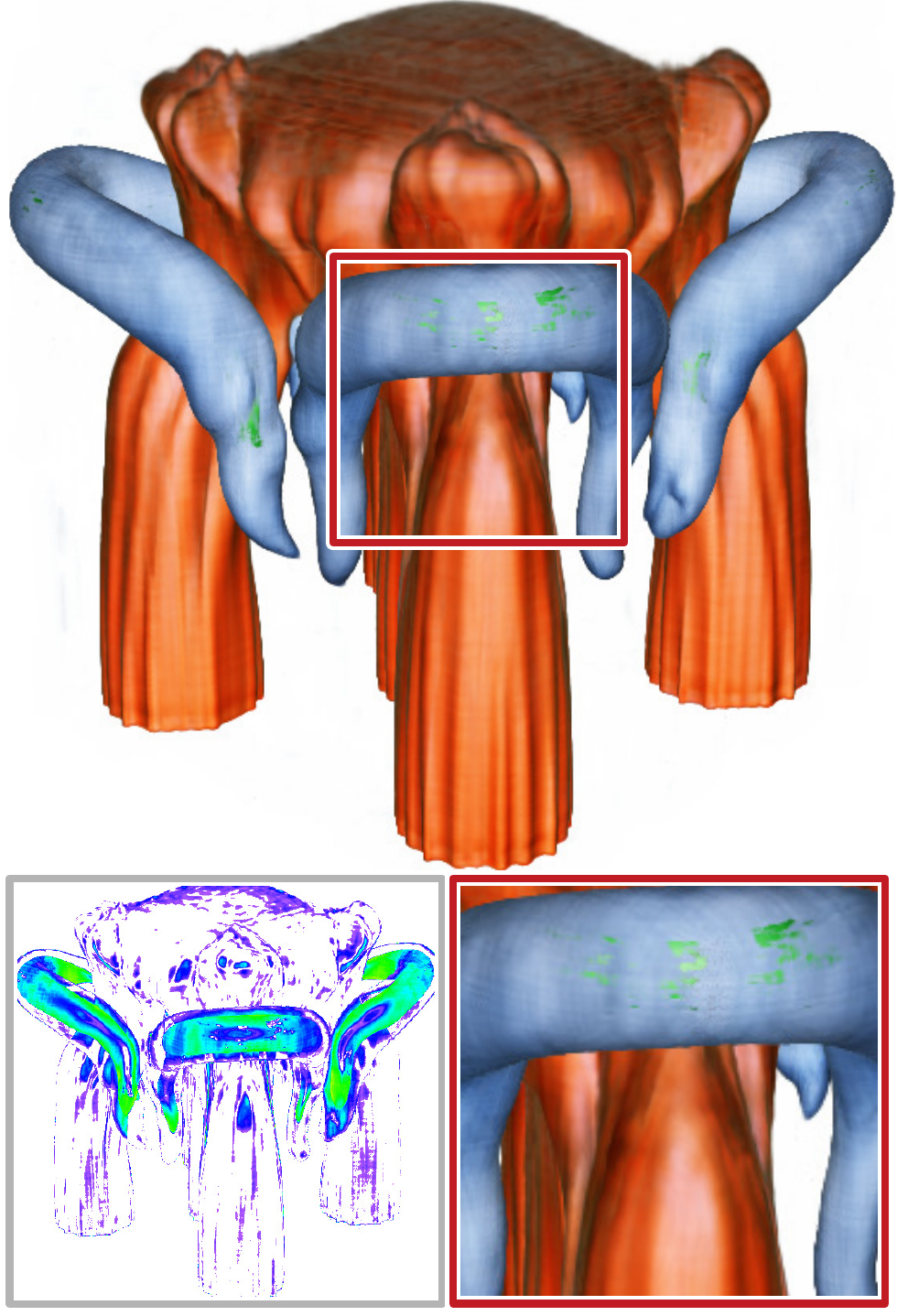}&
 \includegraphics[width=0.185\linewidth]{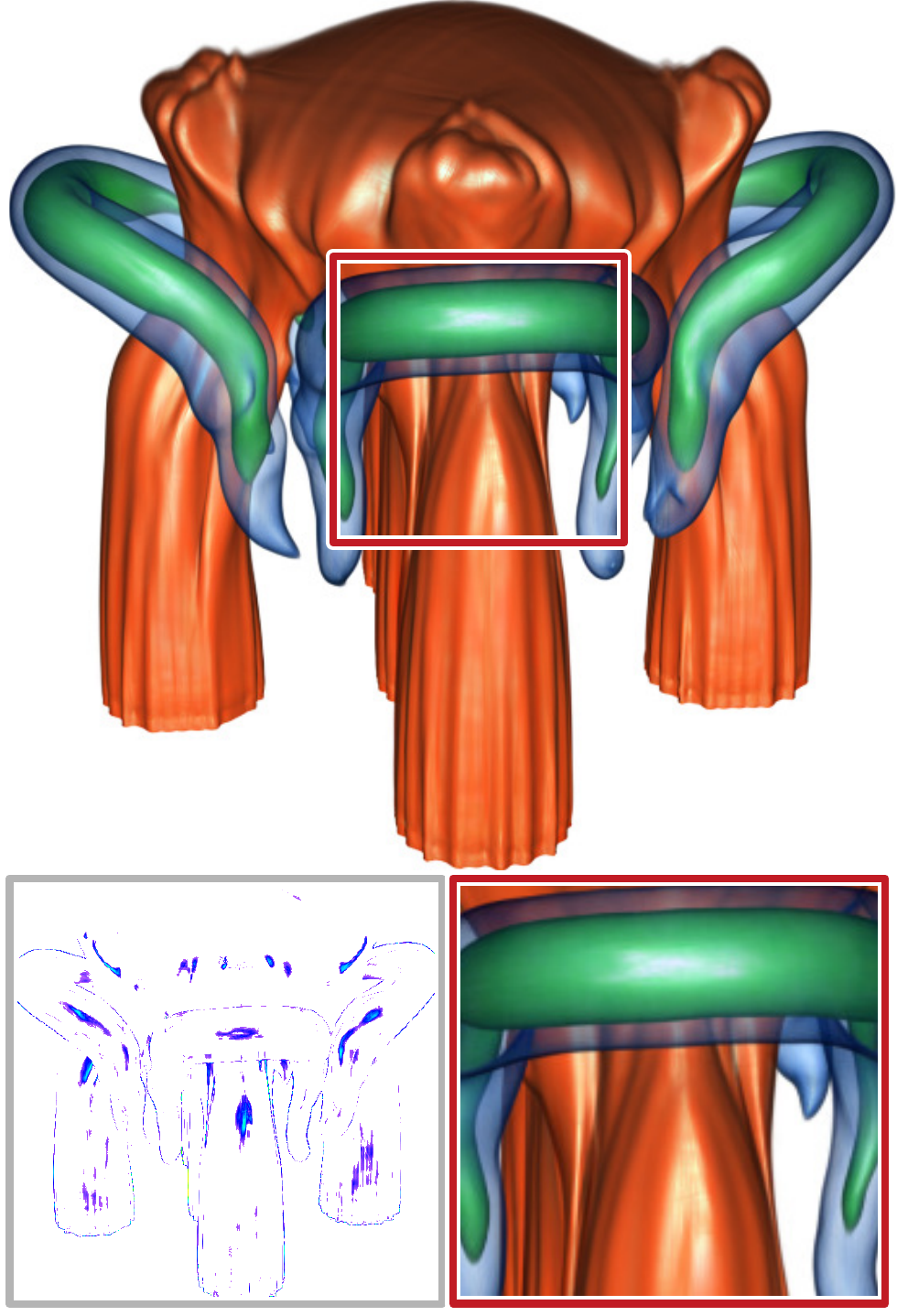}&
  \includegraphics[width=0.185\linewidth]{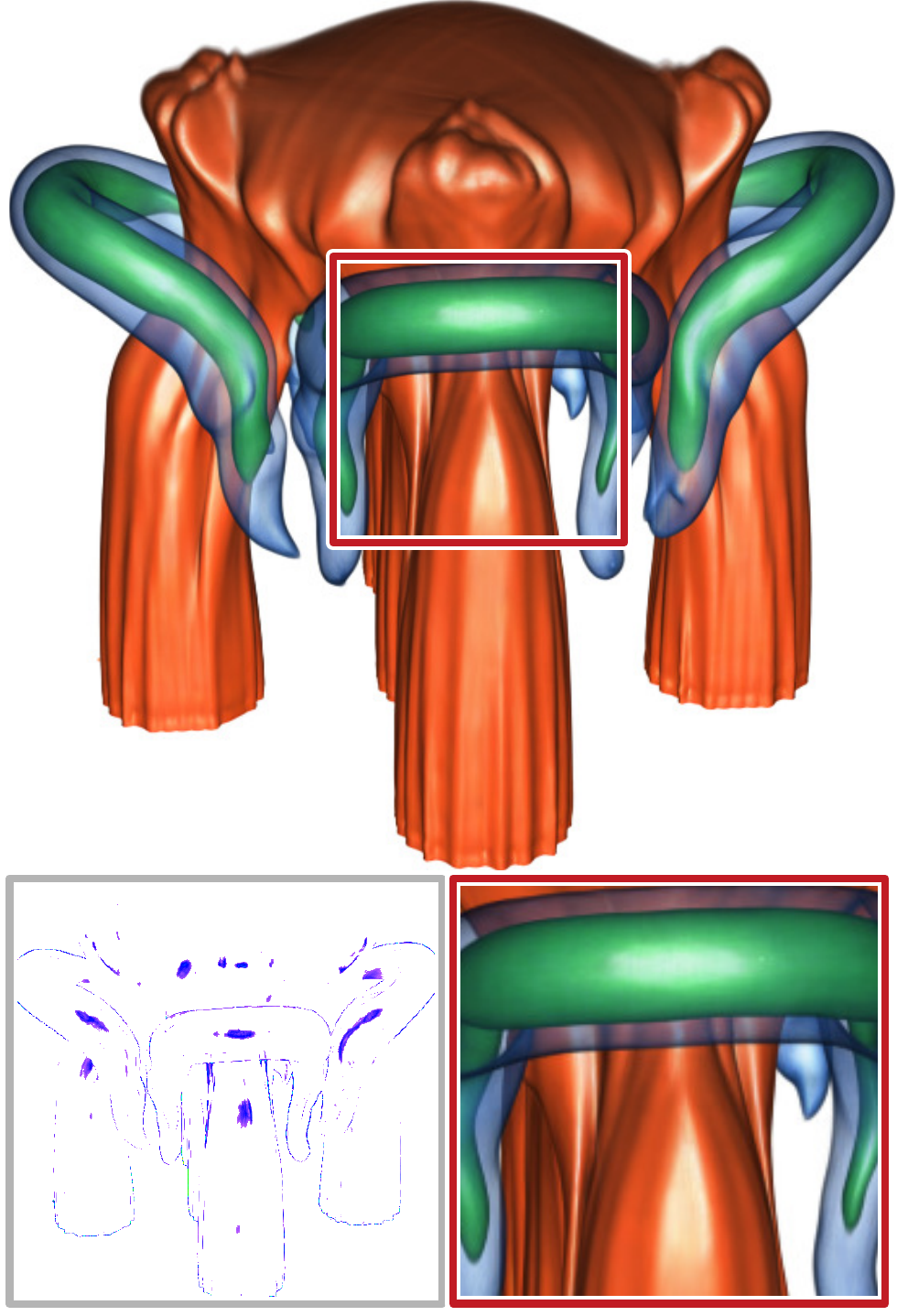}&
\includegraphics[width=0.185\linewidth]{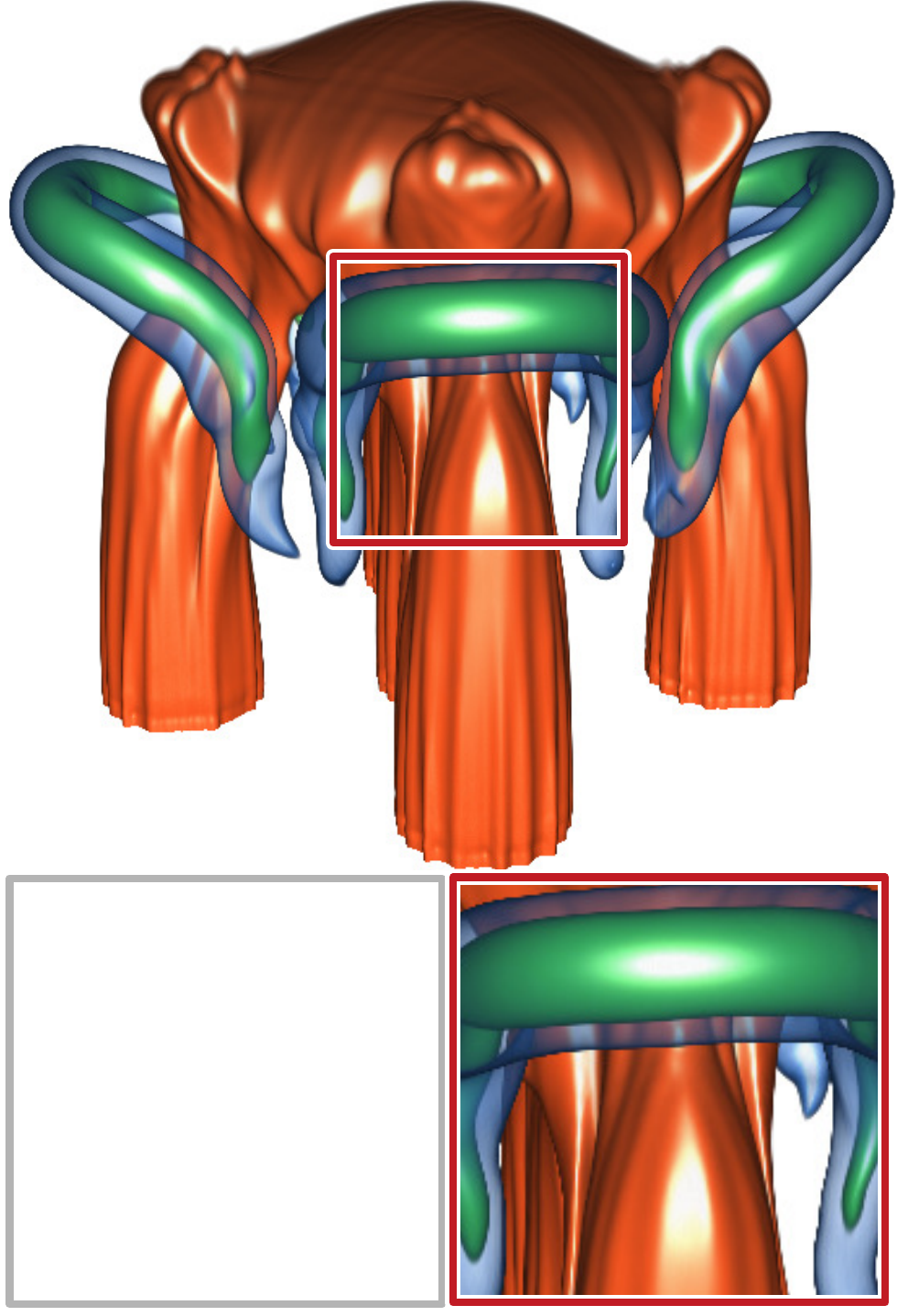}\\
 \includegraphics[width=0.185\linewidth]{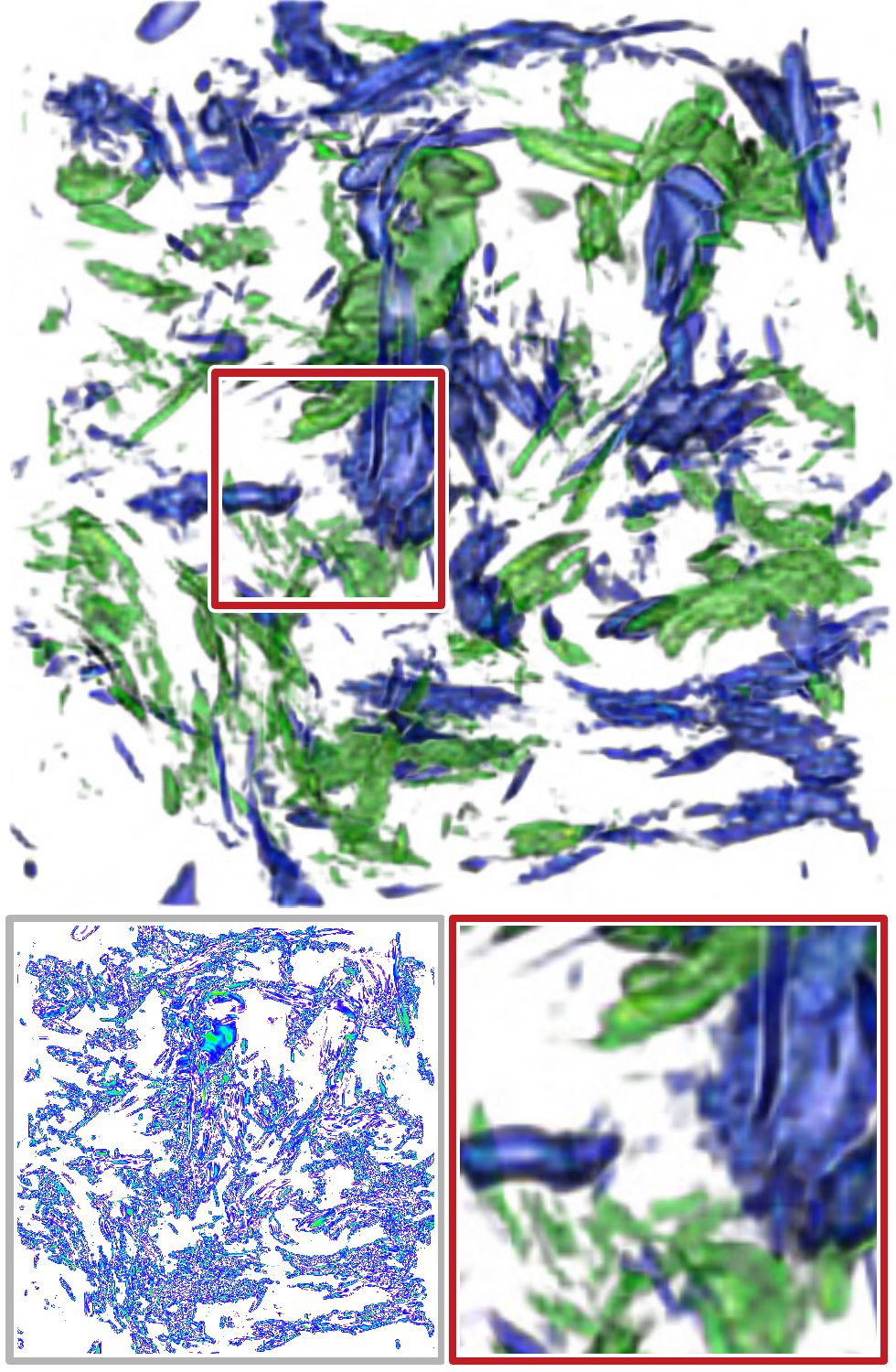}&
 \includegraphics[width=0.185\linewidth]{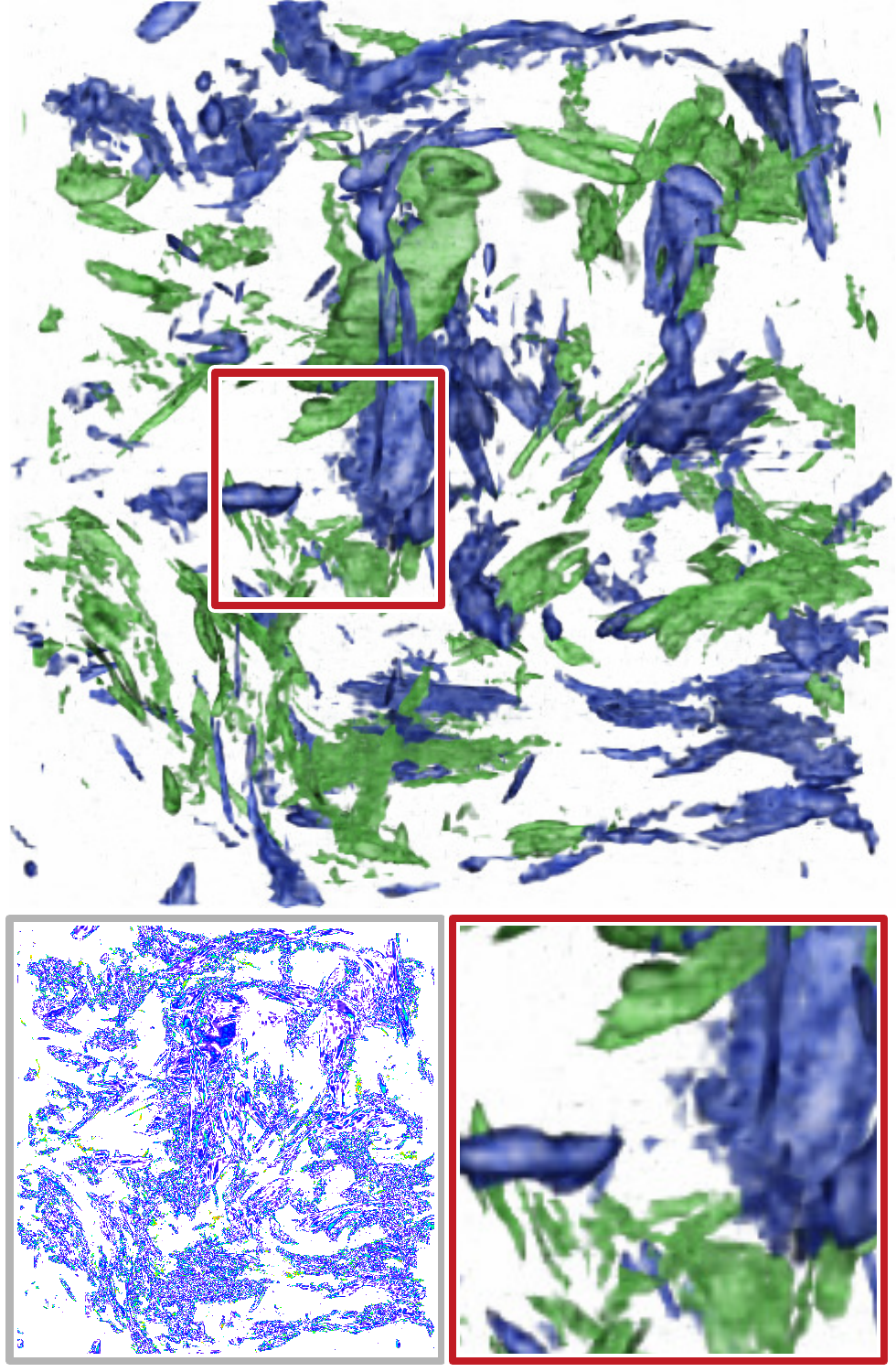}&
 \includegraphics[width=0.185\linewidth]{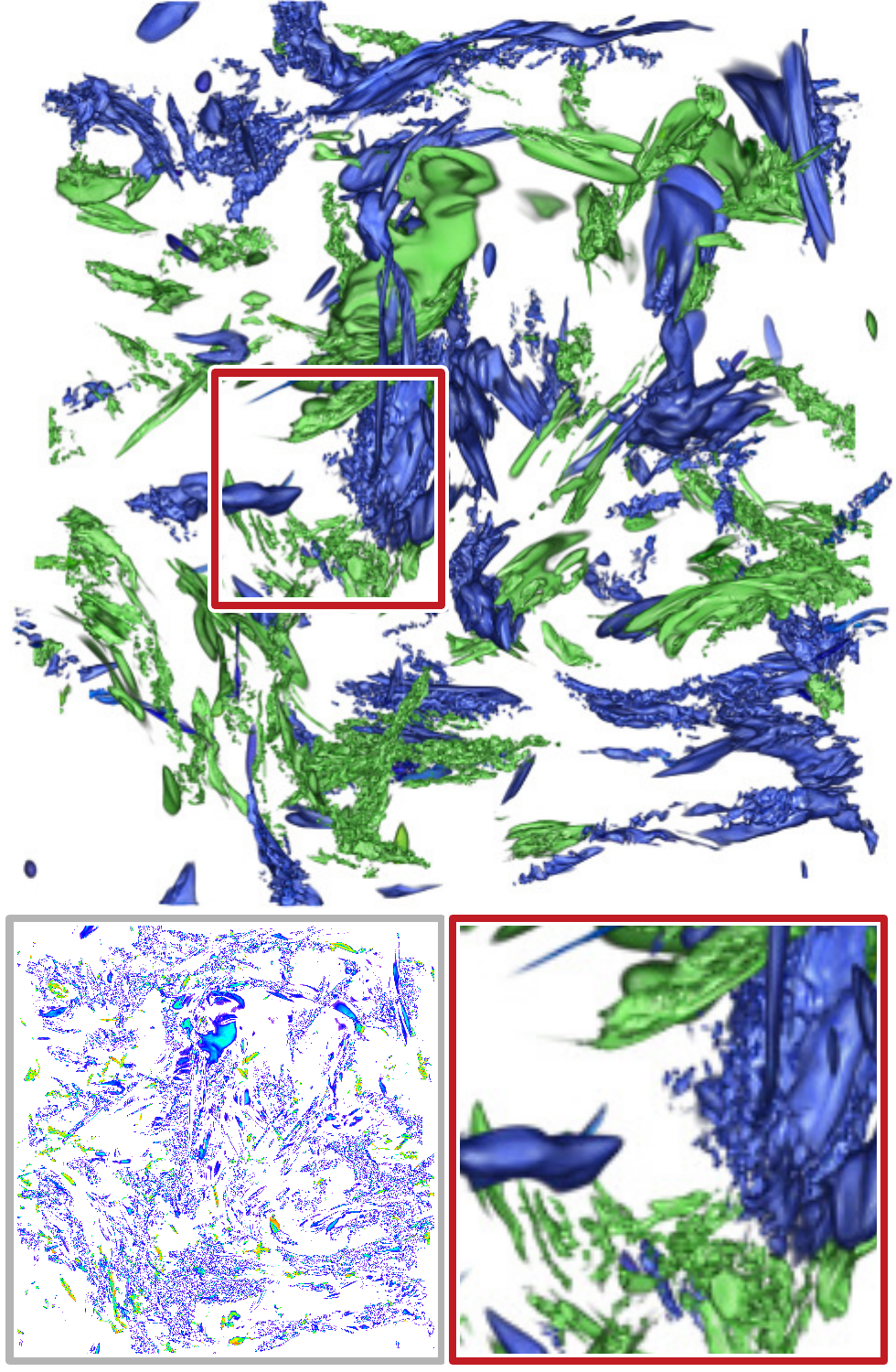}&
  \includegraphics[width=0.185\linewidth]{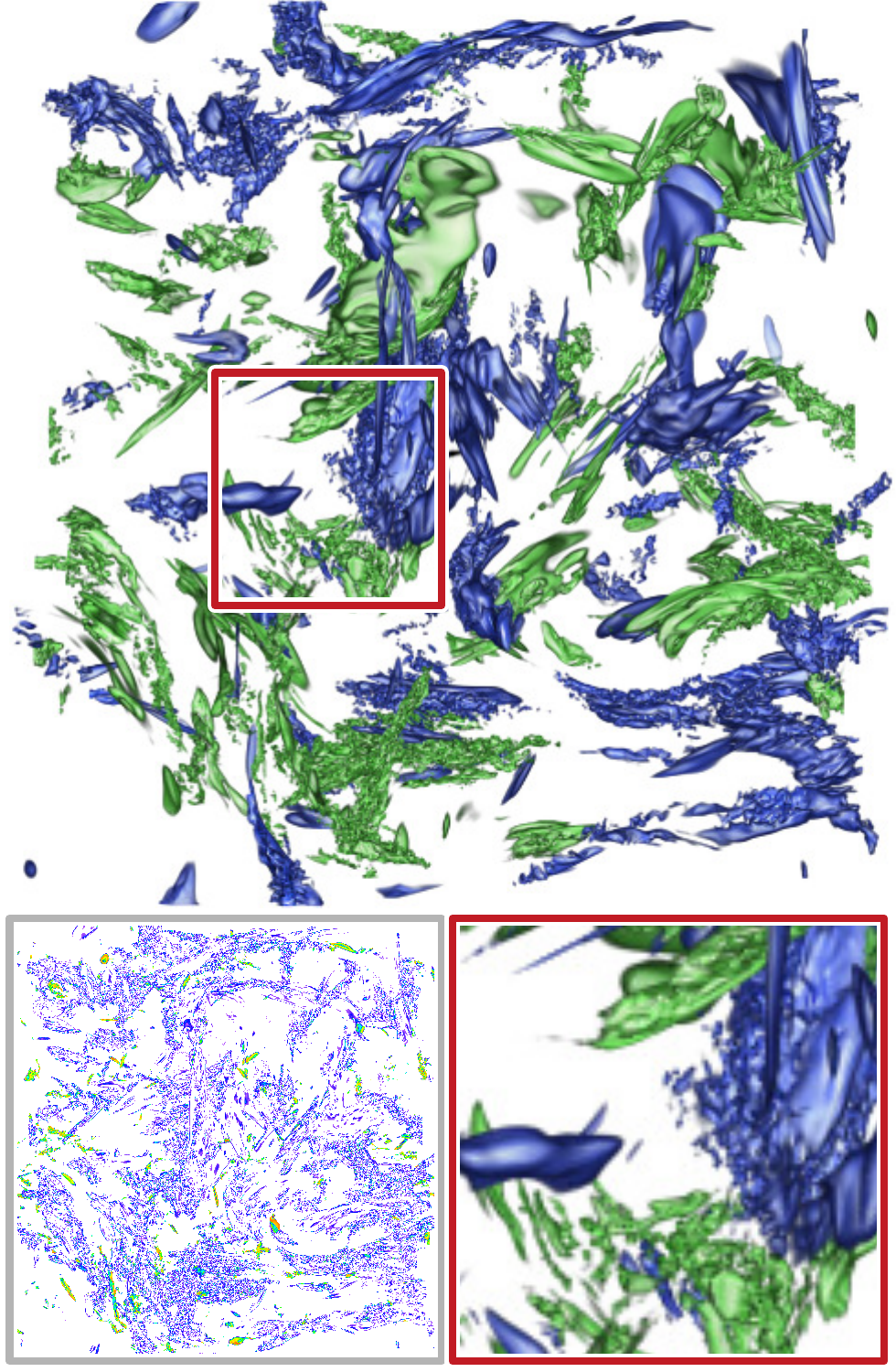}&
\includegraphics[width=0.185\linewidth]{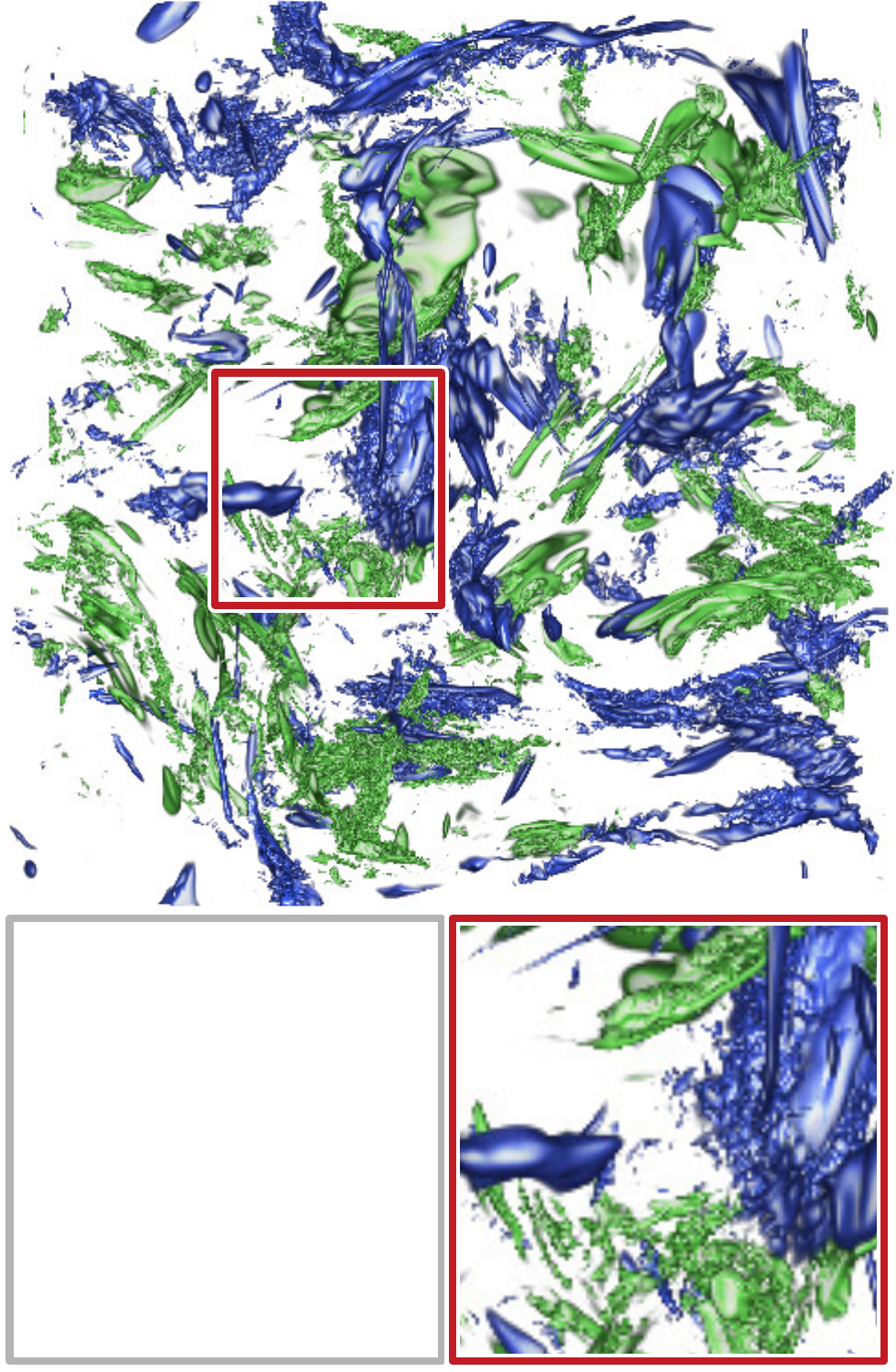}\\
 \includegraphics[width=0.185\linewidth]{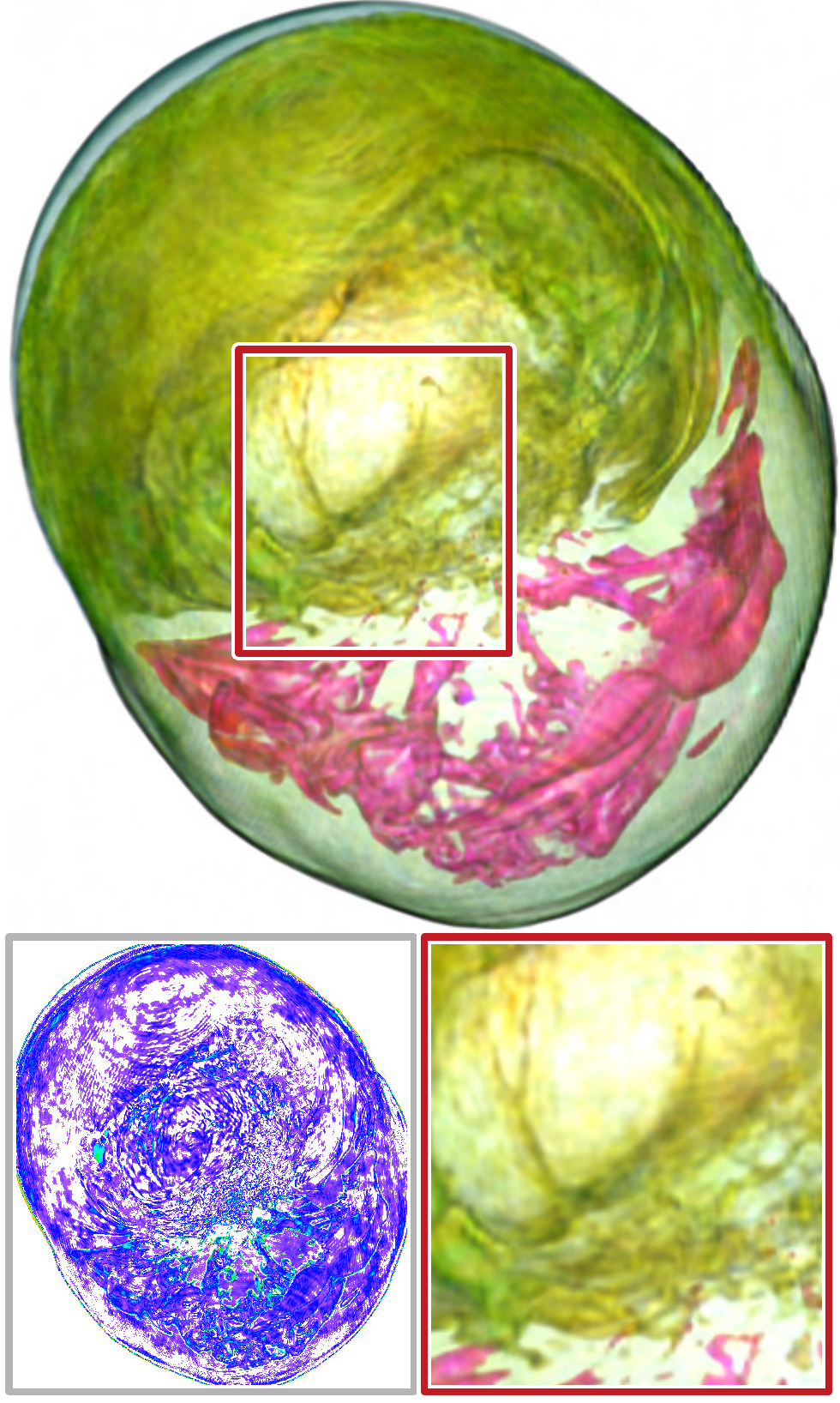}&
 \includegraphics[width=0.185\linewidth]{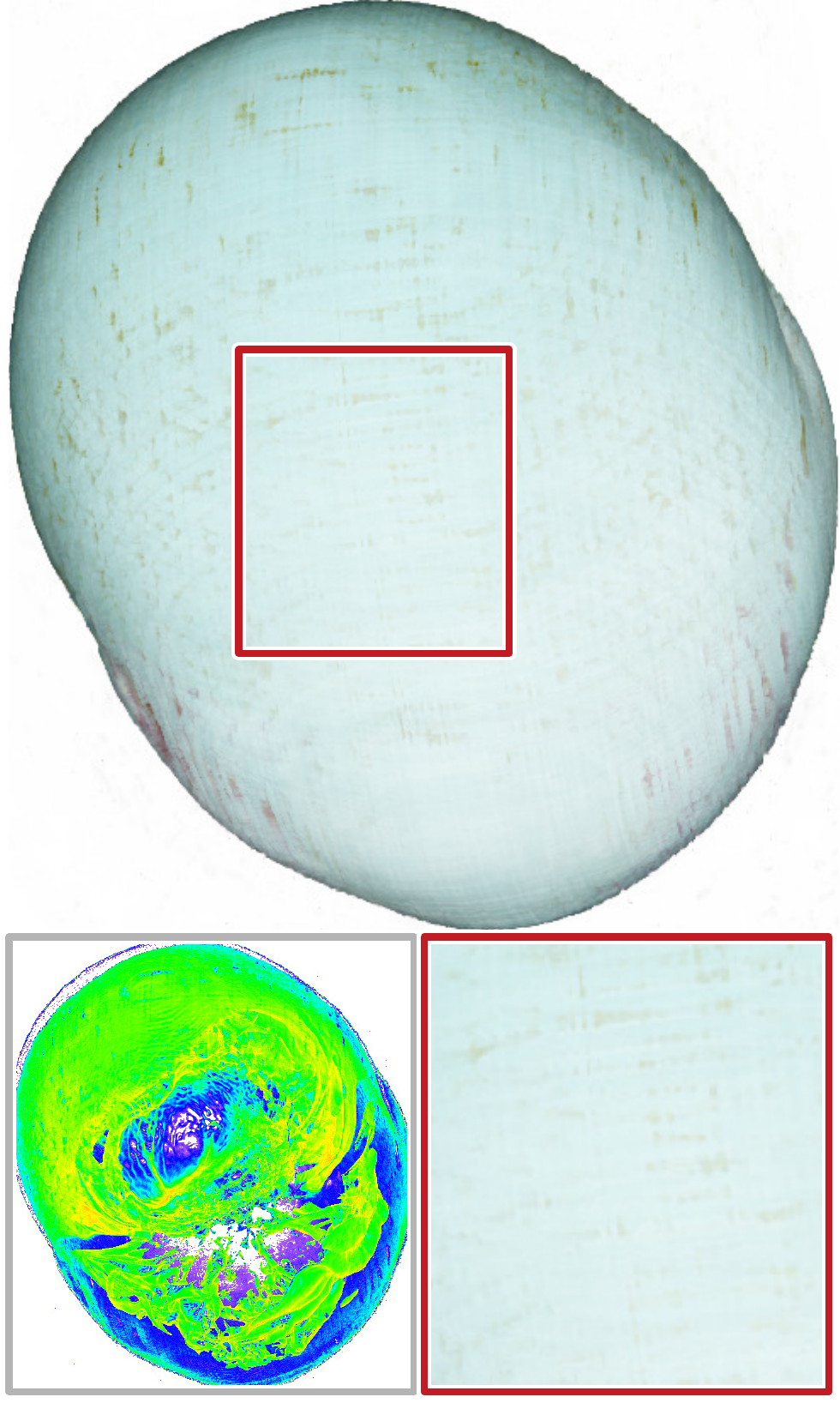}&
 \includegraphics[width=0.185\linewidth]{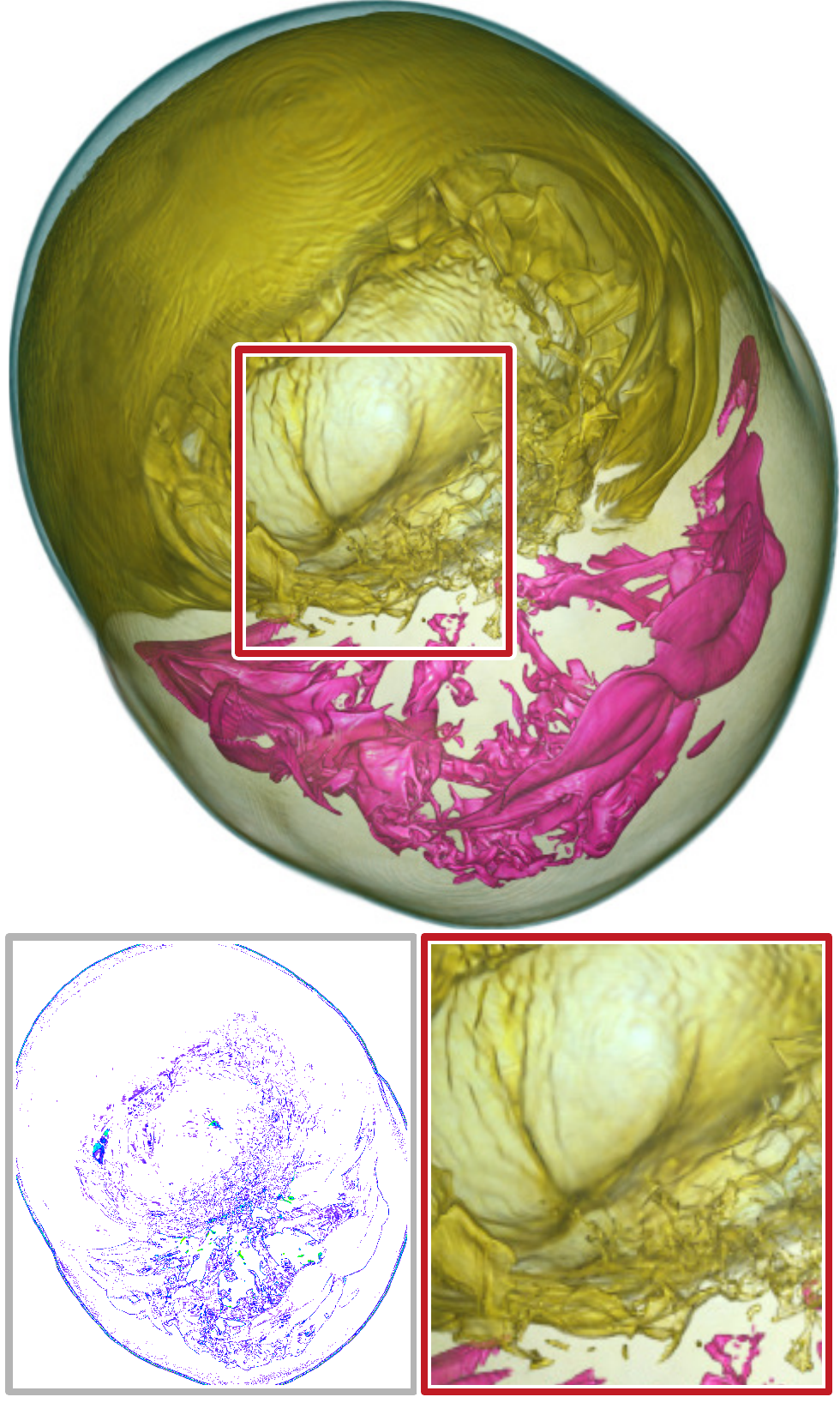}&
  \includegraphics[width=0.185\linewidth]{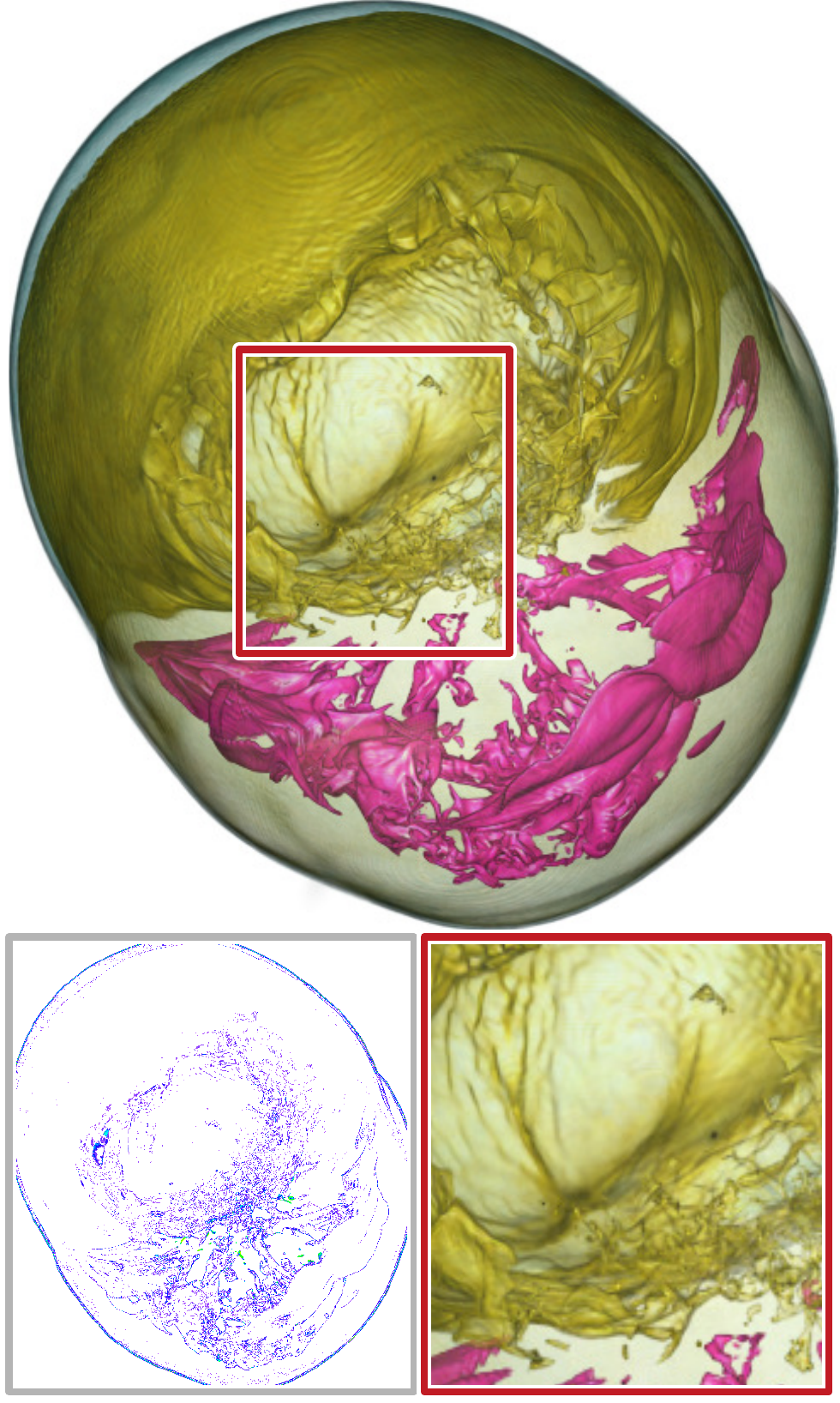}&
\includegraphics[width=0.185\linewidth]{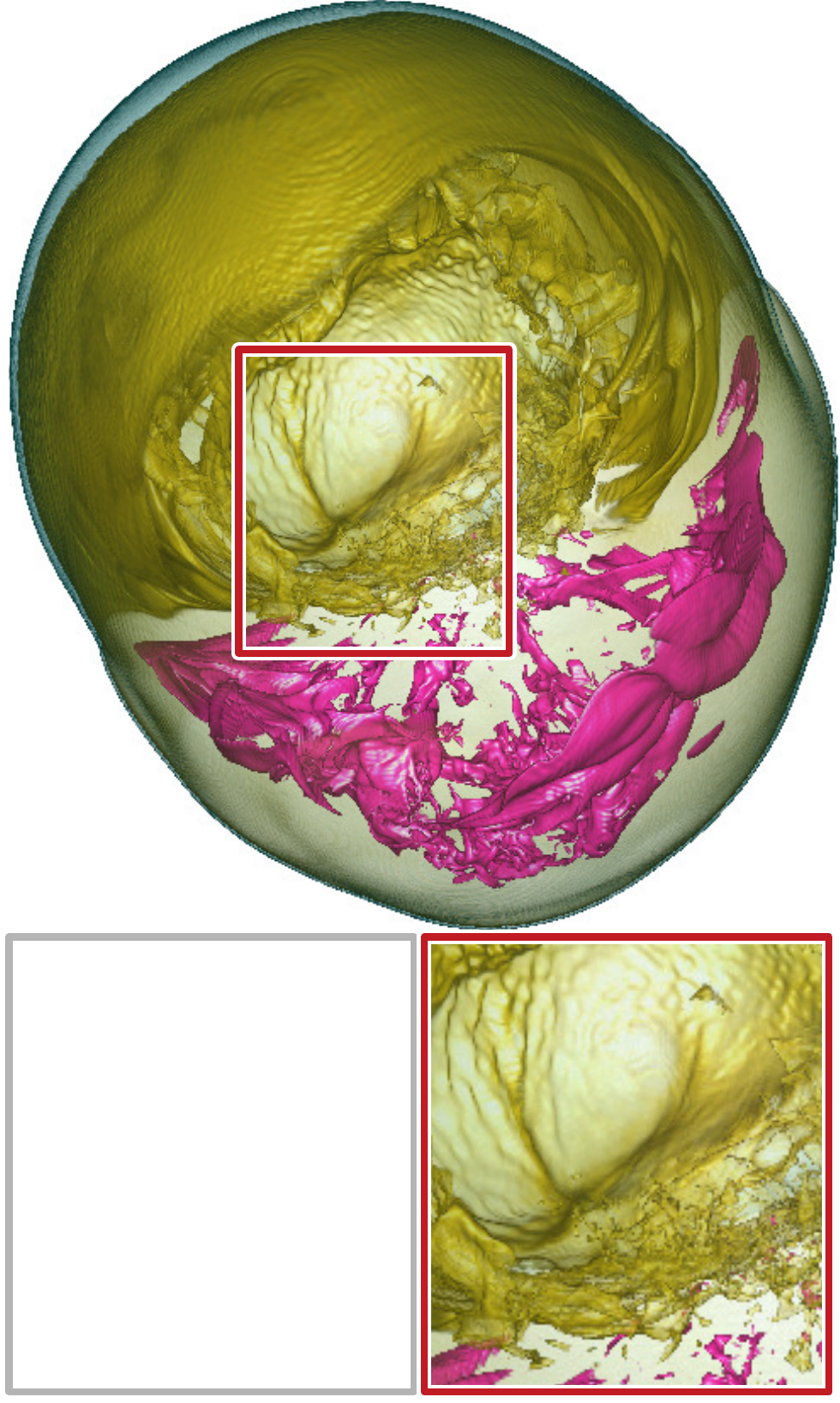}\\
\mbox{\footnotesize (a) Plenoxels} & \mbox{\footnotesize (b) CCNeRF} & \mbox{\footnotesize (c) base 3DGS} &\mbox{\footnotesize (d) iVR-GS} &\mbox{\footnotesize (e) GT} 
\end{array}$
\end{center}
\vspace{-.25in} 
\caption{Comparing scene composing results of four methods w.r.t.\ GT. While base 3DGS and iVR-GS have similar reconstruction quality, iVR-GS supports scene editing, which base 3DGS cannot. Top to bottom: chameleon, five-jet, rotstrat, and supernova.} 
\label{fig:baseline-vr-results}
\end{figure*}

\begin{figure}[!ht]
 \begin{center}
 $\begin{array}{c@{\hspace{0.025in}}c@{\hspace{0.025in}}c@{\hspace{0.025in}}c@{\hspace{0.025in}}c}
 \includegraphics[width=0.185\linewidth]{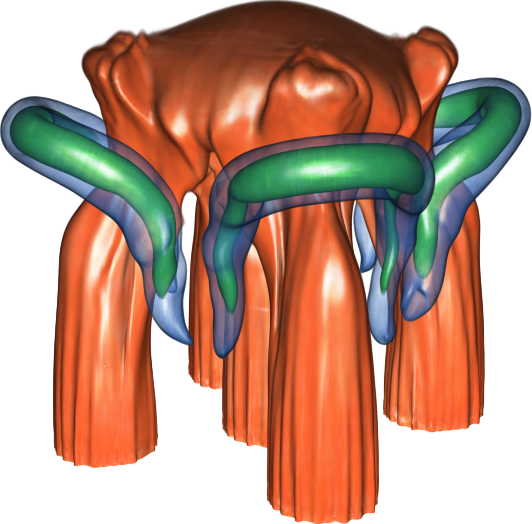}&
 \includegraphics[width=0.185\linewidth]{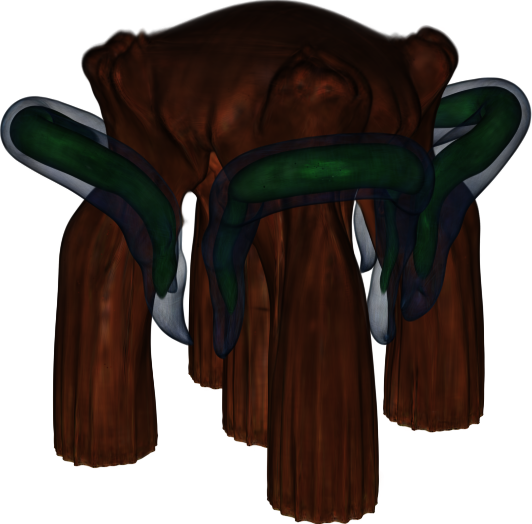}&
 \includegraphics[width=0.185\linewidth]{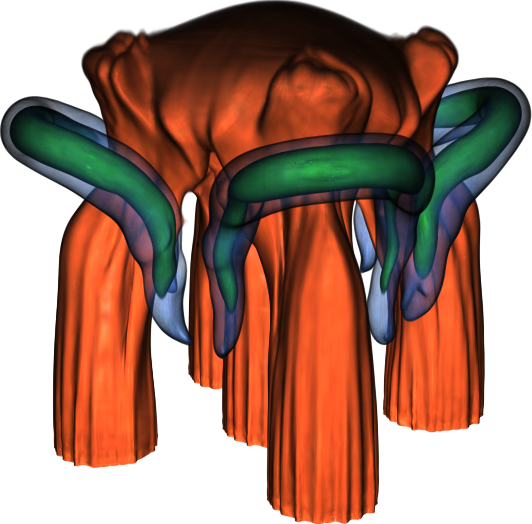}&
  \includegraphics[width=0.185\linewidth]{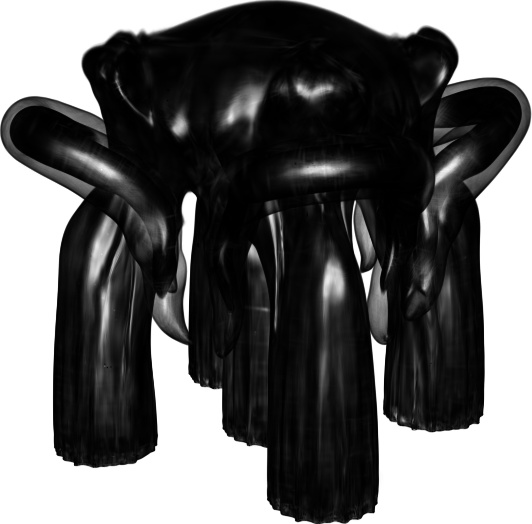}&
\includegraphics[width=0.185\linewidth]{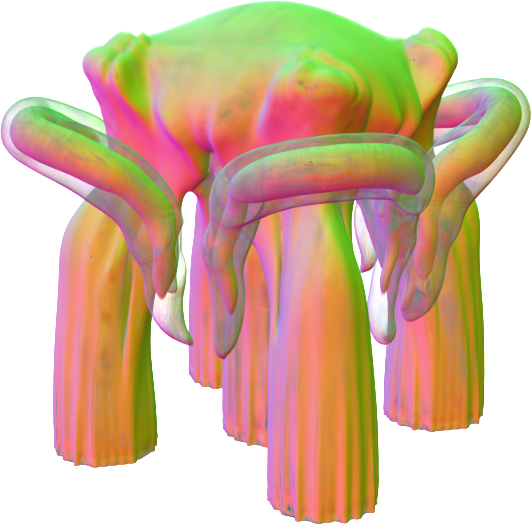}\\
 \includegraphics[width=0.185\linewidth]{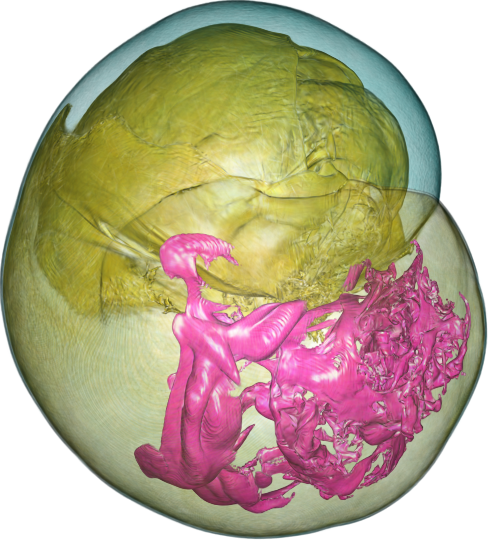}&
 \includegraphics[width=0.185\linewidth]{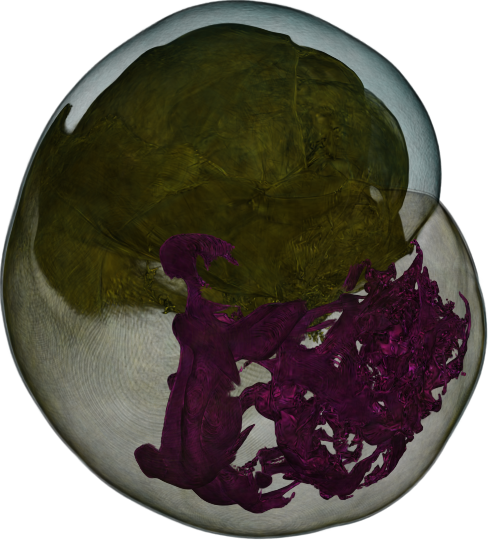}&
 \includegraphics[width=0.185\linewidth]{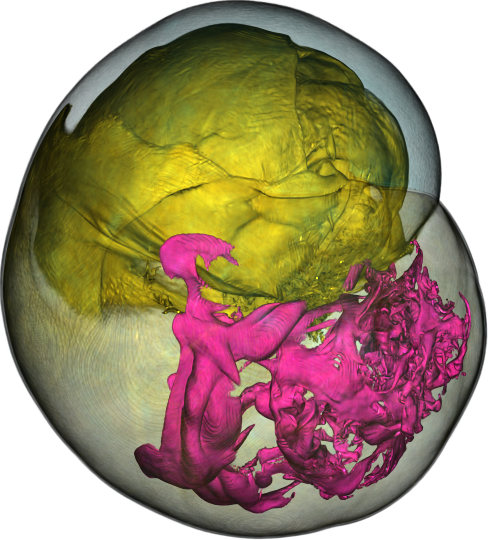}&
  \includegraphics[width=0.185\linewidth]{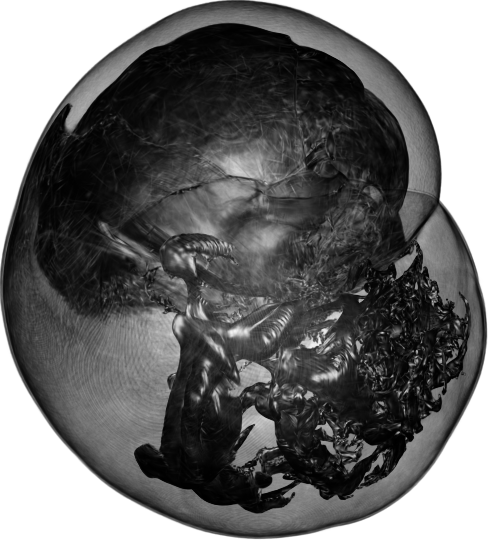}&
\includegraphics[width=0.185\linewidth]{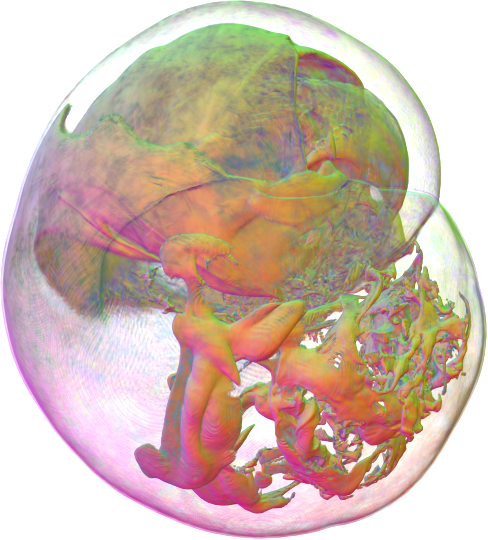}\\
\mbox{\footnotesize (a) lighting} & \mbox{\footnotesize (b) ambient} & \mbox{\footnotesize (c) diffuse} &\mbox{\footnotesize (d) specular} &\mbox{\footnotesize (e) normal} 
\end{array}$
\end{center}
\vspace{-.25in} 
\caption{Qualitative results of iVR-GS estimation on different Blinn-Phong terms under a novel viewpoint. Top and bottom: five-jet and supernova.} 
\label{fig:baseline-decompose}
\end{figure}

\vspace{-0.05in}
\section{Results and Discussion}


\subsection{Evaluation on Scene Composing}
\label{subsec:sceneComposing}

{\bf Datasets and training.}
To demonstrate the superior reconstruction quality and composability of iVR-GS compared with other state-of-the-art NVS methods, we evaluate it with datasets listed in Table~\ref{tab:dataset-partTFs}.
The sizes of these datasets range from small (64 MB) to large (32 GB).
The basic TFs for each dataset are manually selected to extract meaningful objects with appropriate colors and opacities.
For a fair comparison with baseline methods, we do not apply the viewpoint sampling strategy mentioned in Section~\ref{subsec:ViewpointSelection} to reduce the number of images for iVR-GS training.
Instead, we use icosphere sampling to generate 162 multi-view images with Paraview using the NVIDIA IndeX plugin for each basic scene. 
To ensure clarity when viewing the VolVis scene from arbitrary angles, we employ a headlight with different lighting parameters for each volume dataset to render images with 800$\times$800 resolution.
All volume datasets are rendered on a high-performance cluster using an NVIDIA A40 GPU with 48 GB memory.

\begin{table}[htb]
\caption{Datasets for scene composing.}
\vspace{-0.1in}
\centering
\resizebox{\columnwidth}{!}{
\begin{tabular}{c|cc|ccc}
   	     & volume &volume  &\# basic &  CPU/GPU  & rendering     \\ 
 dataset & resolution   &size  &  scenes &  memory  &  time \\ \hline
 chameleon & 1024$\times$1024$\times$1080 &4.2 GB &2 &12.3 GB/7.1 GB &361 ms \\
 five-jet & 256$\times$256$\times$256 &64 MB &3 &1.5 GB/1.6 GB &59 ms \\
 rotstrat & 2048$\times$2048$\times$2048  &32 GB &2 &74.8 GB/45.5 GB &1651 ms \\
 supernova &  864$\times$864$\times$864  &2.4 GB  &3 &7.9 GB/4.6 GB   &212 ms  
\end{tabular}
}
\label{tab:dataset-partTFs}
\end{table}

We use the Adam optimizer to train iVR-GS in two stages: the first with 30,000 iterations for base 3DGS and the second with 10,000 iterations for editable Gaussians.
For both stages, the coefficients for the linear combination of L1 and SSIM losses are 0.8 and 0.2, respectively.
The L1 regularization term in editable Gaussian training is scaled by 0.1, and we scale all other regularization terms and normal consistency loss by 0.01.
For normal and shading attributes, we set their learning rates as 0.01.
Other attributes (such as opacity $o$) are trained in the same way as their optimization in the original 3DGS~\cite{Kerbl-TOG23}.

\begin{table*}[!ht]
\caption{Average PSNR (dB), LPIPS, rendering time, FPS, and total training time for individual basic scenes, as well as average PSNR (dB), LPIPS, rendering time, FPS, and total file size for the composed VolVis scene. The best ones for each dataset are shown in bold.}
\vspace{-0.1in}
\centering
\resizebox{6in}{!}{
\begin{tabular}{c|c|ccccc|cccccc}
 &  & \multicolumn{5}{c|}{individual basic scenes} & \multicolumn{5}{c}{composed VolVis scene}\\ \cline{3-12}
dataset & method & PSNR$\uparrow$ & LPIPS$\downarrow$  & rendering time$\downarrow$ & FPS$\uparrow$ & training time$\downarrow$
& PSNR$\uparrow$ & LPIPS$\downarrow$  & rendering time$\downarrow$ & FPS$\uparrow$ & file size$\downarrow$\\ \hline
\multirow{4}{*}{chameleon} 
&Plenoxels&24.89&0.099&38.1 ms & 26&\textbf{8.8 min} &19.16&0.162&44.5 ms & 22 &356 MB\\
&CCNeRF&25.24&0.063&1.389 s & 0.7&16.6 min &23.23&0.115&1.870 s & 0.5 &\textbf{3.0 MB}\\
&base 3DGS&27.54&0.032&4.8 ms & 208&9.4 min &26.45&0.046&\textbf{4.9 ms} & \textbf{204} &44.8 MB\\
&iVR-GS&\textbf{27.92}&\textbf{0.031}&\textbf{4.5 ms} & \textbf{222}&13.8 min &\textbf{27.19}&\textbf{0.042}&5.2 ms & 192 &6.2 MB\\ \hline
\multirow{4}{*}{five-jet} 	
&Plenoxels&31.44&0.033&34.5 ms &29&\textbf{12.9 min}  &23.20&0.069&47.3 ms &21 &448 MB\\
&CCNeRF&31.36&0.023&1.330 s &0.8 &23.4 min &21.37&0.072 &2.410 s &0.4 &\textbf{4.5 MB}\\
&base 3DGS&34.49&0.013&\textbf{4.6 ms} &\textbf{217}&14.4 min &30.12&0.022&\textbf{5.4 ms} &\textbf{185} &40.0 MB\\
&iVR-GS&\textbf{36.44}&\textbf{0.010}&5.2 ms &192&23.7 min &\textbf{32.22}&\textbf{0.017}&7.3 ms &137 &8.8 MB\\ \hline
\multirow{4}{*}{rotstrat}  
&Plenoxels&24.57&0.105&34.3 ms &29&\textbf{7.6 min} &22.18&0.135&41.2 ms &24 &316 MB\\
&CCNeRF&24.87&\textbf{0.063}&1.539 s &0.6&21.0 min &22.65&\textbf{0.074}&2.278 s &0.4 &\textbf{3.0 MB}\\
&base 3DGS&25.16&0.080&\textbf{4.6 ms} &\textbf{217}&9.0 min &22.95&0.083&\textbf{4.9 ms} &\textbf{204} &51.5 MB\\
&iVR-GS&\textbf{25.22}&0.079&\textbf{4.6 ms} &\textbf{217}&13.6 min &\textbf{23.04}&0.080&5.1 ms &196 &7.3 MB \\ \hline
\multirow{4}{*}{supernova} 
&Plenoxels&30.08&0.064&30.9 ms &32&\textbf{13.2 min} &22.53&0.138&41.4 ms &24 &374 MB\\
&CCNeRF&29.11&0.062&1.549 s &0.6&33.0 min &12.45&0.207&2.806 s &0.4 &\textbf{4.5 MB}\\
&base 3DGS&31.60&0.038&\textbf{4.5 ms} &\textbf{222}&14.4 min &28.73&0.061&\textbf{5.0 ms} &\textbf{200} &48.1 MB\\
&iVR-GS&\textbf{32.20}&\textbf{0.035}&5.0 ms &200&23.4 min &\textbf{29.20}&\textbf{0.054}&7.4 ms &135 &10.0 MB
\end{tabular}
}
\label{tab:compose-eval}
\end{table*}

{\bf Baselines and metrics.}
We compare three baseline methods on reconstruction quality, rendering speed, model size, and composing performance with iVR-GS.
We optimize these baseline methods in the same way as iVR-GS, where we first train multiple individual basic models on each basic scene and then compose them into one model to make the entire VolVis scene visible.
\begin{myitemize}
\vspace{-0.05in}
	\item Plenoxels~\cite{Fridovich-Keil-CVPR22} is a NeRF-based method that represents a scene using an explicit sparse 3D grid with SH coefficients for modeling view-dependent colors. It supports composability by summing multiple grid parameters from each basic model.
	\item CCNeRF~\cite{Tang-neurips22} is a NeRF-based method that models a scene with multiple compact low-rank tensor components without a neural network. The composability is accomplished by concatenating parameters along the rank dimension. Specifically, CCNeRF-CP model architecture is utilized for comparison.
	\item Base 3DGS is the first stage training result of iVR-GS. As a baseline method, the normal attribute of each Gaussian does not influence the rendering result and will be discarded to reduce model size. It is similar to the original 3DGS~\cite{Kerbl-TOG23}, except the alpha channel optimization is added to improve reconstruction quality. We compose multiple base 3DGS models in a way similar to iVR-GS.
\end{myitemize}
For rendering results of iVR-GS and other baseline methods, we evaluate them on 181 images rendered from the volume with gradually increasing polar and azimuthal angles.
Besides evaluating the basic scenes, we compose TFs of all basic scenes in one dataset to render GT images of the composed VolVis scene.
We then evaluate the NVS methods' composability by composing NVS models of all basic scenes. 
We render the scene and compare the reconstruction quality of rendered results with the GT images.
The training and inference of all methods were performed on a local workstation with an NVIDIA RTX 4090 GPU. 
We leverage {\em peak signal-to-noise ratio} (PSNR) and {\em learned perceptual image patch similarity} (LPIPS)~\cite{Zhang-CVPR18} to measure image quality.

{\bf Qualitative comparison.}
As shown in Figure~\ref{fig:baseline-vr-results}, point-based rendering methods iVR-GS and base 3DGS significantly outperform NeRF-based methods regarding rendering quality on scene composing.
The potential reasons are as follows.
For NeRF-based methods, the scene is represented by a continuous grid, where color and density at any point are determined by interpolating neighboring grid parameters. 
In VolVis, however, different basic scenes often form a nested relationship, leading to interference when composing the parameters of multiple NeRF-based models. 
Such interference may disrupt the original grid representation and potentially lead to incorrect results (e.g., CCNeRF in Figure~\ref{fig:baseline-vr-results}).
Although increasing the grid resolution could improve quality, this will also lead to significantly larger model sizes, such as Plenoxels.
Meanwhile, the interpolation process among neighboring grid parameters may also bring blurry rendering 
(refer to zoom-in results of Plenoxels and CCNeRF on rotstrat).
Unlike NeRF-based methods, iVR-GS and base 3DGS employ discrete and independent Gaussian primitives for scene representation to avoid parameter interference in scene composing. 
Although the gap in reconstructed quality between iVR-GS and base 3DGS is small, iVR-GS incorporates the Blinn-Phong reflection model and allows scene editing, whereas base 3DGS bakes a scene that cannot be changed. 

Figure~\ref{fig:baseline-decompose} shows the estimation of iVR-GS on different Blinn-Phong terms, including ambient, diffuse, specular, and normal. 
Additionally, when closely examining the difference images and zoom-in figures in Figure~\ref{fig:baseline-vr-results}, we can observe that iVR-GS achieves more accurate reconstruction of specular highlights than base 3DGS.
Such superior accuracy can be attributed to the representation of light reflection within editable Gaussians.

{\bf Quantitative comparison.}
We report quantitative results when evaluating the individual basic scenes and composed VolVis scenes in Table~\ref{tab:compose-eval}.
iVR-GS achieves the best reconstruction quality for all datasets, only losing to CCNeRF on the LPIPS value of the rotstrat dataset.
Notably, the PSNR and LPIPS value gaps between iVR-GS and base 3DGS are more significant on five-jet than the other three datasets.
This is because the lighting effect in the five-jet scene is more evident, and iVR-GS performs better than base 3DGS in reconstructing the lighting effect.
CCNeRF generally requires the most training and inference time, except for the five-jet dataset's training time. 
This can be attributed to its less explicit structure and missing CUDA accelerated rendering.
Note that all NVS methods do not access the original volume data during training or inference, and the memory usage of NVS models is independent of the volume resolution.
For all methods, the training and inference memory costs remain below 10 GB for GPU, indicating that they all can be trained and inferred on most machines with a single consumer-level GPU.
The rendering speeds of iVR-GS and base 3DGS are comparable among all datasets and are dramatically faster than Plenoxels, CCNeRF, and even DVR used to generate GT images.
Additionally, the file size of iVR-GS is smaller than base 3DGS in terms of the composed VolVis scene, indicating the need for VQ compression when composing multiple models.

\begin{table}[!htb]
\caption{Datasets for scene editing and inverse volume exploration.}
\vspace{-0.1in}
\centering
\resizebox{\columnwidth}{!}{
\begin{tabular}{c|cc|ccc}
   	     & volume   	&volume  &\# basic		&  CPU/GPU  &      rendering \\ 
 dataset & resolution  	 		&size & scenes	&  memory 	&  time    \\ \hline
 combustion & 1920$\times$2880$\times$480  &9.8 GB &38 &24.6 GB/13.1 GB &502 ms    \\
 five-jet & 256$\times$256$\times$256  &64 MB &10 &1.5 GB/1.6 GB &59 ms \\
 vortex		& 512$\times$512$\times$512  &512 MB &29 & 3.3 GB/2.1 GB &61 ms \\

\end{tabular}
}
\label{tab:dataset-allTFs}
\end{table}

\vspace{-0.05in}
\subsection{Evaluation of Scene Editing}
\label{subsec:sceneEditing}

{\bf Datasets and training.}
To evaluate the relighting accuracy and editing flexibility of iVR-GS, we experiment with datasets listed in Table~\ref{tab:dataset-allTFs}.
Compared with experiments in Section~\ref{subsec:sceneComposing}, we leverage numerous disjoint opacity bumps with various opacity values as basic TFs to cover a larger number of value ranges of the volume data, where each range gets narrower.
The color of a basic TF is randomly selected, as the optimized iVR-GS supports color editing.
For a volume dataset under each basic TF, we estimate the complexity of the corresponding basic scene following the entropy score computation in Section~\ref{subsec:ViewpointSelection} and generate a certain number of multi-view images.
In particular, for basic scenes with normalized entropy scores of [0.0, 0.1), [0.1, 0.5], and (0.5, 1.0], we sample 42, 92, and 162  training images by adjusting the subdivision level of icosphere sampling.
All training images are generated with a headlight and 800$\times$800 image resolution.
The training of iVR-GS is the same as that of scene composing experiments, except we train each basic iVR-GS model with a different number of multi-view images.
Note that we did not compare iVR-GS with recent BRDF-based relighting models~\cite{Yao-ECCV22,Zhang-ICCV23,Gao-arXiv23}. 
These models are designed for real-world scenes with environment maps to approximate arbitrary illumination. 
They are not directly applicable to synthetic VolVis scenes where the Blinn-Phong reflection model is commonly used.
\begin{table}[!htb]
\caption{Average PSNR (dB), LPIPS, training time for basic scenes, and total file size.}
\vspace{-0.1in}
\centering
\resizebox{\columnwidth}{!}{
\begin{tabular}{c|cc|cc|cc}
			&\multicolumn{2}{c|}{NVS}&\multicolumn{2}{c|}{relighting}&training&total\\\cline{2-5}
dataset		& PSNR$\uparrow$ 	& LPIPS$\downarrow$ & PSNR$\uparrow$ 	& LPIPS$\downarrow$ &  time & file size\\ \hline
combustion 	&34.06		&0.016	&32.17&0.017	& 7.7 min     	& 124.6 MB			\\ 
five-jet	&33.38		&0.048	&32.99&0.021	& 11.3 min		&38.86 MB			\\ 
vortex 		&32.92			&0.026	&30.59&0.036	&10.5 min  		&162.7 MB 			\\ 
\end{tabular}
}
\label{tab:NVSandRelighting}
\end{table}

\begin{figure*}[!htb]
 \begin{center}
 $\begin{array}{c@{\hspace{0.05in}}c@{\hspace{0.05in}}c@{\hspace{0.05in}}c}
 \includegraphics[height=1.325in]{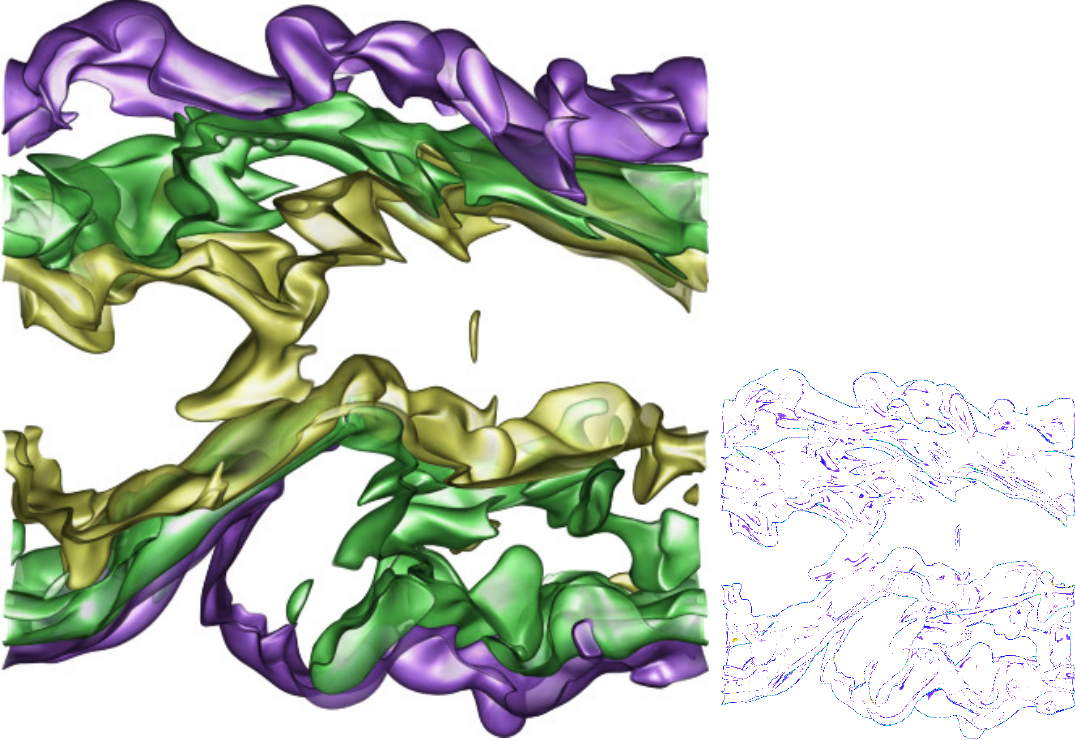}&
 \includegraphics[height=1.325in]{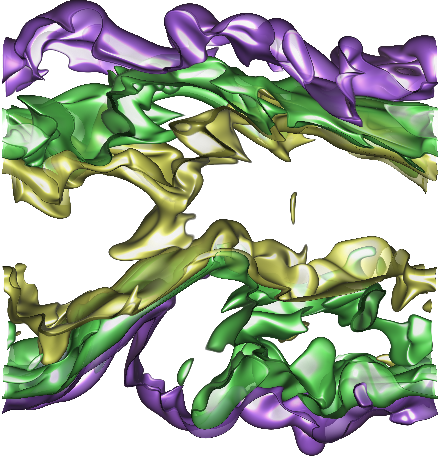}&
 \includegraphics[height=1.325in]{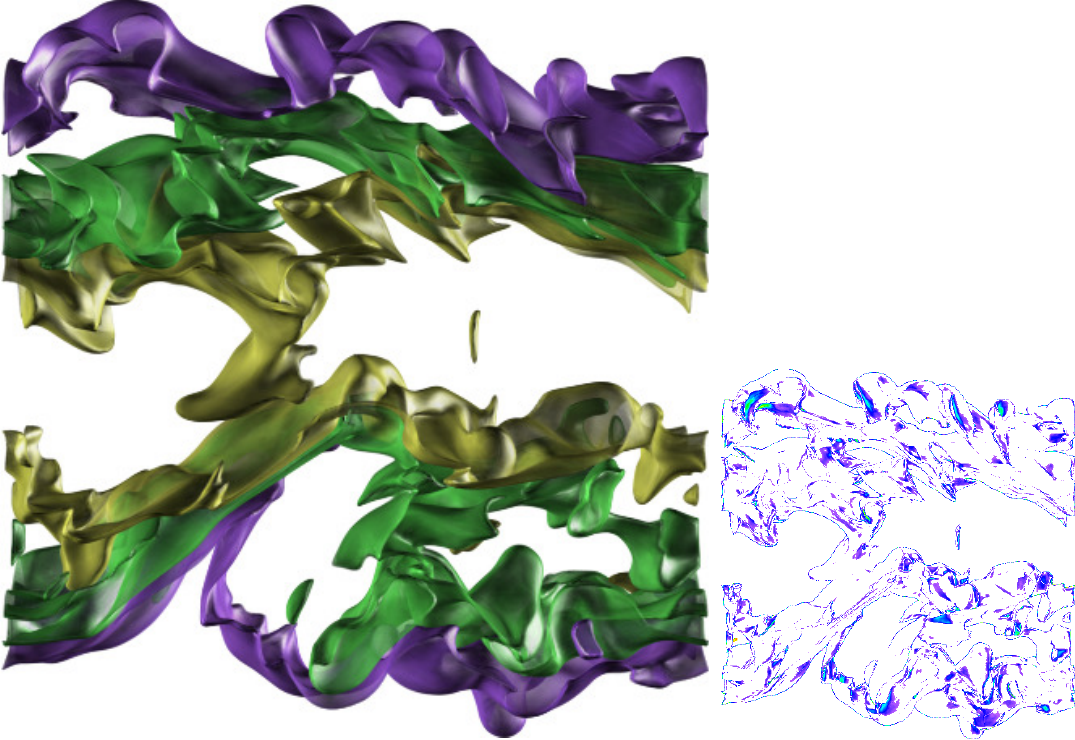}&
\includegraphics[height=1.325in]{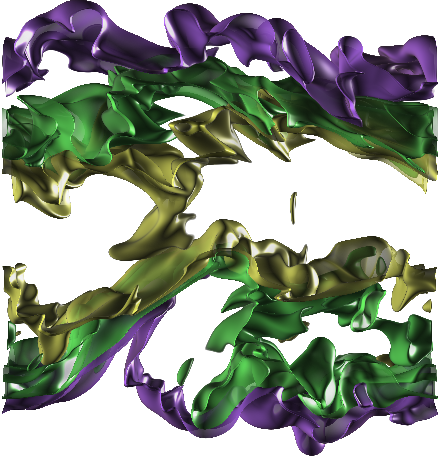}\\
 \includegraphics[height=1.325in]{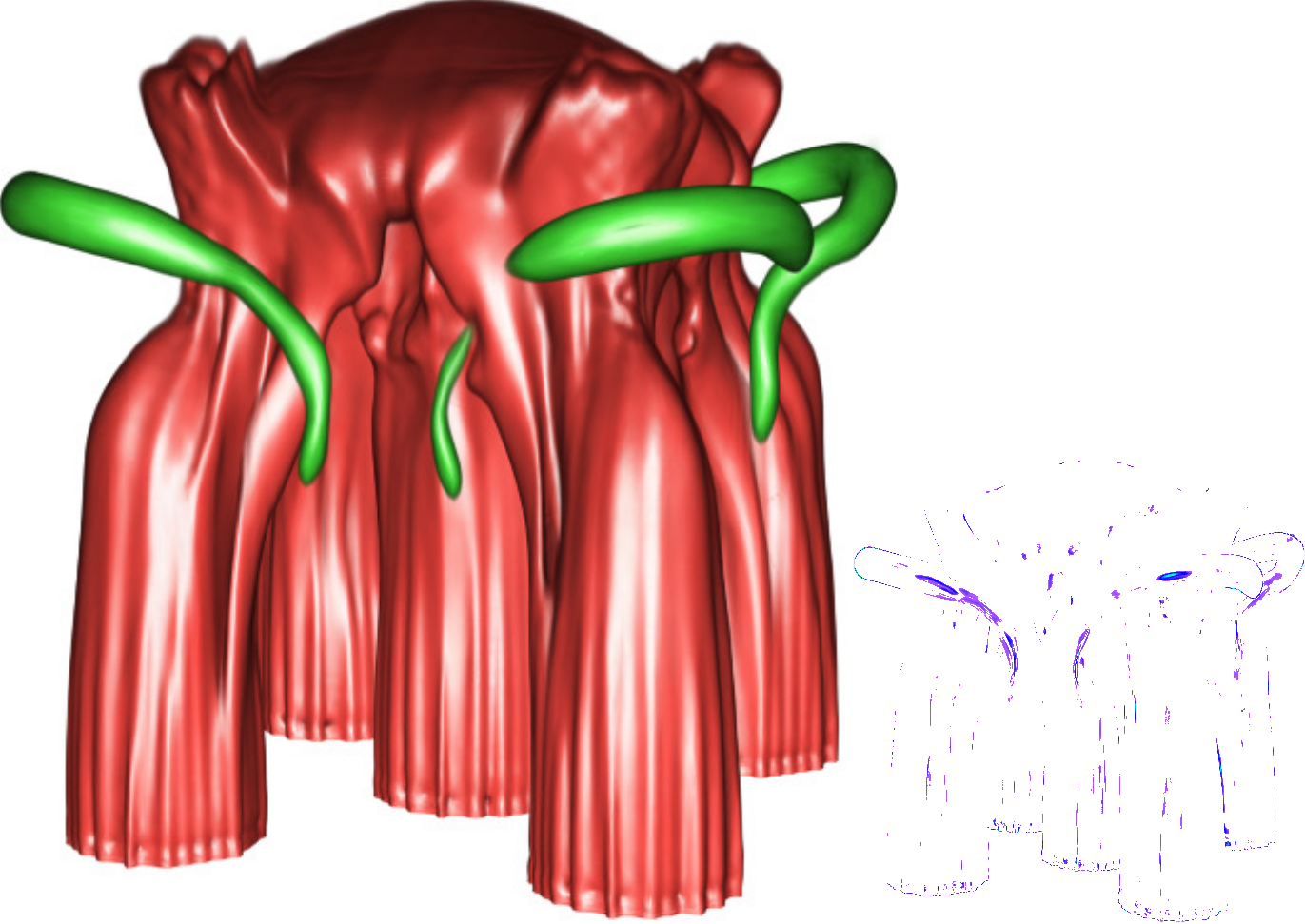}&
 \includegraphics[height=1.325in]{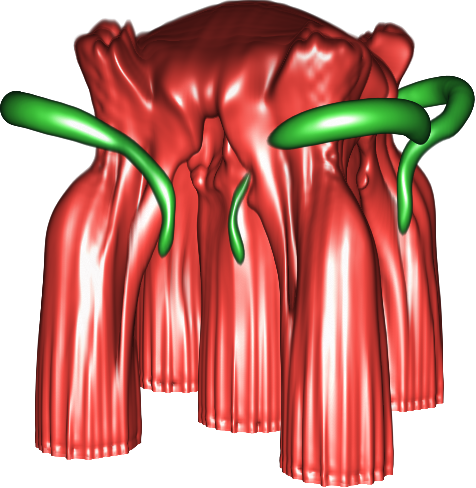}&
 \includegraphics[height=1.325in]{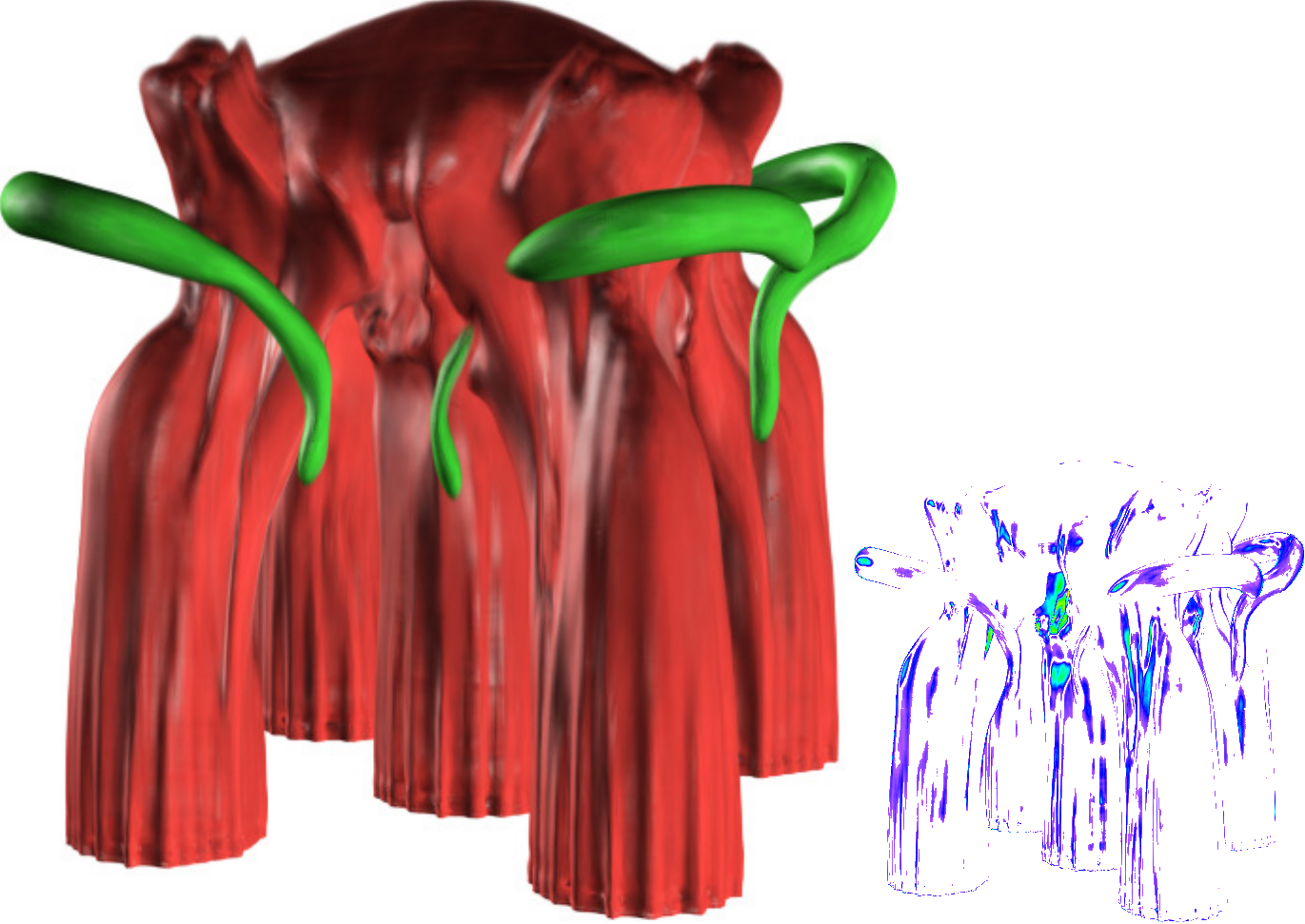}&
\includegraphics[height=1.325in]{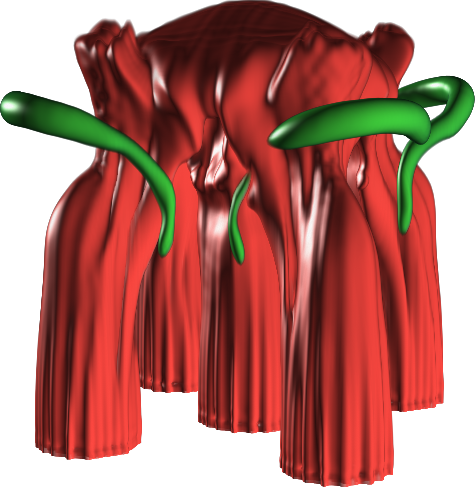}\\
 \includegraphics[height=1.325in]{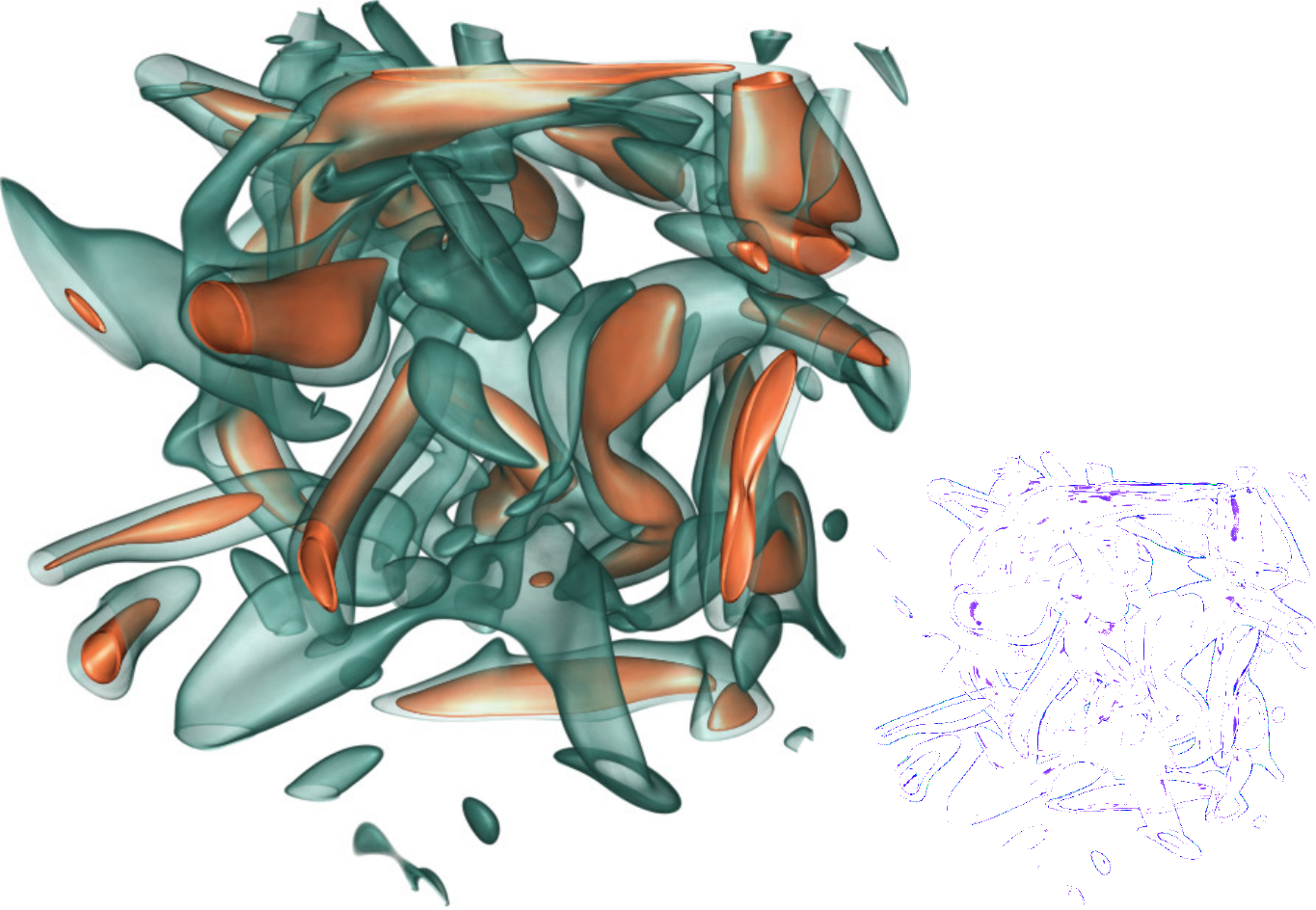}&
 \includegraphics[height=1.325in]{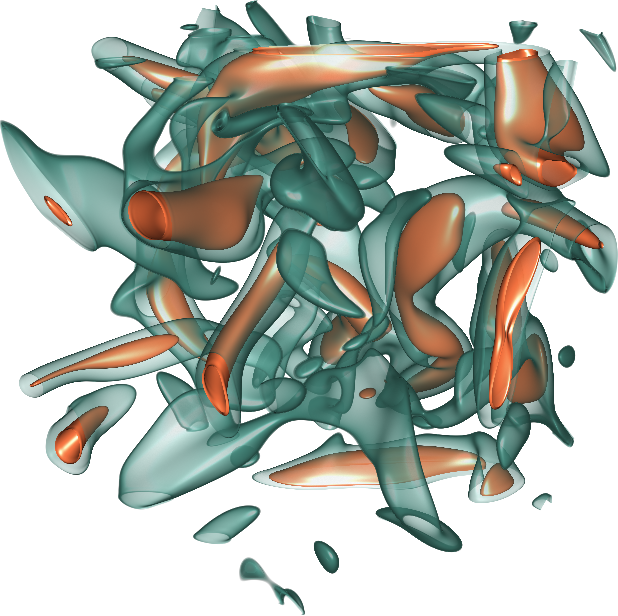}&
 \includegraphics[height=1.325in]{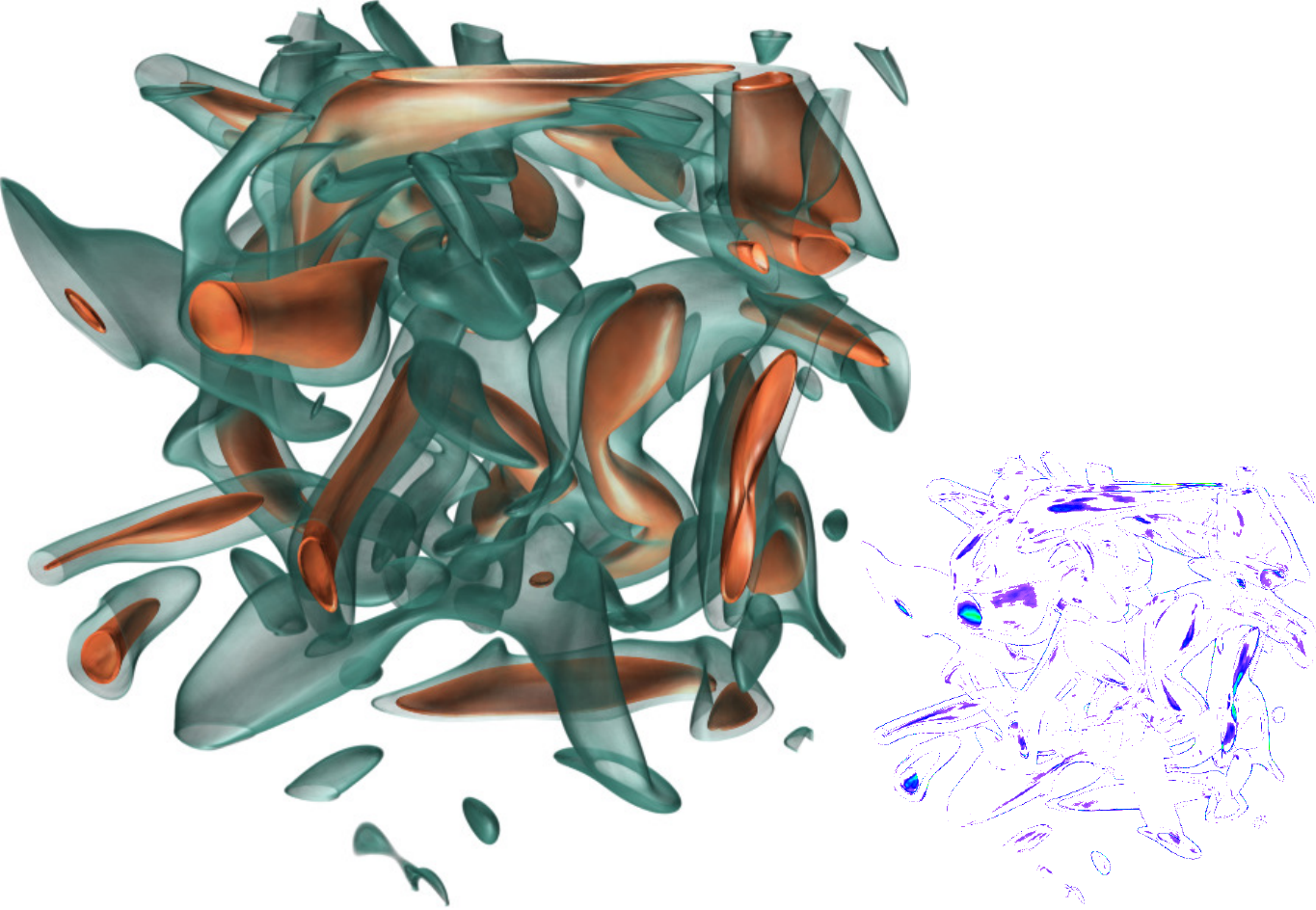}&
\includegraphics[height=1.325in]{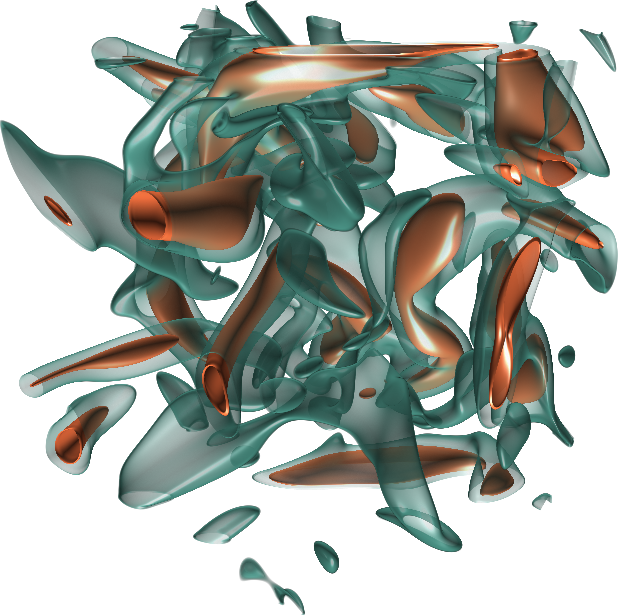}\\
\mbox{\footnotesize (a) NVS} & \mbox{\footnotesize (b) NVS GT} & \mbox{\footnotesize (c) relighting} &\mbox{\footnotesize (d) relighting GT}
\end{array}$
\end{center}
\vspace{-.25in} 
\caption{Comparing NVS and relighting results with GT. Top to bottom: a part of the basic scene composition result for combustion, five-jet, and vortex. The light source moves, respectively, from front to right, front to left, and front to bottom for combustion, five-jet, and vortex.} 
\label{fig:NVS-and-relighting}
\end{figure*}

\begin{figure*}[!htb]
 \begin{center}
 $\begin{array}{c@{\hspace{0.05in}}c@{\hspace{0.05in}}c@{\hspace{0.05in}}c@{\hspace{0.05in}}c}
 \includegraphics[width=0.1825\linewidth]{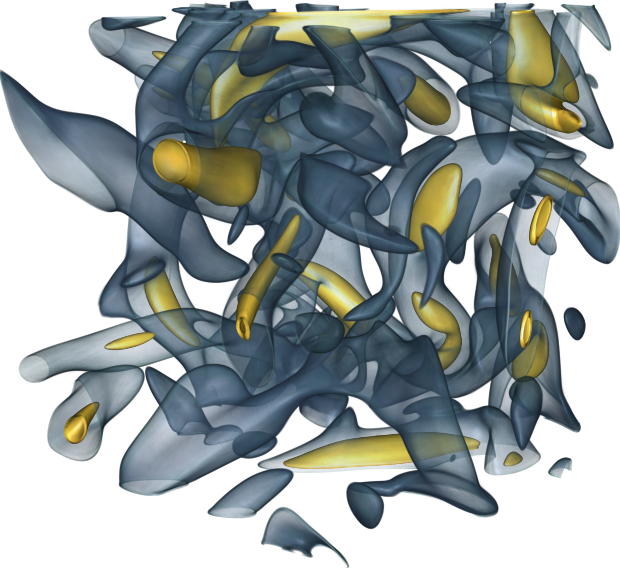}&
 \includegraphics[width=0.1825\linewidth]{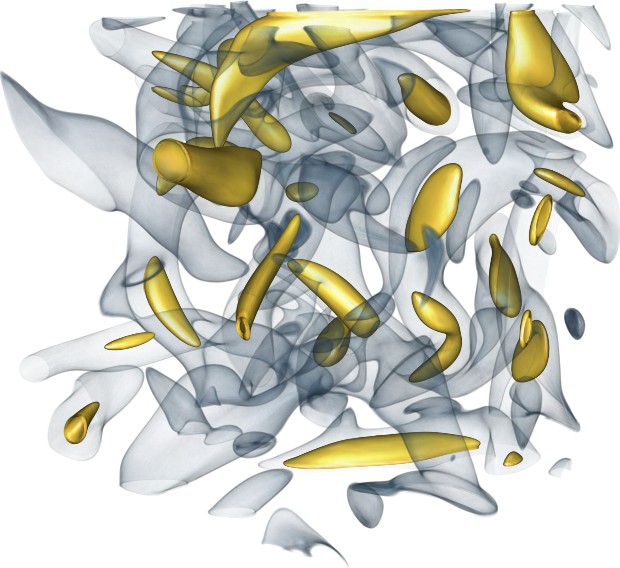}&
 \includegraphics[width=0.1825\linewidth]{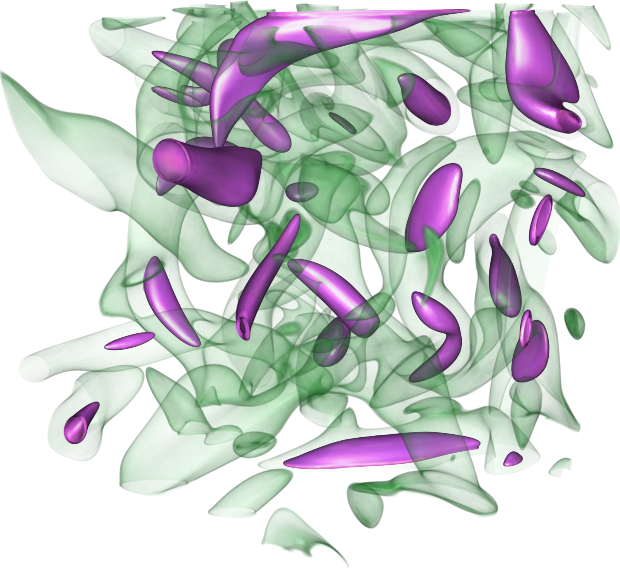}&
 \includegraphics[width=0.1825\linewidth]{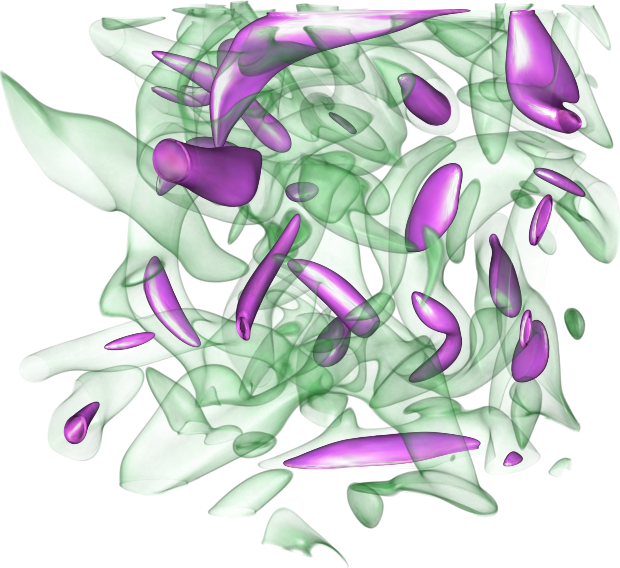}&
\includegraphics[width=0.1825\linewidth]{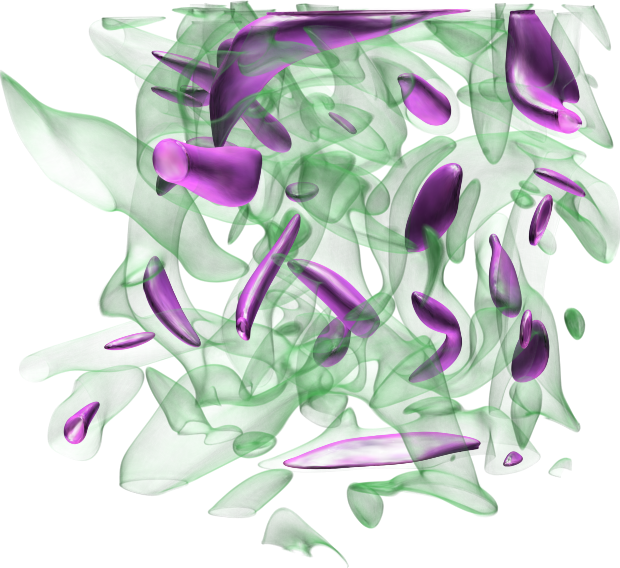}\\
\mbox{\footnotesize (a) selection of basic TFs} & \mbox{\footnotesize (b): (a) + opacity change}& \mbox{\footnotesize (c): (b) + color change} & \mbox{\footnotesize (d): (c) + light mag.\ change} & \mbox{\footnotesize (e): (d) + light dir.\ change}\\
 \includegraphics[width=0.1825\linewidth]{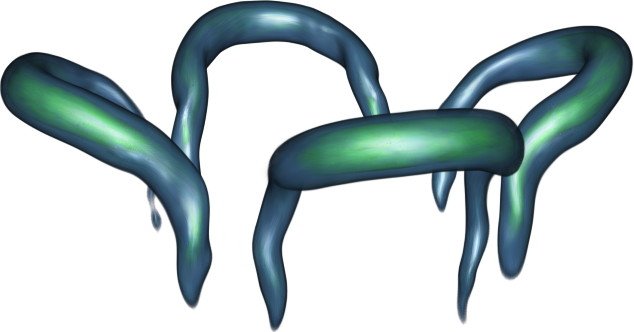}&
  \includegraphics[width=0.1825\linewidth]{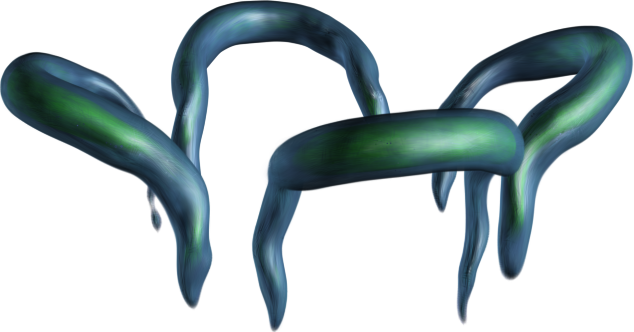}&
  \includegraphics[width=0.1825\linewidth]{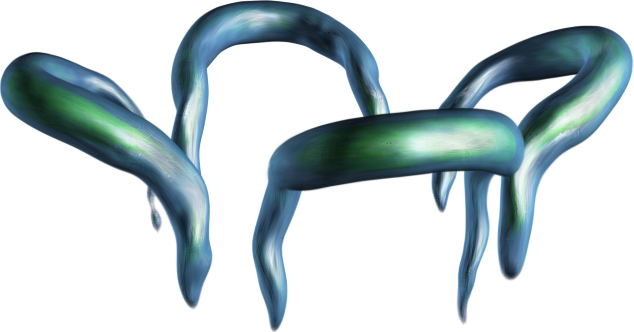}&
 \includegraphics[width=0.1825\linewidth]{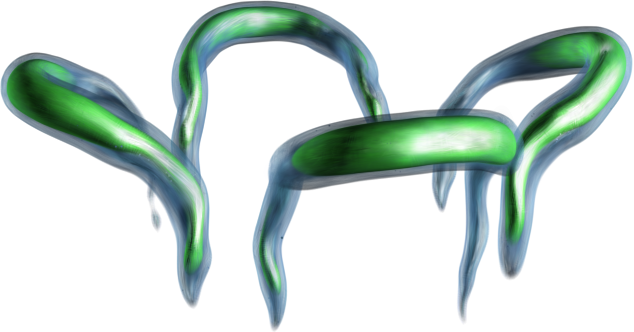}&
\includegraphics[width=0.1825\linewidth]{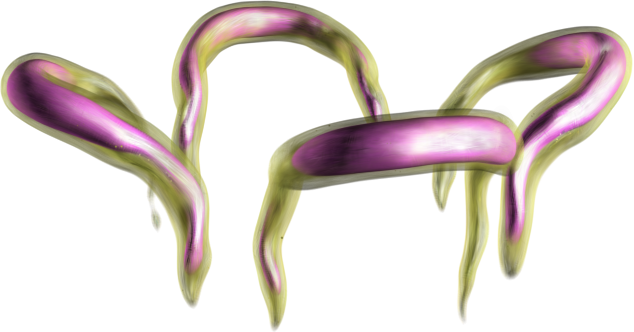}\\
\mbox{\footnotesize (f) selection of basic TFs} & \mbox{\footnotesize (g): (f) + light dir.\ change }& \mbox{\footnotesize (h): (g) + light mag.\ change} & \mbox{\footnotesize (i): (h) + opacity change} & \mbox{\footnotesize (j): (i) + color change}\\
\end{array}$
\end{center}
\vspace{-.25in} 
\caption{Examples of iterative scene editing results with composed iVR-GS models on the vortex and five-jet datasets. 
The light source moves from front to top of the scene from (d) to (e) and from front to right of the scene from (f) to (g).} 
\label{fig:scene-editing}
\end{figure*}

{\bf NVS and relighting on basic scenes.} 
We evaluate all basic iVR-GS models of each dataset on NVS and relighting tasks.
For NVS, we evaluate the model on 181 novel view images, which are uniformly sampled following a camera trajectory from polar and azimuthal angles (-90$^{\circ}$, -180$^{\circ}$) to (90$^{\circ}$, 180$^{\circ}$).
All evaluation images are rendered with one headlight.
For relighting, we use orbital light and change its direction following the same camera trajectory in NVS tasks. 
We evenly sampled three camera angles in between: (-90$^{\circ}$, -180$^{\circ}$), (-45$^{\circ}$,-90$^{\circ}$), (0$^{\circ}$, 0$^{\circ}$), (45$^{\circ}$, 90$^{\circ}$), and (90$^{\circ}$, 180$^{\circ}$), resulting in a total of 905 (181$\times$5) evaluation images for each dataset. 
The quantitative results are reported in Table~\ref{tab:NVSandRelighting}.
Since the relighting task involves inferring images from unseen viewpoints and unseen light directions, 
as expected, PSNR and LPIPS values of the relighting task are inferior to NVS. 
In Figure~\ref{fig:NVS-and-relighting}, we compose a part of the entire scene to show the output image quality of the NVS and relighting tasks.
Even though the specular highlights of relighting results are noticeably worse than NVS results (especially for five-jet) due to the changes in light direction, the overall reconstruction of iVR-GS is faithful for both tasks. 

\begin{figure}[!htb]
 \begin{center}
 $\begin{array}{c@{\hspace{0.025in}}c@{\hspace{0.025in}}c@{\hspace{0.025in}}c}
\includegraphics[height=0.75in]{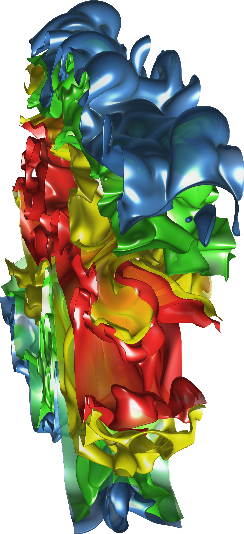}&
 \includegraphics[height=0.75in]{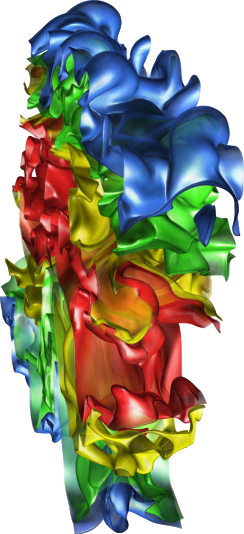}&
 \includegraphics[height=0.75in]{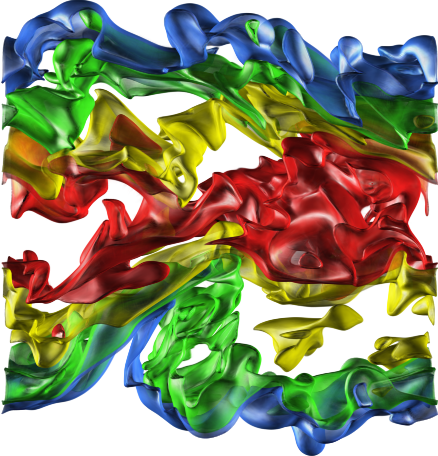}&
\includegraphics[height=0.75in]{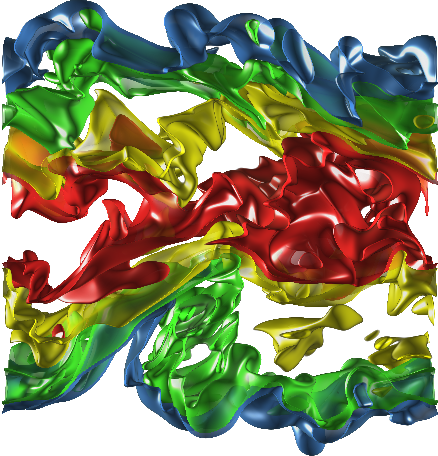}\\
 \includegraphics[height=0.75in]{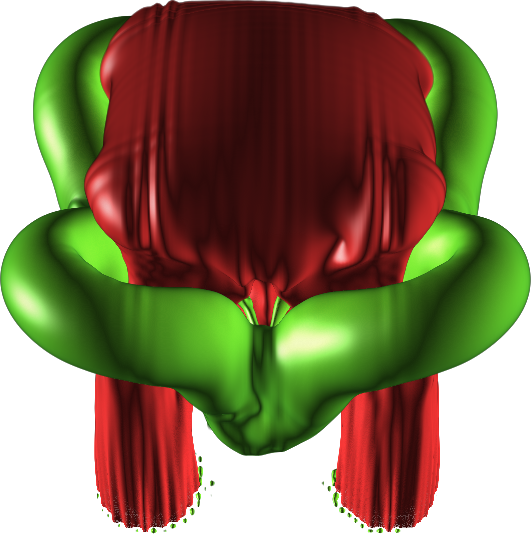}&
 \includegraphics[height=0.75in]{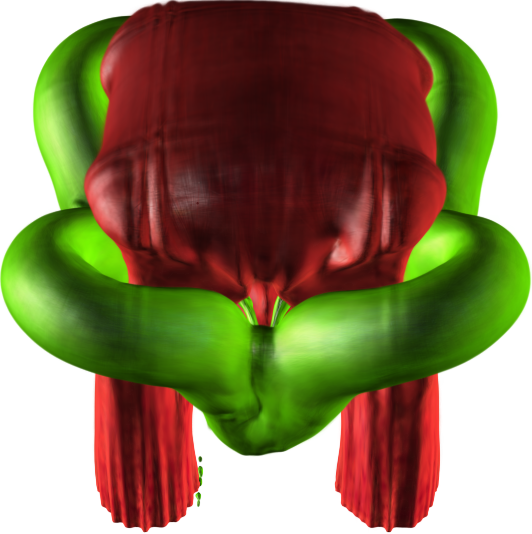}&
 \includegraphics[height=0.75in]{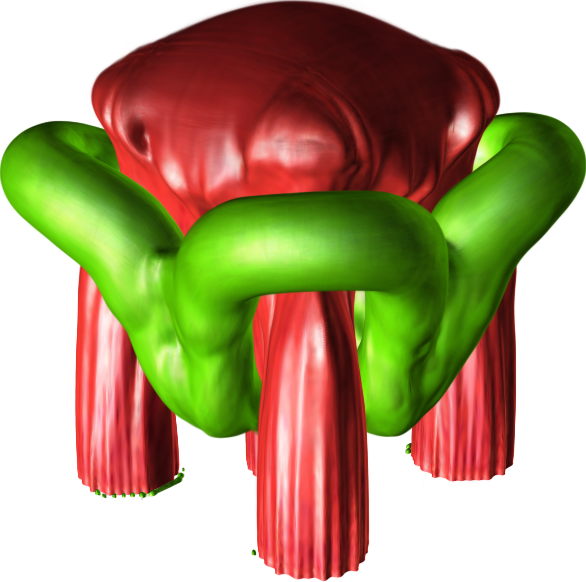}&
\includegraphics[height=0.75in]{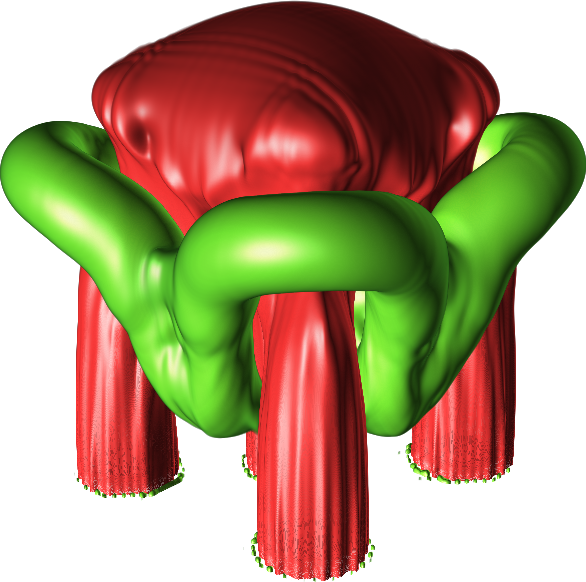}\\
 \includegraphics[height=0.75in]{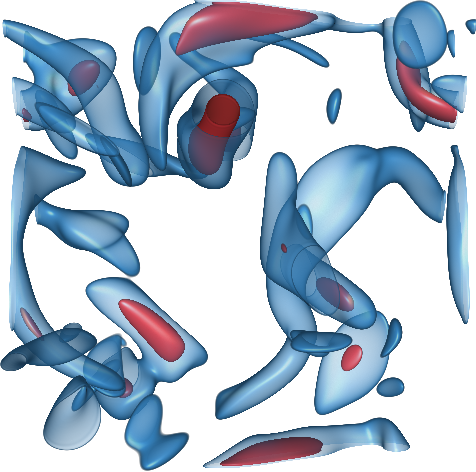}&
 \includegraphics[height=0.75in]{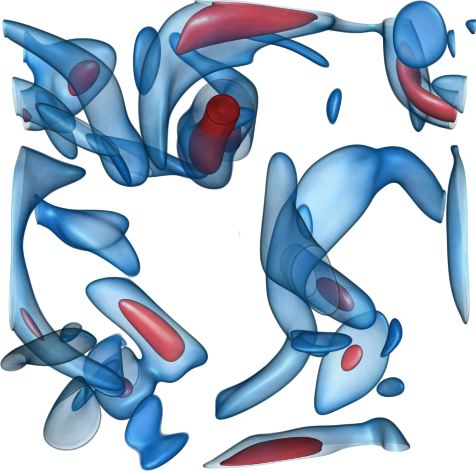}&
 \includegraphics[height=0.75in]{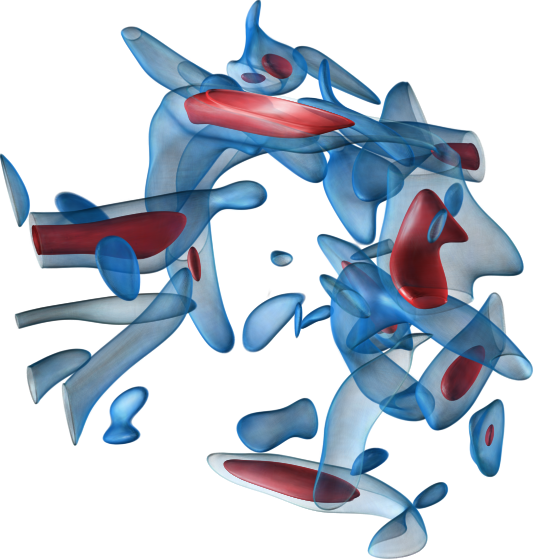}&
\includegraphics[height=0.75in]{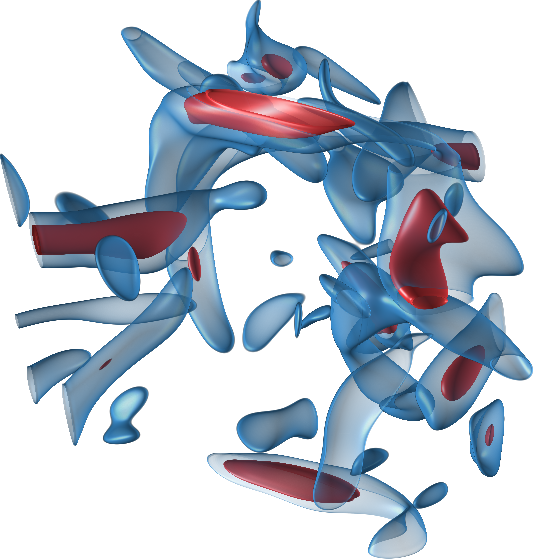}\\
\mbox{\footnotesize (a) reference} & \mbox{\footnotesize (b) reconstructed} & \mbox{\footnotesize (c) NVS} &\mbox{\footnotesize (d) NVS}\\
\mbox{\footnotesize image} & \mbox{\footnotesize result} & \mbox{\footnotesize result} &\mbox{\footnotesize GT}
\end{array}$
\end{center}
\vspace{-.25in}  
\caption{Inverse volume exploration results of iVR-GS on the reference image and novel viewpoint. iVR-GS can accurately approximate desired rendering settings even for a poor viewpoint, such as the combustion example. Top to bottom: combustion, five-jet, and vortex.}
\label{fig:inverse-volume-exploration}
\end{figure}

{\bf Scene composing and editing.} 
iVR-GS supports flexible scene editing on various value ranges simultaneously by composing multiple basic iVR-GS models into one composed iVR-GS.
To demonstrate the flexibility of scene editing for a composed model, we show iterative scene editing results of the vortex and five-jet datasets in Figure~\ref{fig:scene-editing}.
We start by selecting desired basic TFs and then composing corresponding basic iVR-GS models.
Different opacity and color editings can be employed for various basic models within the composed iVR-GS.
Editings of light magnitude and direction are performed globally on the composed scene representation.
Users can explore the VolVis scene and achieve desired rendering results using diverse scene editing options.

\begin{table}[!htb]
\caption{Total number of primitives, average PSNR (dB), LPIPS, and training time for inverse volume exploration.}
\vspace{-0.1in}
\centering
{\scriptsize
\begin{tabular}{c|c|cc|c}
dataset		& \# of primtives& PSNR$\uparrow$ 	& LPIPS$\downarrow$ &  optimization time\\ \hline
combustion 	&5,432,519 &27.62	&0.033		& 88 s\\ 
five-jet	&1,484,642 &25.86   &0.046		& 48 s				\\ 
vortex 		&7,104,441 &30.49	&0.031		& 122 s  		 			\\ 
\end{tabular}
}
\label{tab:inverse-volume-exploration}
\end{table}

\vspace{-0.05in}
\subsection{Inverse Volume Exploration}

The composed iVR-GS model could automatically approximate the intended rendering setting following a reference image with the desired effect.
To demonstrate the reconstruction accuracy of inverse volume exploration, we evaluate datasets listed in Table~\ref{tab:dataset-allTFs} and basic iVR-GS models optimized in Section~\ref{subsec:sceneEditing}.
For each dataset, we first compose all related basic iVR-GS models into a composed model.
We then render an appropriate image as the target reference image with all settings (i.e., viewpoint, color, opacity, and light direction and magnitude) that differ from the training images used in Section~\ref{subsec:sceneEditing}.
We ensure the visible voxel value ranges of the reference image are covered by the composed iVR-GS.
Finally, we freeze all Gaussian primitive attribute values and optimize multiple transformation parameters to modify the scene's color, opacity, and light settings.
Setting the learning rate to 0.01, we optimize all transformation parameters for 1,000 iterations on the reference image until the training converges.
We then render 181 images from the volume under novel viewpoints with the same settings as the reference image to evaluate reconstruction quality.
In Table~\ref{tab:inverse-volume-exploration}, we report the total number of primitives of composed models, average PSNR and LPIPS values of NVS results, and optimization time for each dataset.
As the total number of primitives increases, the training takes longer. 
Figure~\ref{fig:inverse-volume-exploration} compares the synthesized image results of iVR-GS given a reference image, followed by a novel viewpoint. 
Generally speaking, as seen from iVR-GS reconstructed results, the color, opacity, and light settings are transformed quite well for all three examples.
The synthesized results show perceptual fidelity even when the reference image is associated with a poor viewpoint (i.e., the combustion example).
Furthermore, the subsequent NVS results of iVR-GS with the rendering settings learned from the reference image closely approximate the GT results under a new viewpoint. 


\begin{table}[!htb]
\caption{Number of primitives, average PSNR (dB) and LPIPS, as well as the file size of the chameleon dataset.}
\vspace{-0.1in}
\centering
{\scriptsize
\begin{tabular}{c|c|cc|c}
 method &  \# of primitives&  PSNR$\uparrow$&  LPIPS$\downarrow$&file size$\downarrow$  \\ \hline
 base 3DGS&190,323  &26.45  &0.046  &44.8 MB  \\ 
 iVR-GS w/o VQ&254,151  &\textbf{27.25}  &0.043  &21.3 MB  \\ 
 iVR-GS w/ VQ&254,151  &27.19  &\textbf{0.042}  &\textbf{6.2 MB}  \\ 
\end{tabular}
}
\label{tab:ablation-VQ}
\end{table}

\begin{figure}[!htb]
 \begin{center}
 $\begin{array}{c@{\hspace{0.025in}}c@{\hspace{0.025in}}c}
 \includegraphics[width=0.315\linewidth]{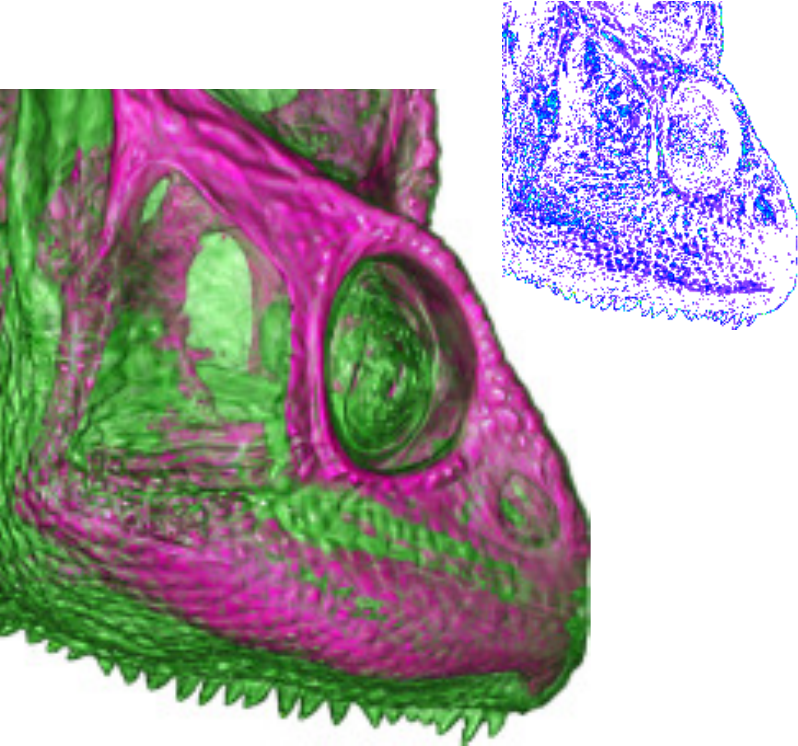}&
 \includegraphics[width=0.315\linewidth]{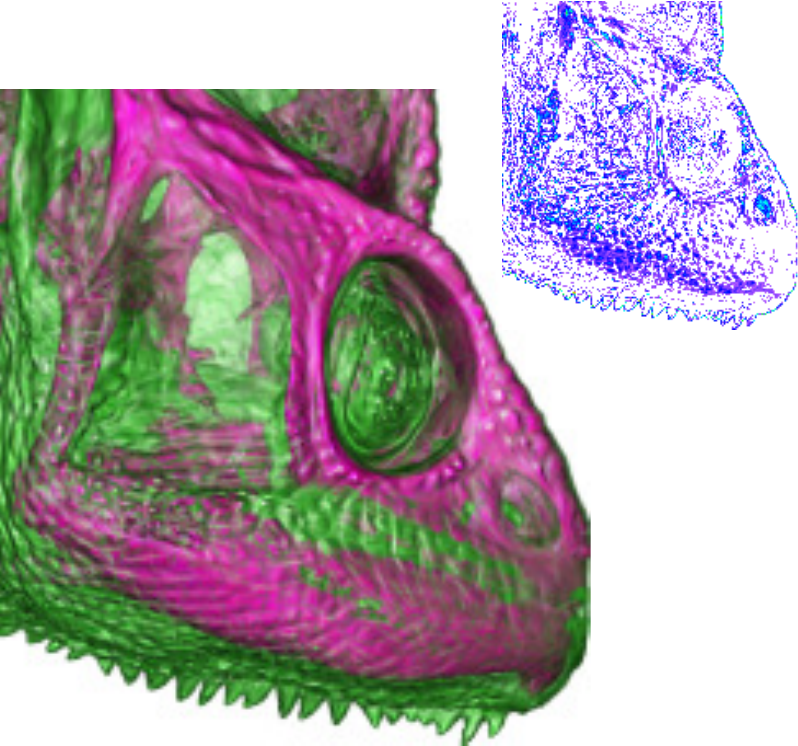}&
\includegraphics[width=0.315\linewidth]{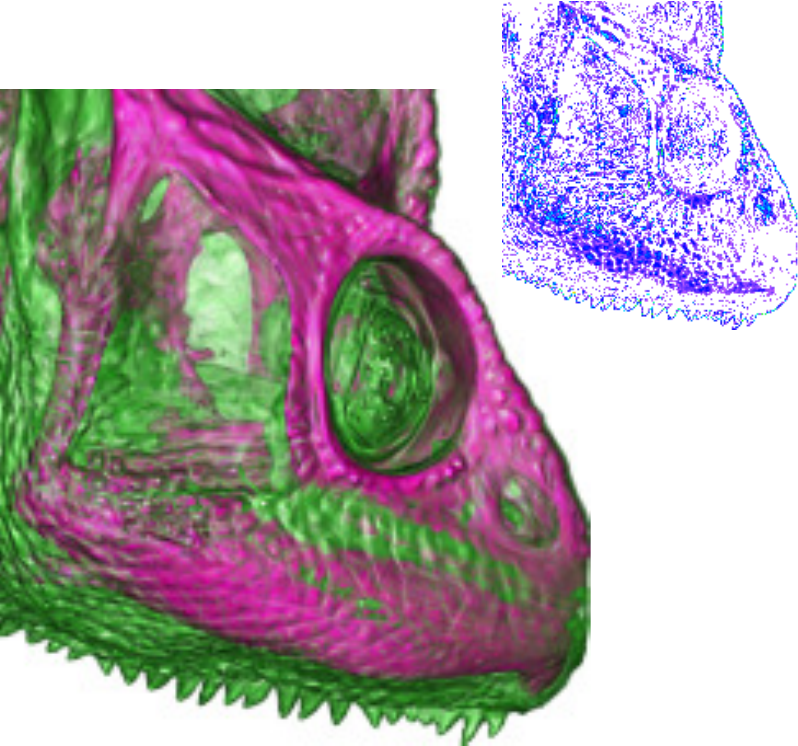}\\
\mbox{\footnotesize (a) base 3DGS} & \mbox{\footnotesize (b) iVR-GS w/o VQ} & \mbox{\footnotesize (c) iVR-GS w/ VQ}
\end{array}$
\end{center}
\vspace{-.25in} 
\caption{Zoom-in NVS results with pixel-wise perceivable differences of the chameleon dataset. The file size of iVR-GS w/ VQ is 7.2$\times$ smaller than base 3DGS while achieving similar results.} 
\label{fig:ablation-VQ}
\end{figure}

\vspace{-0.05in}
\subsection{Ablation Study}

{\bf Evaluation of model compression.} 
Using the chameleon dataset, we train base 3DGS and iVR-GS with and without VQ to investigate model compression.
Quality metrics in Table~\ref{tab:ablation-VQ} and zoom-in comparisons in Figure~\ref{fig:ablation-VQ} indicate close reconstruction quality for all methods.
Since iVR-GS needs additional primitives to construct compact surfaces for correct color representation, it needs more primitives than base 3DGS.
However, even without applying VQ, iVR-GS gets a smaller file size than base 3DGS.
This is because, compared to a set of SH coefficients, using normals and shading attributes within editable Gaussians is a more compact color representation with fewer parameters.
With VQ, iVR-GS can achieve 7.2$\times$ smaller file size than base 3DGS without significant time or accuracy loss while enabling scene editing.

\vspace{-0.05in}
\subsection{Limitations}

Even though iVR-GS achieves efficient rendering and editing of VolVis scenes, it still has several limitations.
First, the size of the composed iVR-GS model is directly related to the number of basic scenes represented. 
VQ compression allows us to increase the number of basic scenes within the same model size. 
When representing one VolVis scene composed of hundreds of basic scenes, the final composed iVR-GS model may still reach the GB level.
Second, we employ uniform viewpoint sampling for iVR-GS, which may not be optimal as different viewpoints can reduce the reconstruction uncertainty of iVR-GS to varying degrees. 
Learning-based incremental sampling could improve accuracy, such as~\cite{Niedermayr-arxiv24, Pan-ECCV22}. 
Still, such a strategy incurs extra training costs and may require a remote workstation or cluster to send newly generated DVR images to the local machine during model training. 
Third, iVR-GS models the lighting effect using the Blinn-Phong model. 
It may encounter challenges when handling VolVis scenes with a different lighting model. 
In such cases, iVR-GS may need to redesign the lighting representation to fit the new lighting model employed.


\vspace{-0.05in}
\section{Conclusions and Future Work}

We have presented iVR-GS, an NVS framework designed for the VolVis scene to support flexible exploration without access to volume data.
iVR-GS overcomes existing NVS limitations by leveraging numerous composable and editable Gaussian primitives for scene representation.
Compared with Plenoxels, CCNeRF, and base 3DGS, it produces higher-quality reconstruction results and supports real-time scene editing and inverse volume exploration.

In the future, we will extend iVR-GS to perform style transfer of the VolVis scene.
By taking advantage of the efficient rendering of Gaussian representation, we aim to achieve real-time style transfer.
Moreover, we will adopt iVR-GS for time-varying or ensemble VolVis scenes with time-dependent Gaussian primitives.
For large-scale volume datasets, the rendering cost of iVR-GS is much smaller than conventional DVR, opening the door for real-time rendering of VolVis scenes on low-end local devices.
Deploying iVR-GS on VR devices and integrating novel interactions can provide users with an immersive large-volume exploration experience. 

\vspace{-0.05in}
\acknowledgments{This research was supported in part by the U.S.\ National Science Foundation through grants IIS-1955395, IIS-2101696, OAC-2104158, and IIS-2401144, and the U.S.\ Department of Energy through grant DE-SC0023145. The authors would like to thank the anonymous reviewers for their insightful comments.}


\setcounter{section}{0}
\setcounter{figure}{0}
\setcounter{table}{0}

\section*{Appendix}

\section{Implementation Details}

\begin{figure}[!htb]
\centering
 \begin{center}
 $\begin{array}{c@{\hspace{0.05in}}c@{\hspace{0.05in}}c}
 \includegraphics[width=0.315\linewidth]{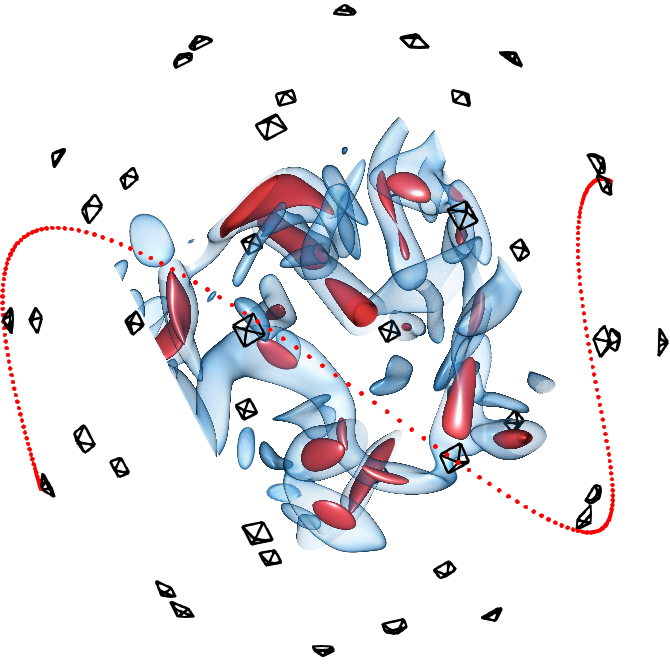}&
  \includegraphics[width=0.315\linewidth]{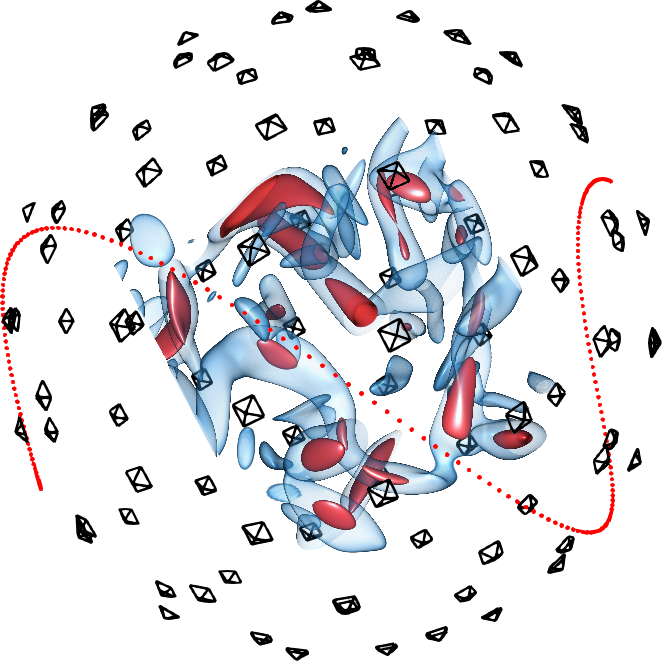}&
 \includegraphics[width=0.315\linewidth]{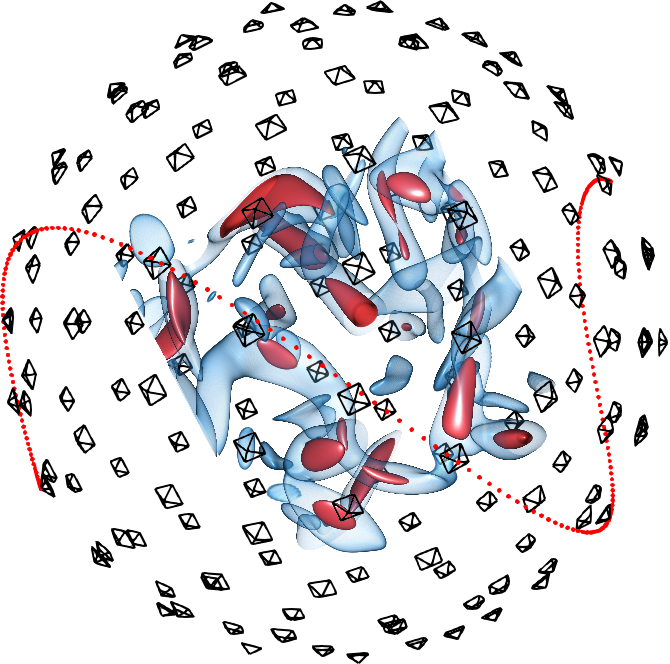}\\
\mbox{\footnotesize (a) 42 training views} & 
\mbox{\footnotesize (b) 92 training views} & \mbox{\footnotesize (c) 162 training views}
\end{array}$
\end{center}
\vspace{-.25in}
\caption{The distribution of 42, 92, and 162 training views. The black pyramids denote the camera poses for capturing training images, and the red dots represent the camera positions for generating testing images.}
\label{fig:viewsampling}
\end{figure}

We implemented iVR-GS in Python using the PyTorch framework and CUDA kernels for efficient rasterization and VQ compression.
We estimated the scene complexity using entropy and generated 42, 92, or 162 multi-view images from the volume for training each basic iVR-GS model. 
The trained model was then evaluated on 181 novel viewpoints. 
Figure~\ref{fig:viewsampling} shows the distribution of viewpoints for training and testing.
For both stages in iVR-GS, we used the Adam optimizer with $\epsilon = 1e^{-15}$.
We set the same learning rate as vanilla 3DGS~\cite{Kerbl-TOG23} for SH coefficients and attributes $\{\bm{\mu}, \mathbf{q}, \mathbf{s}, o\}$, while setting a learning rate of 0.01 for other attributes $\{\mathbf{n},\Delta\mathbf{c}, k^{a}, k^{d}, k^{s}, \beta\}$.
We updated our model parameters on one randomly selected training image per iteration.
During optimization, iVR-GS utilizes a similar densification and pruning strategy as vanilla 3DGS to achieve adaptive density control.
Specifically, we densified a Gaussian primitive if its position or normal attributes exhibited a large gradient magnitude and pruned a Gaussian primitive when it was essentially transparent. 
We refer the readers to the released code for more details on implementation and hyperparameter settings.

\vspace{-0.05in}
\section{Additional Results}

\begin{figure}[!htb]
\centering
\includegraphics[width=\columnwidth]{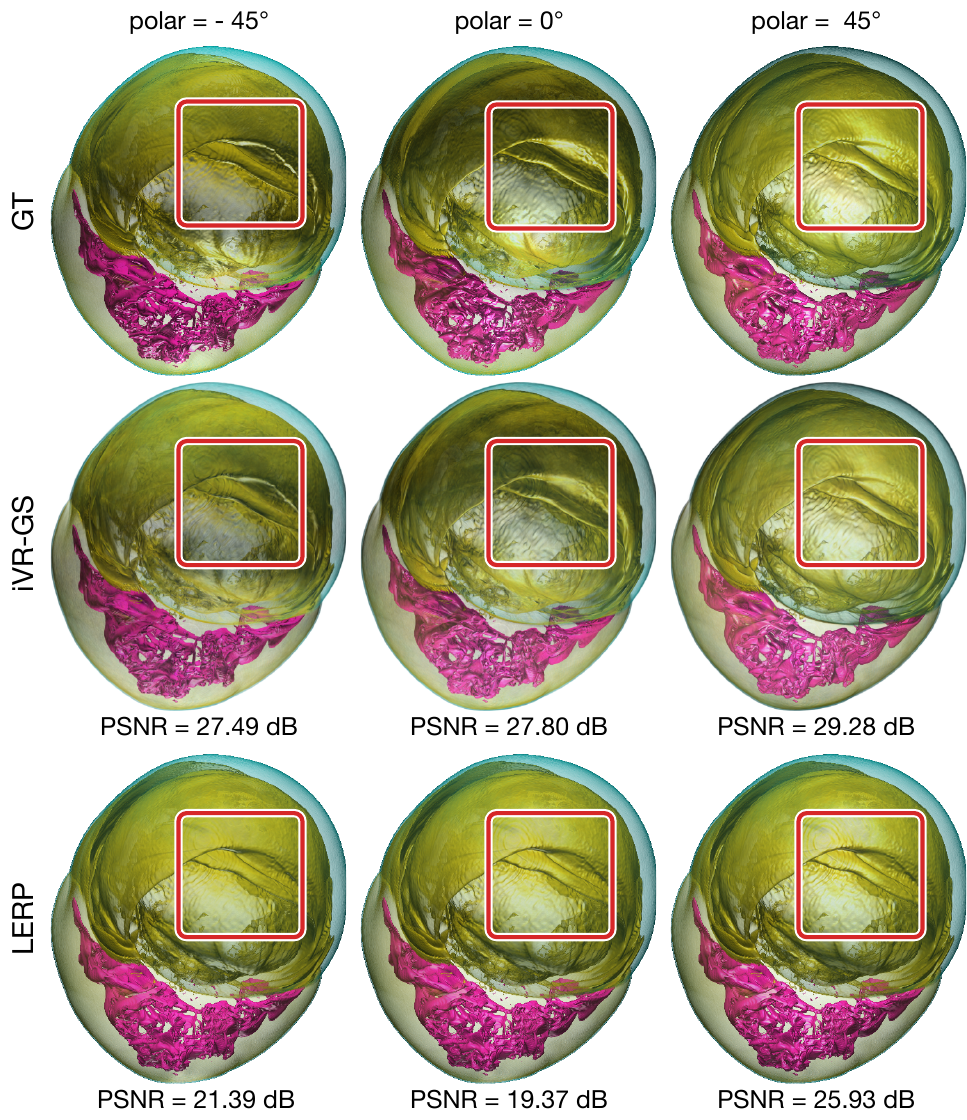}
\vspace{-.25in}
\caption{Comparing relighting results of iVR-GS and LERP. Left to right: the light source moves from bottom to top. 
}
\label{fig:iVRGS-vs-LERP}
\end{figure}

{\bf Comparison of relighting.}
To highlight the performance of iVR-GS on the relighting task, we compare the relighting results of iVR-GS and linear interpolation (LERP) on the supernova dataset.
Given one viewpoint, we keep the azimuthal angle of the light source the same as the camera while adjusting the polar angle from -90$^{\circ}$ to 90$^{\circ}$.
Following the light source trajectory from bottom to top, we uniformly sample 180 images.
LERP takes the first and the last image as input and outputs intermediate images through pixel-wise interpolation.
Figure~\ref{fig:iVRGS-vs-LERP} shows PSNR values and relighting results of three selected angles.
We can see that straightforward image-space operations such as LERP cannot produce faithfully relighting results.
It is necessary to model the geometry information for accurate relighting.

\begin{figure*}[!htb]
 \begin{center}
 $\begin{array}{c@{\hspace{0.05in}}c@{\hspace{0.05in}}c@{\hspace{0.05in}}c@{\hspace{0.05in}}c}
 \includegraphics[height=1.125in]{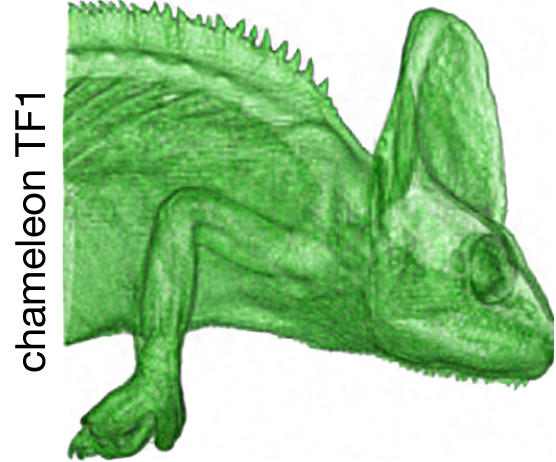}&
 \includegraphics[height=1.125in]{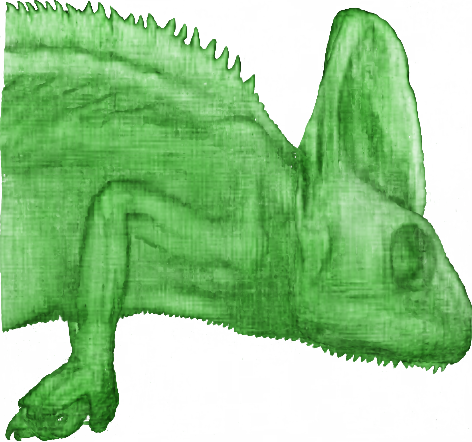}&
 \includegraphics[height=1.125in]{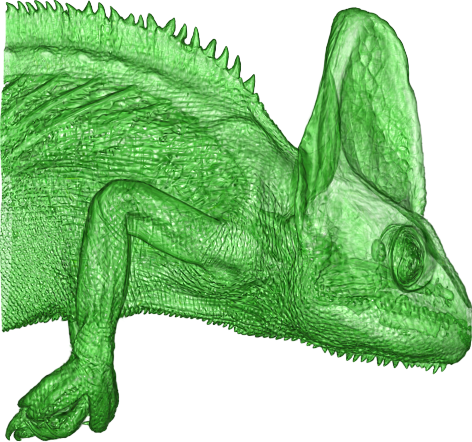}&
  \includegraphics[height=1.125in]{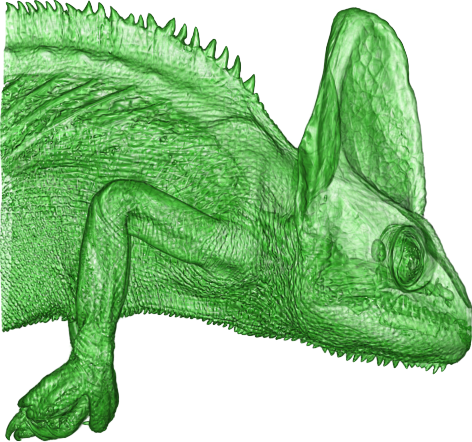}&
\includegraphics[height=1.125in]{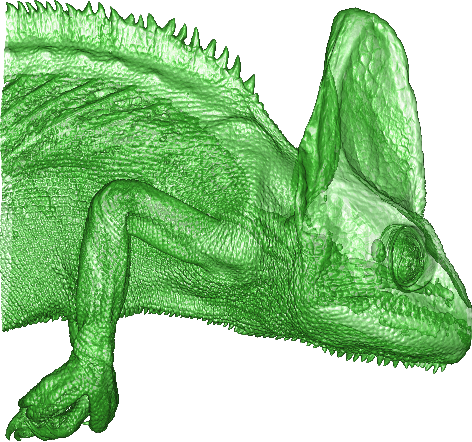}\\

 \includegraphics[height=1.125in]{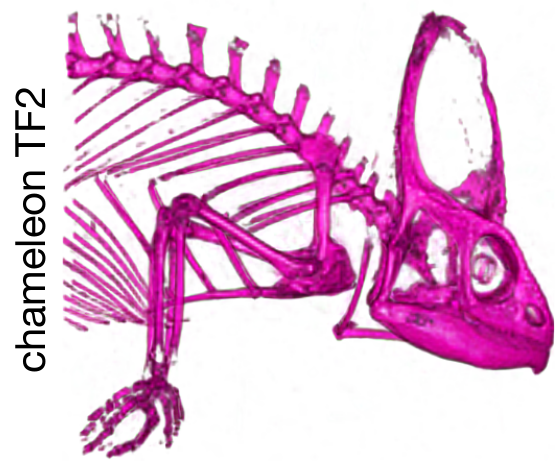}&
 \includegraphics[height=1.125in]{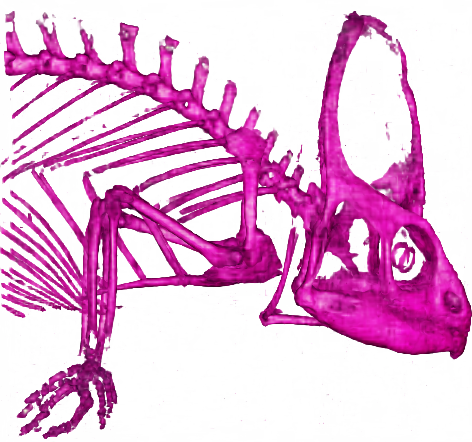}&
 \includegraphics[height=1.125in]{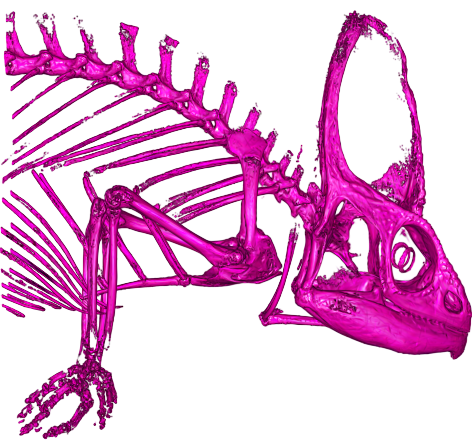}&
  \includegraphics[height=1.125in]{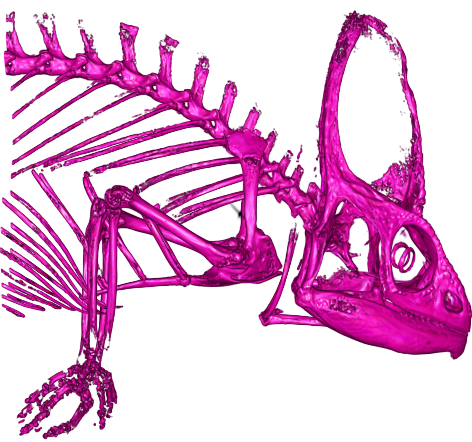}&
\includegraphics[height=1.125in]{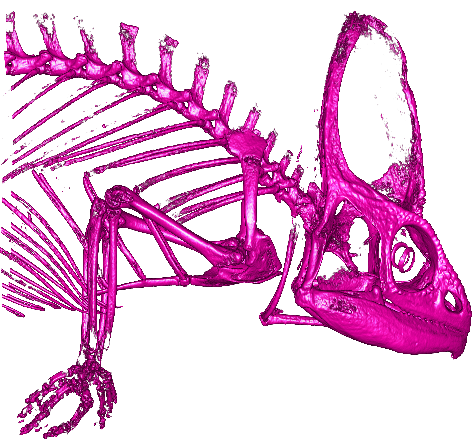}\\

 \includegraphics[height=1.335in]{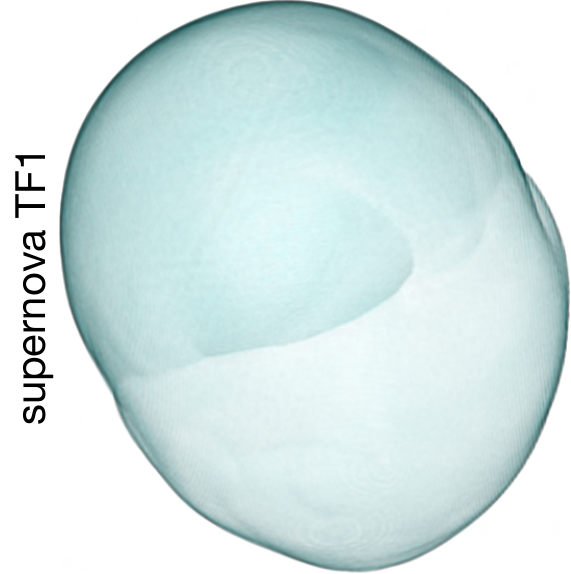}&
 \includegraphics[height=1.335in]{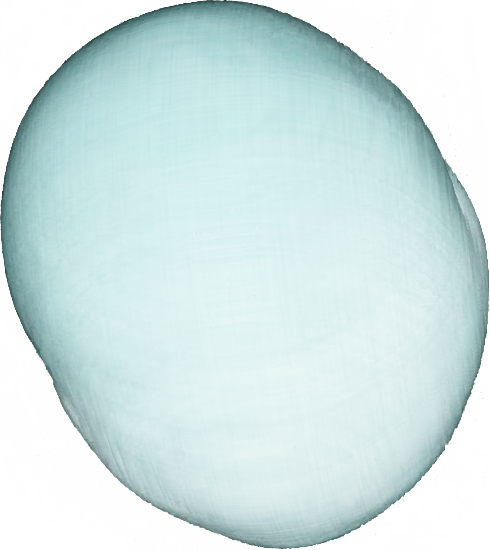}&
 \includegraphics[height=1.335in]{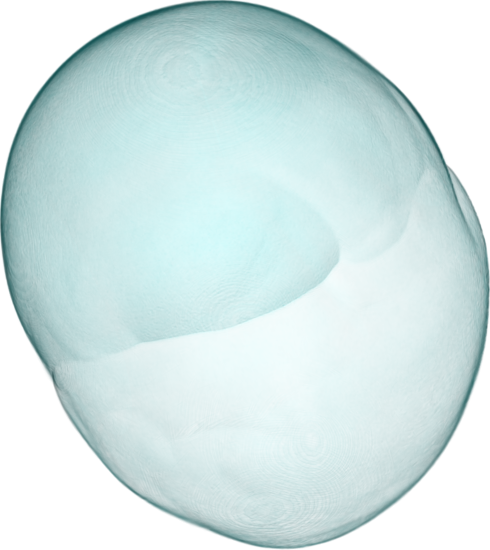}&
  \includegraphics[height=1.335in]{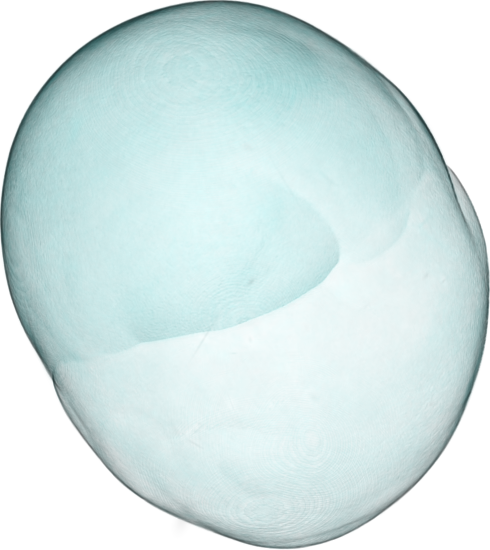}&
\includegraphics[height=1.335in]{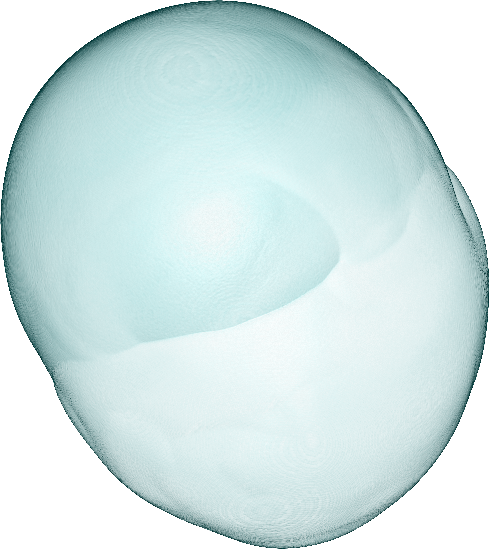}\\

 \includegraphics[height=1.335in]{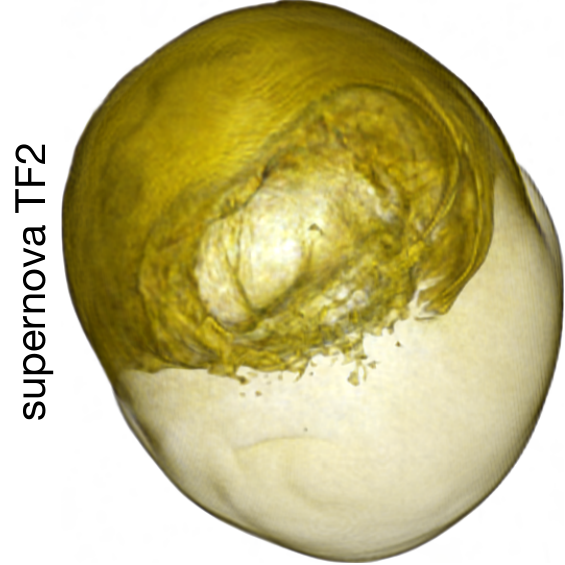}&
 \includegraphics[height=1.335in]{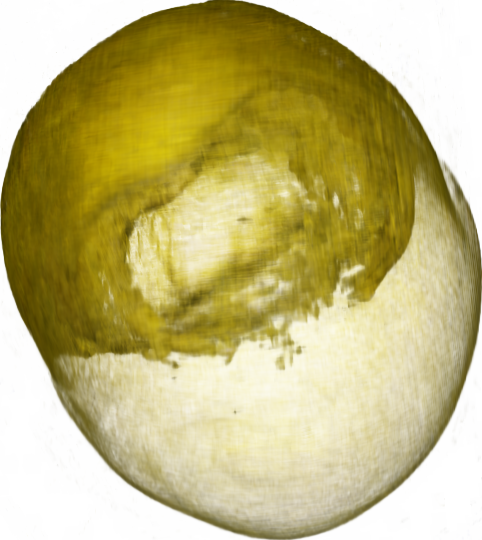}&
 \includegraphics[height=1.335in]{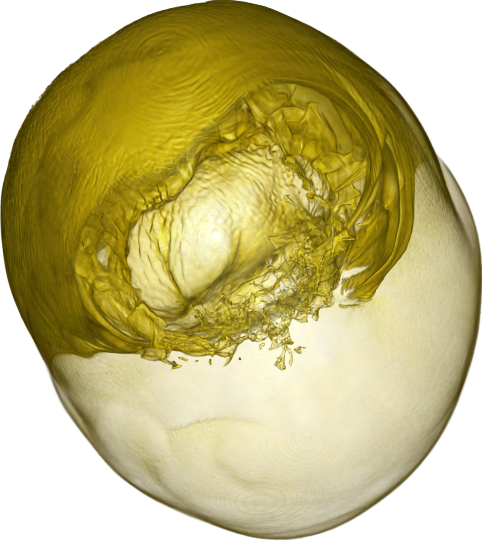}&
  \includegraphics[height=1.335in]{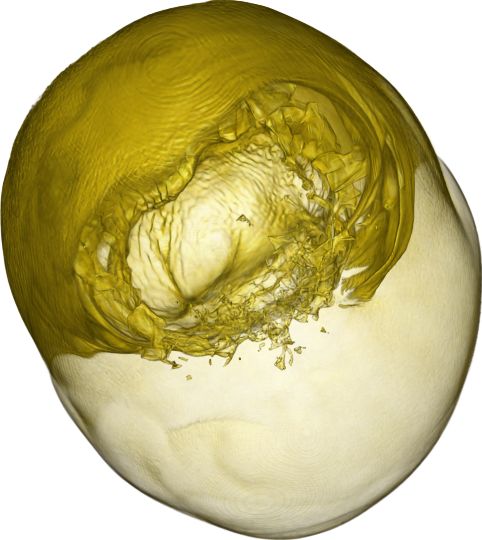}&
\includegraphics[height=1.335in]{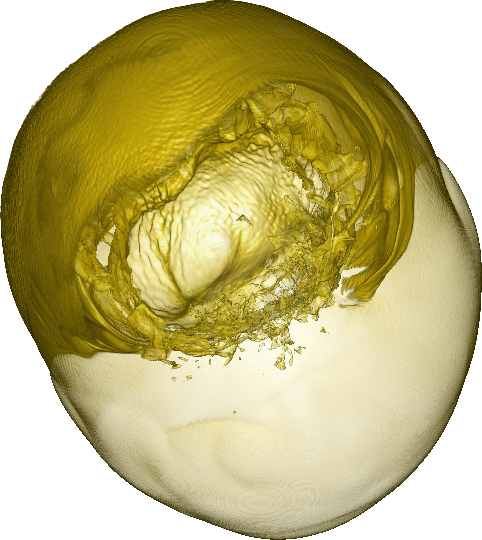}\\

 \includegraphics[height=1.0in]{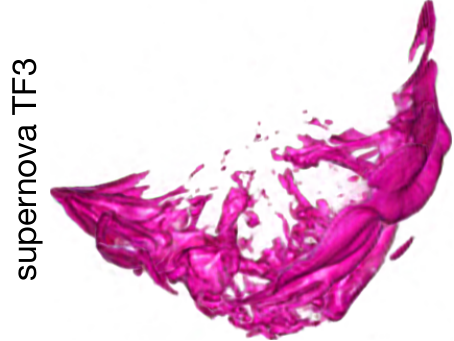}&
 \includegraphics[height=1.0in]{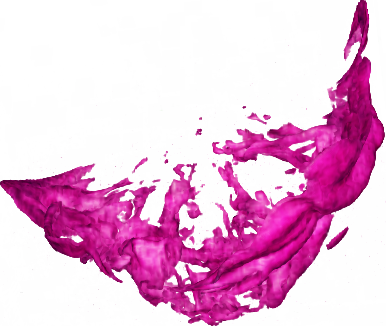}&
 \includegraphics[height=1.0in]{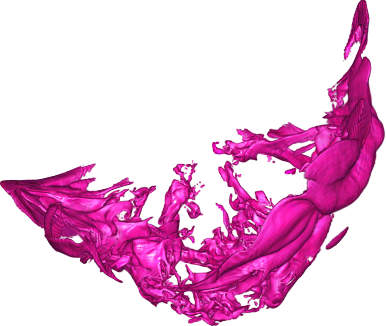}&
  \includegraphics[height=1.0in]{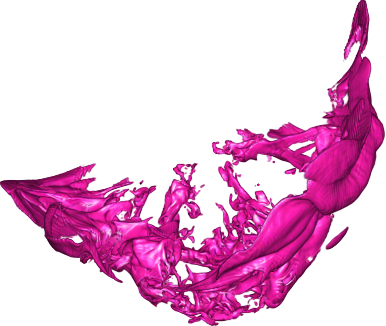}&
\includegraphics[height=1.0in]{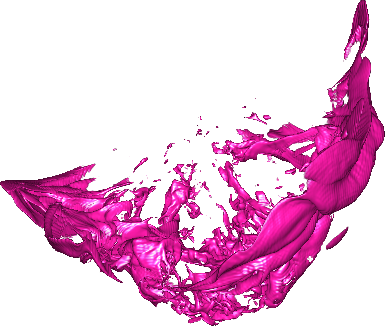}\\
\mbox{\footnotesize (a) Plenoxels} & \mbox{\footnotesize (b) CCNeRF} & \mbox{\footnotesize (c) base 3DGS} &\mbox{\footnotesize (d) iVR-GS} &\mbox{\footnotesize (e) GT} 
\end{array}$
\end{center}
\vspace{-.25in} 
\caption{Comparing basic scenes results of four methods w.r.t.\ GT.} 
\label{fig:appendix-BasicScenes}
\end{figure*}

{\bf Comparison of basic scenes.}
In addition to scene composing results shown in Figure~\ref{fig:baseline-vr-results} in the paper, we provide basic scene results of chameleon and supernova datasets for all four methods in Figure~\ref{fig:appendix-BasicScenes}.
Although NeRF-based methods (Plenoxels and CCNeRF) generate relatively lower-quality images than point-based methods (base 3DGS and iVR-GS), all methods can generate reasonable results when reconstructing only one basic scene.
However, as shown in Figure~\ref{fig:baseline-vr-results} in the paper, composing multiple Plenoxels and CCNeRF basic models will cause parameter interference and lead to inaccurate or incorrect rendering results.
When rendering one volume dataset with multiple basic TFs, the resulting basic scenes will likely exhibit nested structures, which is rare for real-world scenes.
In this regard, point-based methods fit the model composition for the VolVis scenes better than their NeRF-based counterparts.

\begin{table}[htb]
\caption{Average PSNR (dB) under different numbers of training images and normalized entropy scores for each basic TF of the vortex dataset. 
Based on normalized entropy scores, we sample 162, 92, and 42 images for TF1 to TF5, TF6 and TF7, and TF8 to TF10, respectively.}
\vspace{-0.1in}
\centering
\resizebox{\columnwidth}{!}{
\begin{tabular}{c|c|c|c|c|c|c}
&voxel&\multicolumn{4}{|c|}{PSNR$\uparrow$}&normalized \\ \cline{3-6}
 basic TF  	&value range		& 162 images 	& 92 images 		& 42 images 		& 12 images & entropy score   \\ \hline
 TF1 		&-1.0 $\sim$ -0.8	&28.01				&26.03				&22.80				&15.50			&0.86 \\
 TF2 		&-0.8 $\sim$ -0.6	&30.64				&27.90				&23.98				&15.72			&1.00 \\
 TF3 		&-0.6 $\sim$ -0.4	&29.33				&26.95				&23.83				&16.11			&0.97 \\
 TF4 		&-0.4 $\sim$ -0.2	&31.10				&29.78				&25.61				&16.37			&0.85 \\
 TF5 		&-0.2 $\sim$ 0.0	&33.10				&31.99				&29.18				&21.13			&0.60 \\ \hline
 TF6 		&0.0 $\sim$ 0.2		&35.13				&34.64				&32.29				&25.93			&0.37 \\
 TF7 		&0.2 $\sim$ 0.4		&38.39				&38.20				&36.06				&30.38			&0.18 \\ \hline
 TF8 		&0.4 $\sim$ 0.6		&41.58				&41.51				&39.61				&34.28			&0.08 \\
 TF9 		&0.6 $\sim$ 0.8		&45.11				&45.03				&43.93				&39.70			&0.04 \\
 TF10 		&0.8 $\sim$ 1.0		&48.55				&48.54				&47.88				&44.75			&0.01 \\
\end{tabular}
}
\label{tab:NumberOfViews}
\end{table}

{\bf Number of views.}
We compute entropy scores to estimate the number of multi-view images necessary for reconstructing each basic scene.
We leverage ten disjoint basic TFs to generate basic scenes of the vortex dataset.
One iVR-GS model is then optimized on each basic scene with different numbers of multi-view training images.
Table~\ref{tab:NumberOfViews} reports the average PSNR value of each basic scene under various numbers of training images, along with normalized entropy scores.
In practice, we sample 42, 92, and 162 images for normalized entropy scores [0.0, 0.1), [0.1, 0.5], and (0.5, 1.0], respectively.
We sample 92 images for TF6 and TF7 as their normalized entropy scores are in [0.1, 0.5], and increasing training images from 42 to 92 can improve 2 dB for both TFs while reaching 162 images only offers negligible quality improvement compared with 92 images.
For this example, we can reduce 30.8\% training images without significant accuracy loss compared to 162 images for all basic TFs.

\begin{table*}[!htb]
\caption{Average PSNR (dB), LPIPS, and total training time for individual basic scenes, as well as average PSNR (dB), LPIPS, total number of primitives, and total file size for the composed VolVis scene.}
\vspace{-0.1in}
\centering
{\scriptsize
\begin{tabular}{c|c|ccc|cccc}
dataset & method & \multicolumn{3}{c|}{individual basic scenes} & \multicolumn{4}{c}{composed VolVis scene}\\ \cline{3-9}
&& PSNR$\uparrow$ & LPIPS$\downarrow$  & training time$\downarrow$
& PSNR$\uparrow$ & LPIPS$\downarrow$ & \# of  primitives$\downarrow$ & file size$\downarrow$\\ \hline
\multirow{2}{*}{five-jet} 
&one-stage&36.50&0.011&59.4 min &32.42&0.015 &711,169 &16.7 MB\\
&two-stage&36.44&0.010&23.7 min &32.22&0.017 &375,352 &8.8 MB\\ \hline
\multirow{2}{*}{supernova} 
&one-stage&32.42&0.032&51.1 min &29.44&0.047 &684,146 &16.4 MB\\
&two-stage&32.20&0.035&23.4 min &29.20&0.054 &416,211 &10.0 MB\\ 
\end{tabular}
}
\label{tab:vs-one-stage}
\end{table*}


\begin{figure}[!htb]
\centering
\includegraphics[width=0.8\columnwidth]{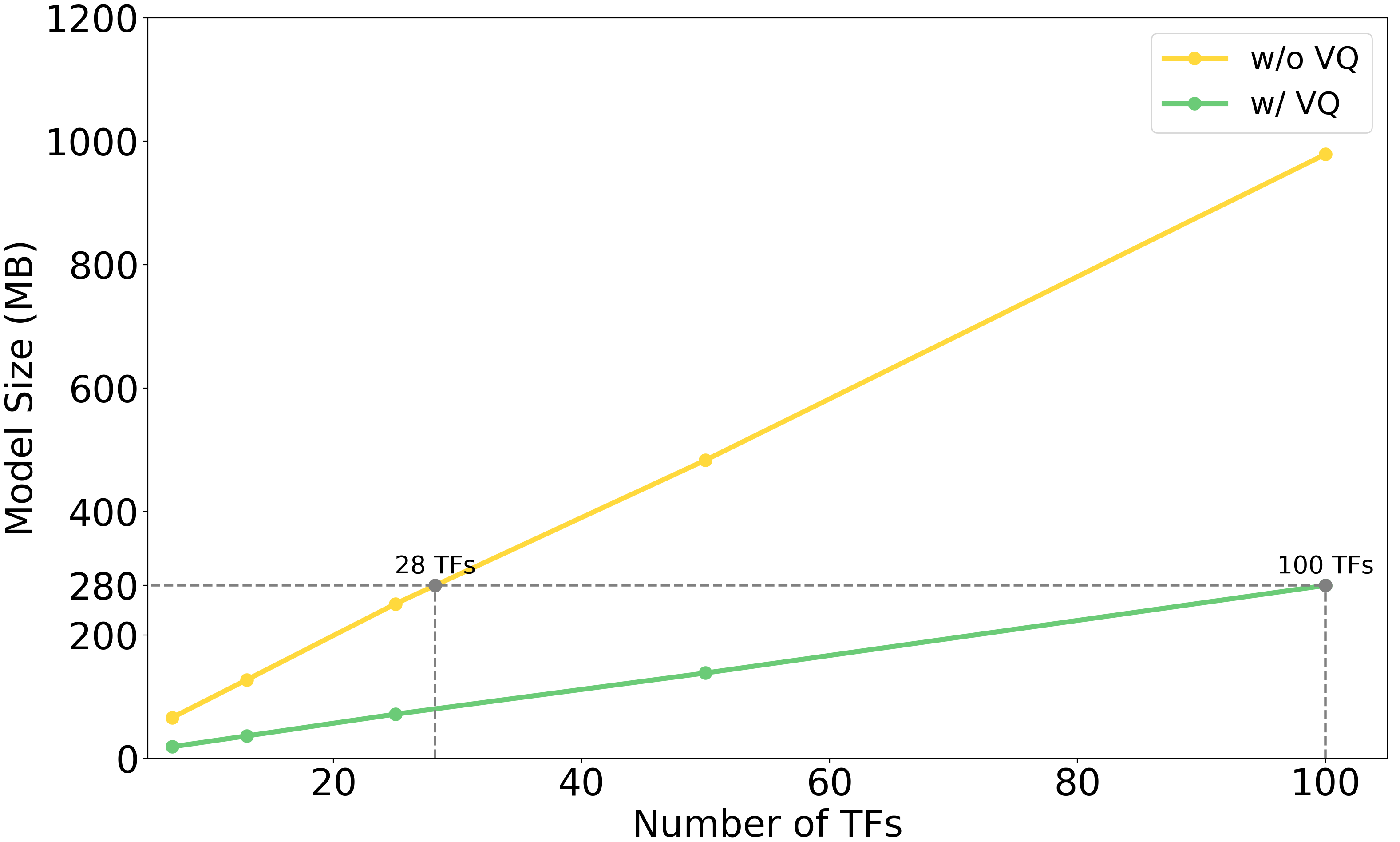}
\vspace{-.1in}
\caption{Model sizes (MB) of the composed iVR-GS under different numbers of TFs of the combustion dataset.}
\label{fig:scalability}
\end{figure}

\begin{figure}[!htb]
 \begin{center}
 $\begin{array}{c@{\hspace{0.05in}}c@{\hspace{0.05in}}c}
 \includegraphics[height=0.85in]{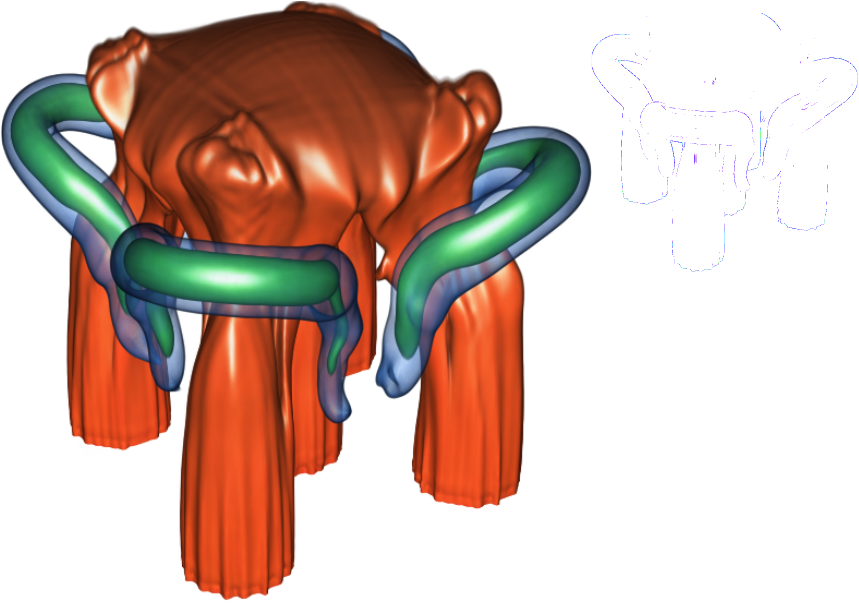}&
 \includegraphics[height=0.85in]{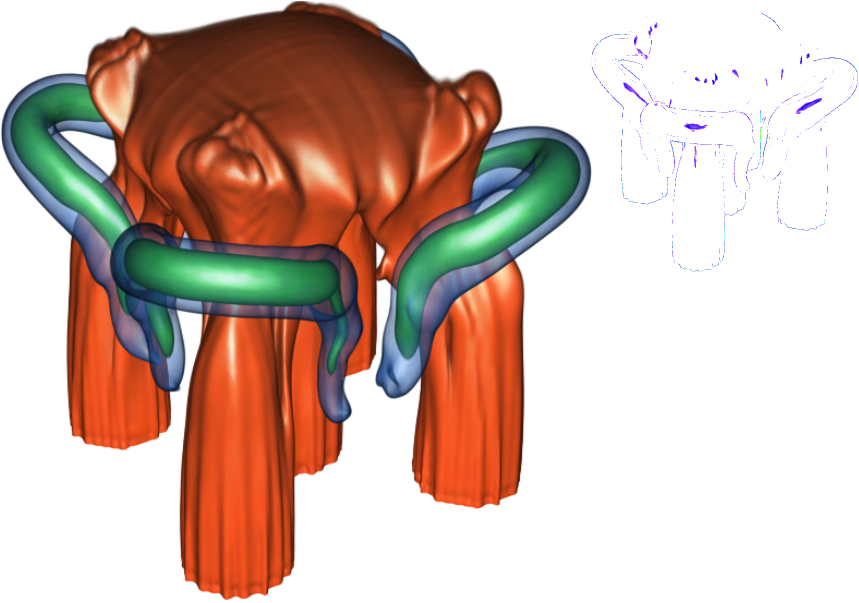}&
 \includegraphics[height=0.85in]{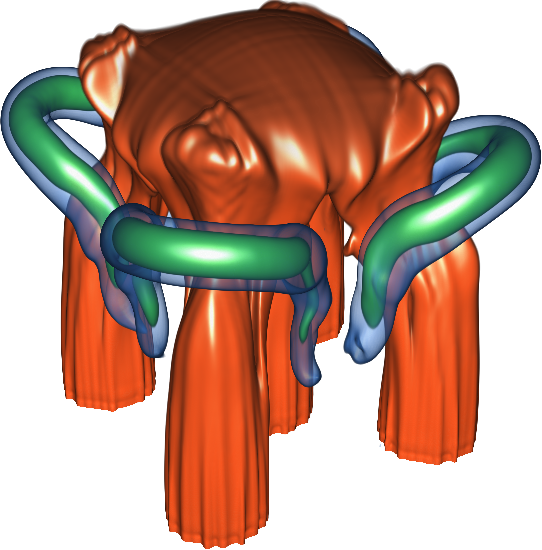}\\
 \includegraphics[height=0.85in]{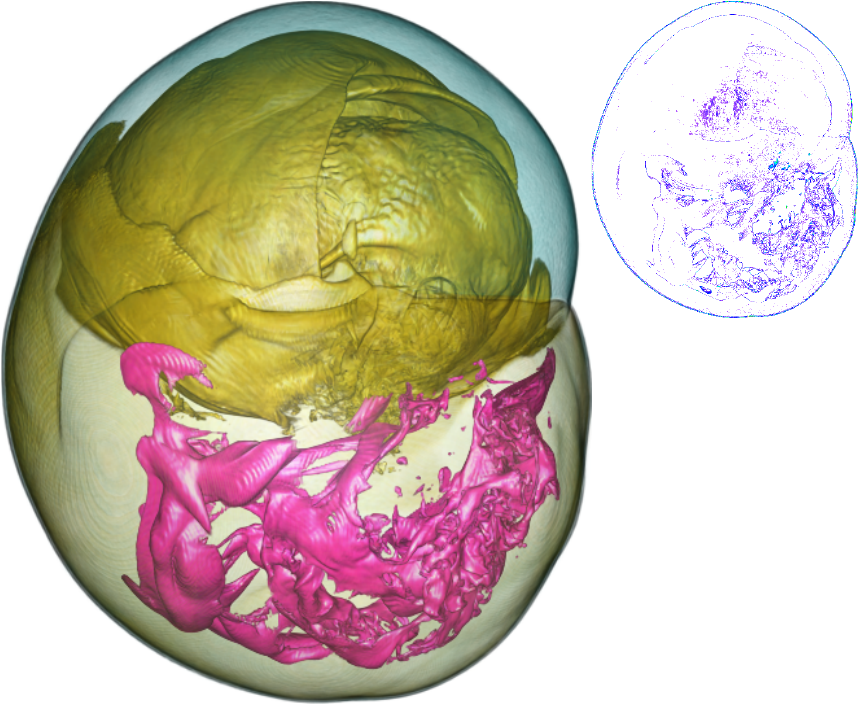}&
 \includegraphics[height=0.85in]{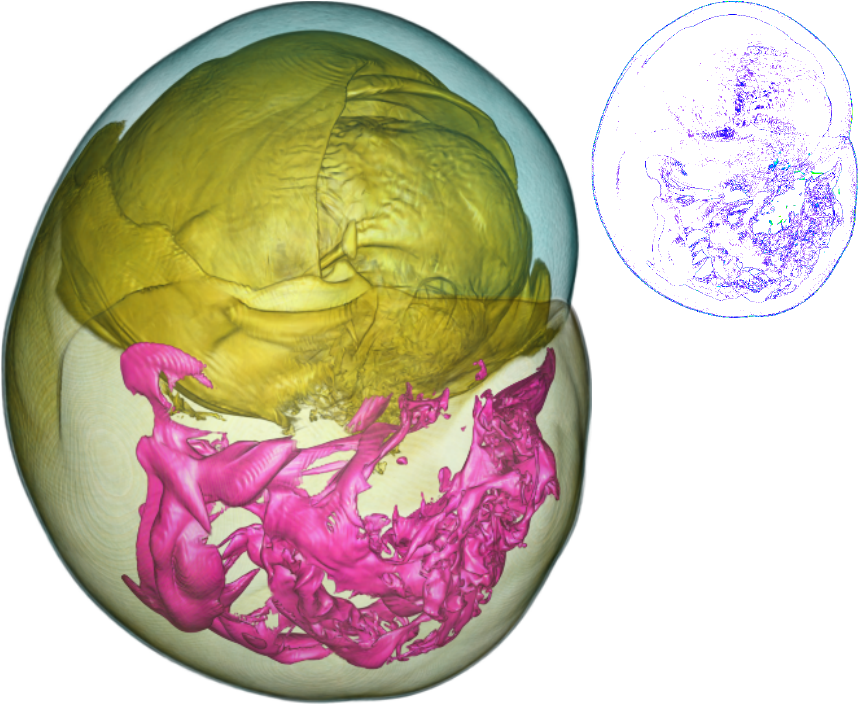}&
 \includegraphics[height=0.85in]{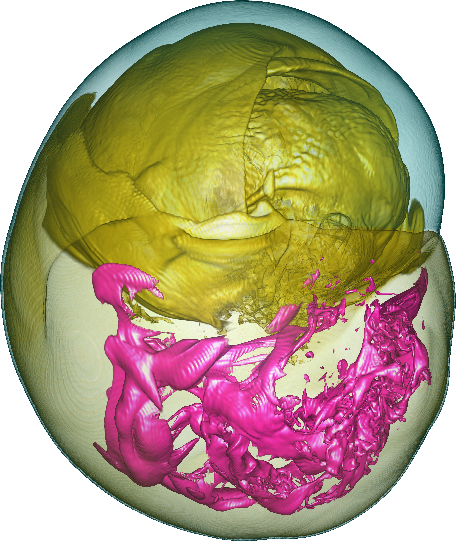}\\
\mbox{\footnotesize (a) one-stage} & \mbox{\footnotesize (b) two-stage} & \mbox{\footnotesize (c) GT}
\end{array}$
\end{center}
\vspace{-.25in} 
\caption{Comparing rendering results of iVR-GS using one-stage or two-stage training strategy. Top and bottom: five-jet and supernova.}
\label{fig:vs-one-stage}
\end{figure}

\begin{figure}[!htb]
 \begin{center}
 $\begin{array}{c@{\hspace{0.05in}}c}
 \includegraphics[width=0.425\columnwidth]{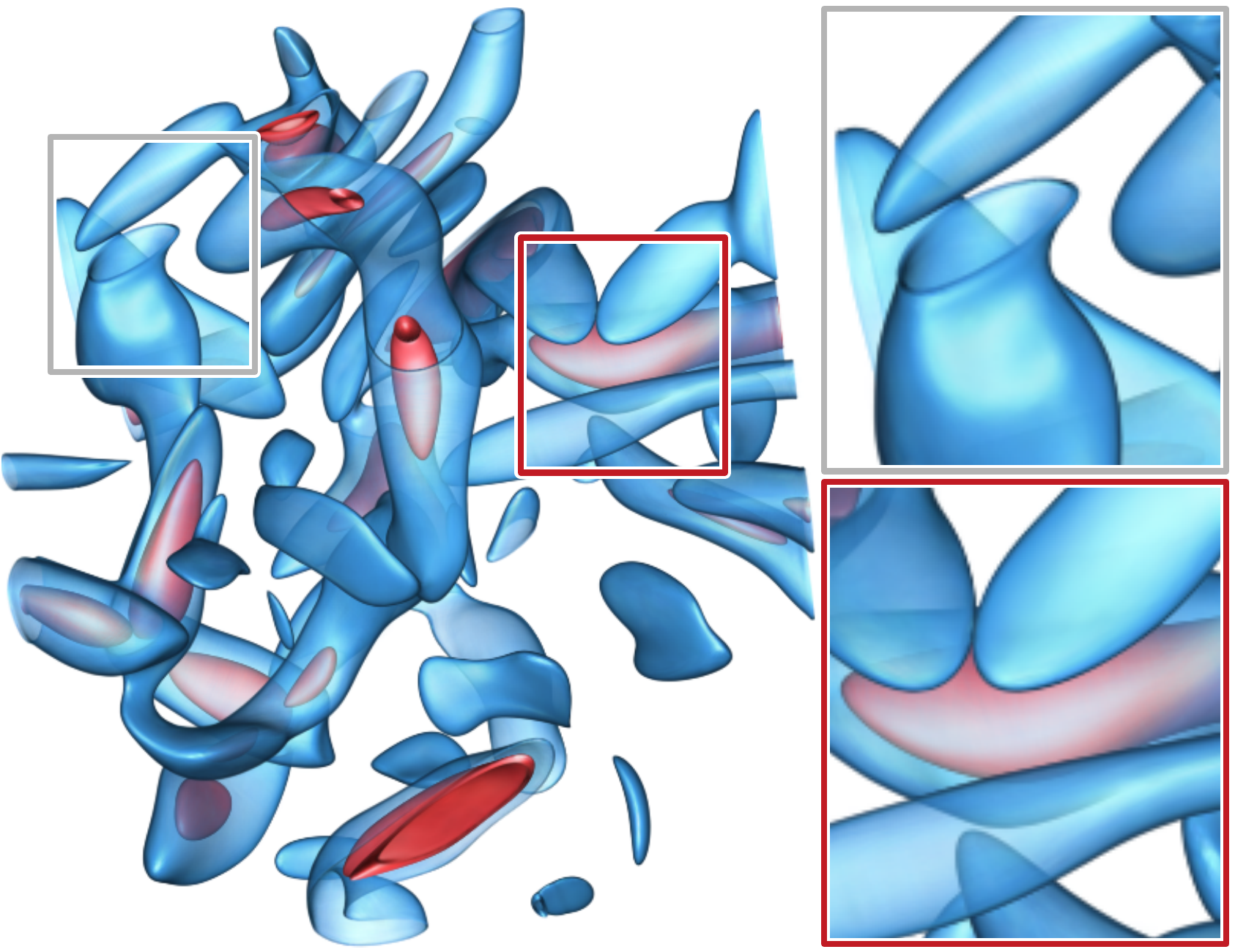}&
 \includegraphics[width=0.425\columnwidth]{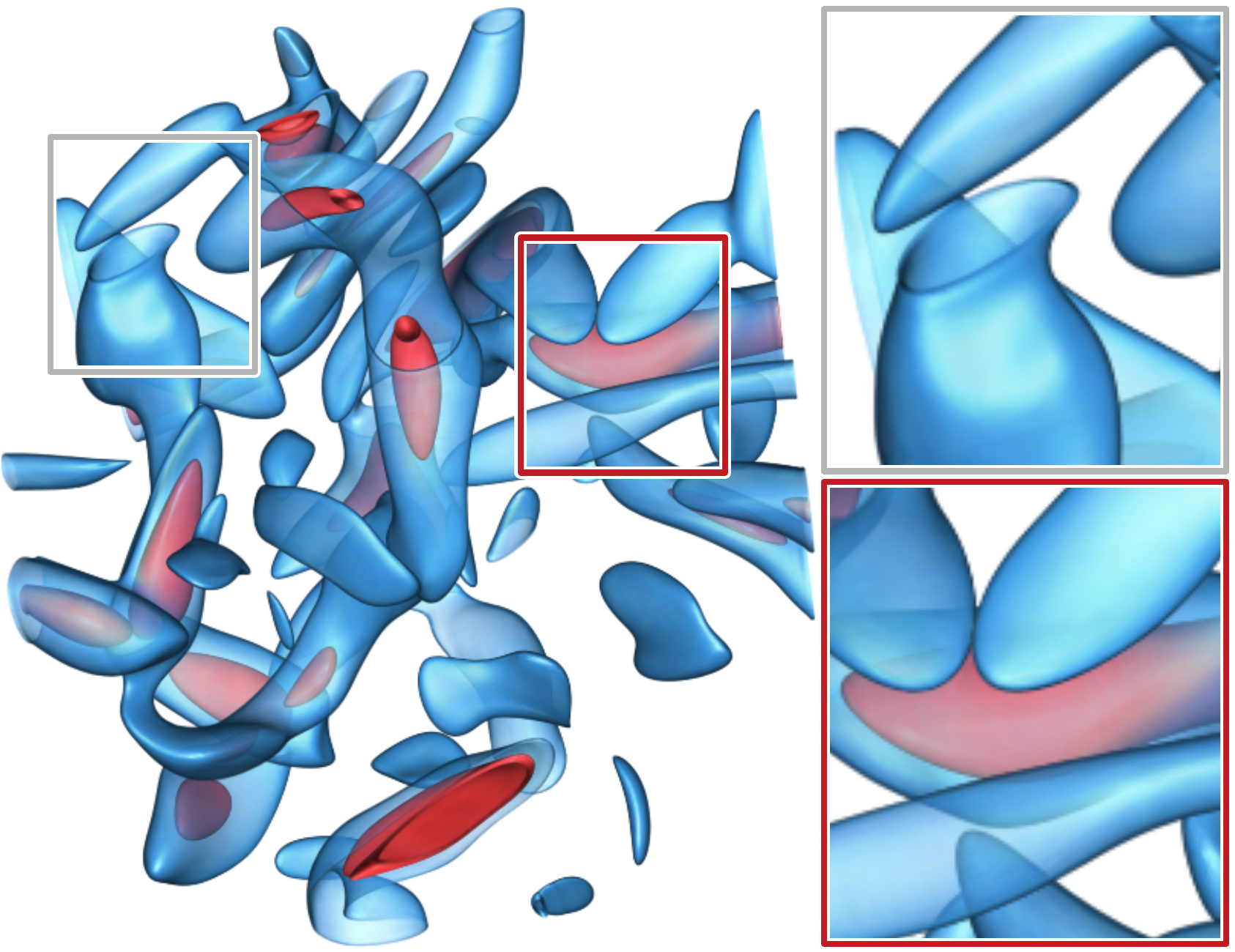}\\
 \mbox{\footnotesize (a) blue: $M_1$; red: $M_1$} & \mbox{\footnotesize (b) blue: $M_1$; red: $M_2$}\\
 \includegraphics[width=0.425\columnwidth]{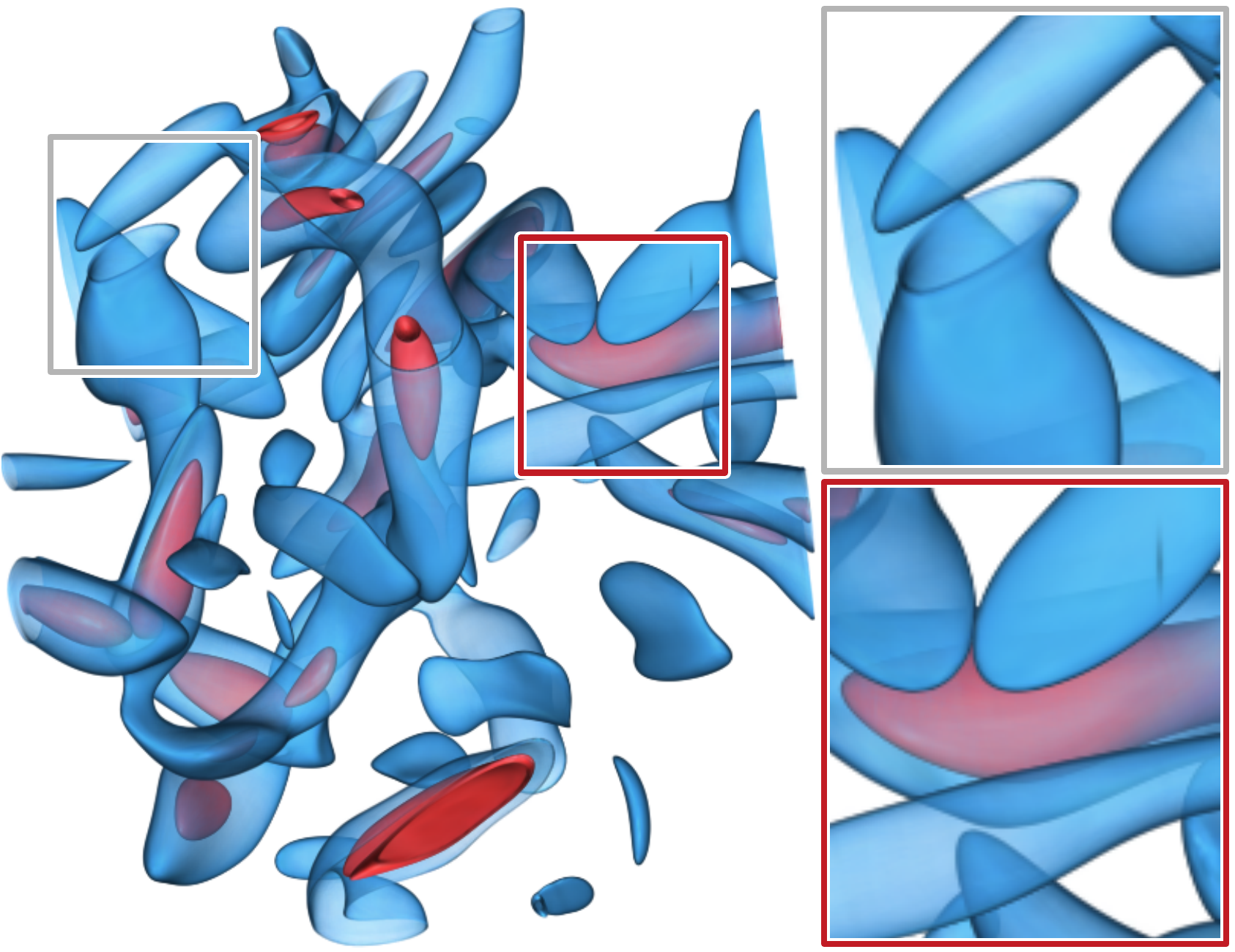}&
 \includegraphics[width=0.425\columnwidth]{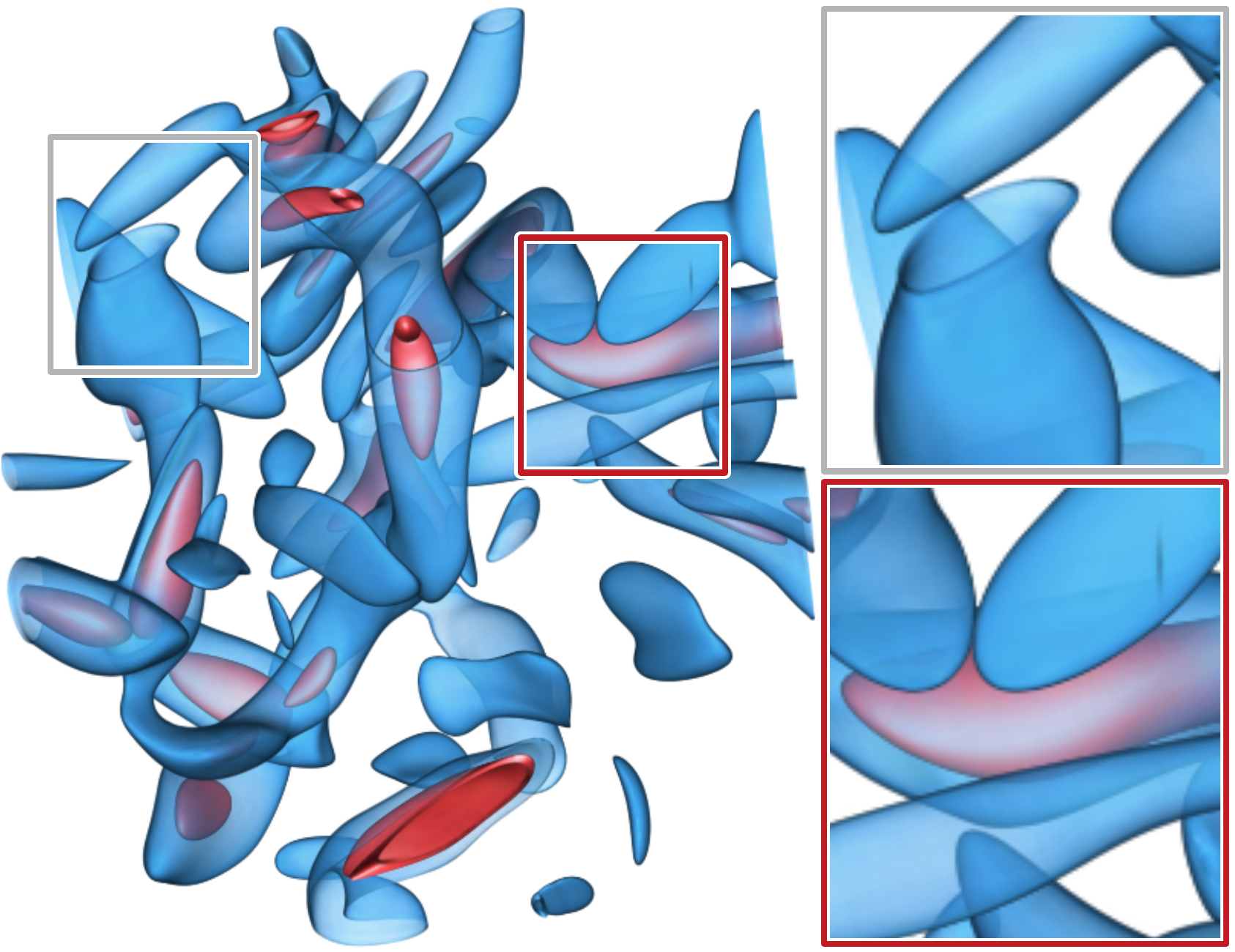}\\
\mbox{\footnotesize (c) blue: $M_2$; red: $M_2$} & \mbox{\footnotesize (d) blue: $M_2$; red: $M_1$}
\end{array}$
\end{center}
\vspace{-.25in} 
\caption{Different iVR-GS models rendering results. 
(a) and (c) use the same material for each basic scene. 
(b) and (d) use different materials for each basic scene.}
\label{fig:different-materials}
\end{figure}

{\bf Number of TFs.}
Increasing the number of basic TFs will not influence the rendering quality of the VolVis scene, as each basic iVR-GS model is trained and inferred independently.
However, increasing the number of basic models will result in a larger storage for the composed iVR-GS model.
To address this issue, we employ a VQ-based method to compress each basic model without sacrificing reconstruction accuracy, enhancing the overall scalability of our approach (i.e., including more basic TFs under the same storage budget).
To demonstrate the effectiveness of VQ compression and the scalability of the iVR-GS, we evaluate the model size of the composed iVR-GS model under different numbers of TFs using the combustion dataset.
For each number of basic TFs, we sample those TFs uniformly over the whole voxel value range to ensure both content-rich and content-poor parts are included.
Figure~\ref{fig:scalability} shows the model size curves.
We can observe that employing VQ can significantly improve the scalability of iVR-GS.
When keeping the model size around 280 MB, iVR-GS with VQ can include nearly 4$\times$ the number of TFs than iVR-GS without VQ.

{\bf Comparison with one-stage training.}
During the optimization of basic iVR-GS, we leverage a two-stage training strategy, where we first train a base 3DGS model to capture the basic scene geometry and then optimize editable Gaussian primitives to enable editing of the basic scene.
Compared with directly optimizing editable Gaussian primitives from scratch (i.e., one-stage training) for a basic scene, fitting editable primitives on the basic scene geometry represented by a base 3DGS model can significantly accelerate the optimization.
To demonstrate the superior training speed of this two-stage scenario, we compare it with the one-stage training strategy using the five-jet and supernova datasets.
During optimization, two-stage training utilizes 30,000 iterations for base 3DGS model training and 10,000 for editable Gaussian optimization, while one-stage training leverages 40,000 iterations to fit editable Gaussian primitives from scratch.
As shown in Figure~\ref{fig:vs-one-stage} and Table~\ref{tab:vs-one-stage}, although both one-stage and two-stage can reconstruct the VolVis scene with similar quality, one-stage training tends to use more primitives and incur larger file sizes to represent the scene while requiring over 2$\times$ training time compared to two-stage training. 
This is because one-stage training needs to fit scene geometry and appearance simultaneously, making it difficult for the model to represent the scene with fewer primitives, which leads to a larger file size and longer training time.

{\bf Basic scenes with different materials.}
To further demonstrate the flexibility of iVR-GS, we use the vortex dataset to showcase its ability to represent VolVis scenes where different materials are assigned to each basic scene. 
Such VolVis scenes with different materials cannot be directly rendered in ParaView using the NVIDIA IndeX plugin. 
For each basic TF, we render two basic scenes with two different materials (material $M_1$ with low shininess coefficient for high specular highlight and material $M_2$ with high shininess coefficient for low specular highlight) and optimize each basic scene with one iVR-GS model.
We then compose the two basic models corresponding to different basic TFs into one composed model to make the entire VolVis scene visible.
Figure~\ref{fig:different-materials} shows the novel view rendering results of different composed models.
In particular, Figure~\ref{fig:different-materials} (b) and (d) are VolVis scenes represented by the iVR-GS model, where different basic scenes are assigned different materials with various lighting effects.

\vspace{-0.05in}
\bibliographystyle{abbrv-doi-hyperref}
\bibliography{template}

\end{document}